\documentclass[twocolumn,linenumbers]{aastex62}

\graphicspath{{./}{figures/}}

\received{\today}
\revised{TBA}
\accepted{TBA}
\submitjournal{ApJ}

\shorttitle{Globular Clusters in BCGs}
\shortauthors{W. E. Harris}

\usepackage{bm}
\usepackage{amsmath}
\usepackage{threeparttable}
\usepackage[figuresright]{rotating}
\usepackage{tablefootnote}
\usepackage{mathrsfs}
\usepackage{graphicx} 
\usepackage{lineno}
\nolinenumbers
\usepackage{color}
\usepackage{fancyvrb}

\DefineVerbatimEnvironment{Highlighting}{Verbatim}{commandchars=\\\{\}}
\usepackage{framed}
\definecolor{shadecolor}{RGB}{248,248,248}

\begin{document}

\title{A Photometric Survey of Globular Cluster Systems in Brightest Cluster Galaxies} 

\correspondingauthor{William E. Harris}
\author[0000-0001-8762-5772]{William E. Harris}
\email{harris@physics.mcmaster.ca}
\affil{Department of Physics and Astronomy, McMaster University, Hamilton, ON L8S 4M1, Canada}

\begin{abstract}
\nolinenumbers
Hubble Space Telescope imaging for 26 giant early-type galaxies, all drawn from the MAST archive,  is used to carry out photometry of their surrounding globular cluster (GC) systems.  Most of these targets are Brightest Cluster Galaxies (BCGs) and their distances range from 24 to 210 Mpc.  The catalogs of photometry, completed with DOLPHOT, are publicly available.  The GC color indices are converted to [Fe/H] through a combination of 12-Gyr SSP (Single Stellar Population) models and direct spectroscopic calibration of the fiducial color index (F475W-F850LP).  All the resulting metallicity distribution functions (MDFs) can be accurately matched by bimodal Gaussian functions. The GC systems in all the galaxies also exhibit shallow metallicity gradients with projected galactocentric distance that average $Z \sim R_{gc}^{-0.3}$.  Several parameters of the MDFs including the means, dispersions, and blue/red fractions are summarized.  Perhaps the most interesting new result is the trend of blue/red GC fraction with galaxy mass, which connects with predictions from recent simulations of GC formation within  hierarchical assembly of large galaxies.  The observed trend reveals two major transition stages:  for low-mass galaxies, the metal-rich (red) GC fraction $f(red)$ increases steadily with galaxy mass, until halo mass $M_h \simeq 3 \times 10^{12} M_{\odot}$.  Above this point, more than half the metal-poor (blue) GCs come from accreted satellites and $f(red)$ starts declining.  But above a still higher transition point near $M_h \simeq 10^{14} M_{\odot}$, the data hint that $f(red)$ may start to increase again because the metal-rich GCs also become dominated by accreted systems.
\end{abstract}


\section{Introduction}\label{sec:intro}

Globular clusters (GCs), the old massive star clusters formed
in the early stages of galaxy assembly, are routinely 
found in all galaxies except the smallest dwarfs
\citep{harris2010,forbes+2018,beasley2020,eadie+2022}.  Thanks to the rapid development of theoretical models and simulations in recent years, a basic framework has now been established that follows GC formation in the gas-rich halos that build up during hierarchical assembly of galaxies
 \citep[among others, see][]{cote+1998,kravtsov_gnedin2005,muratov_gnedin2010,kruijssen2012,tonini2013,boylan-kolchin2017,pfeffer+2018,el-badry+2019,choksi+2018,choksi_gnedin2019,kruijssen+2019,reina-campos+2019,halbesma+2020}.

Confronting the impressive state of current theory with observations requires high quality data. Challenging observational questions about GC populations remain at both the low-mass and high-mass ends of the galaxy mass spectrum.
Much recent interest has focussed on the lowest-mass galaxies and their surprising diversity of GC populations \citep[e.g.][]{forbes+2018a,lim+2018,amorisco+2018,saifohalli+2021,trujillo-gomez+2021,eadie+2022,danieli+2022,carlsten+2022,vandokkum+2022}.  It is not yet clear, for example, if the near-linear proportionality between GC system mass and total galaxy mass that has been established from large galaxies \citep{blakeslee1997,spitler_forbes2009,hudson+2014,harris+2015,harris+2017,burkert_forbes2020} extends downward into the dwarf regime even in the presence of much scatter \citep{forbes+2018a,prole+2019,doppel+2021,eadie+2022,zaritsky2022}.  

But there is also a need for improved observational data aimed at the GC systems in the rare highest-mass galaxies.  These have the most complex history of assembly through mergers and accretions that happened intensely at high redshift but continue right up to (and beyond) the present day.  Traces of that history have been left behind in their GC systems including the GC mass distribution, their radial distribution throughout the halo, and their metallicity distribution including the proportions of `blue' (metal-poor) and `red' (metal-rich) clusters.  At the very highest galaxy masses, it is also not yet clear if the $M_{halo}-M_{GCS}$ relation is genuinely linear or displays curvature \citep[e.g.][]{harris+2017,boylan-kolchin2017,el-badry+2019,choksi_gnedin2019}.    

The Hubble Space Telescope (HST) MAST archive has been an extraordinarily valuable resource for work of this type.  Many giant galaxies lie in the distance range $\lesssim 200$ Mpc within which GC populations are readily imaged with reasonable exposure times with the ACS and WFC3 cameras.
Photometry of the GC systems for many such targets has been published in previous HST programs \citep{harris+2006,harris2009a,harris+2014,harris+2016,harris+2017a}, but more can be found in the Archive.  These are not all in the same color indices, so all the photometry needs to be put on an internally homogeneous basis in the process of data analysis.

The purpose of the present study is to provide more of the necessary data for an investigation of the largest galaxies.  The MAST Archive was searched for target galaxies that were at distances $\gtrsim 25$ Mpc (guaranteeing that their GCs would be near-starlike in morphology and permitting stellar photometry codes to be used, as described below); with imaging in at least two optical/near-IR filters; and with exposure times deep enough to cover at least the bright half of the standard GC luminosity function with $SNR \gtrsim 5$ (effectively, one or more orbits per filter).  This search yielded the 26 galaxies listed in Table \ref{tab:galaxies}, all imaged with the ACS Wide Field Camera.  Most of these are BCGs (Brightest Cluster Galaxies), but some are in smaller galaxy groups.\footnote{There are roughly 20 more in a similar distance range imaged with WFC3/UVIS, and these will be the subject of an upcoming study.}  In this paper, the photometry of the GC populations around these galaxies is presented along with a first look at their metallicity distributions in particular.

Organization of the paper is as follows:  Section 2 describes details of the photometric measurement and completeness tests, and Section 3 the resulting color-magnitude diagrams for the GCs in each galaxy and the color distributions. Section 4 describes tests of the effect of any marginal resolution of the GCs on their measured colors.  Section 5 lays out a preliminary conversion of the different color indices into [Fe/H] metallicities, derived from a combination of SSP models and observational spectroscopic calibration, while Section 6 shows the metallicity distributions for all galaxies now put onto a common system. A brief analysis of the distribution shapes and parameters is shown, with perhaps the most intriguing result showing up in the correlation of the relative numbers of blue vs.\ red GCs as a function of galaxy mass.  Section 7 ends with a summary of the findings and prospects for future work.


\begin{table*}[ht]
\centering
\caption{List of Program Galaxies} 
\label{tab:galaxies}
\begin{tabular}{llccccccc}
  \hline \hline
  Galaxy & Environment & $d$ & $M_V^T$ & $A_V$ & $R_e$ & HST & Blue Filter & Red Filter \\
   & & (Mpc) & (mag) & (mag) & (kpc) & Program & (sec) & (sec) \\
  (1) & (2) & (3) & (4) & (5) & (6) & (7) & (8) & (9) \\
   \hline
   \\
   NGC1129 & AWM7 & 71 & $-22.3:$ & 0.309 & 24.2 & 13698 & F435W (2416s) & F606W (2288s) \\
   NGC1132 & N1132 Group & 101 & $-22.87$ & 0.176 & 16.5 & 10558 & F475W (7800s) & F850LP (9630s) \\
   NGC1272 & Perseus/Abell 426 & 74 & $-23.09$ & 0.176 & 20.5 & 10201 & F555W (2368s) & F814W (2260s)\\
   NGC1275 & Perseus/Abell 426 & 74 & $-22.74$ & 0.447 & 6.1 & 15235 & F475W (2436s) & F814W (2325s) \\
   NGC1278 & Perseus/Abell 426 & 74 & $-21.46$ & 0.452 & 8.8 & 15235 & F475W (2577s) & F814W (2429s) \\
   NGC1407 & Eridanus/N1407 & 23 & $-22.04$ & 0.187 & 7.8 & 9427 & F435W (1500s) & F814W (680s) \\
   NGC3258 & Antlia/Abell S0636 & 45 & $-21.95$ & 0.221 & 6.5 & 9427 & F435W (5360s) & F814W (2280s) \\
   NGC3268 & Antlia/Abell S0636 & 45 & $-21.95$ & 0.281 & 7.8 & 9427 & F435W (5360s) & F814W (2280s) \\
   NGC3348 & CfA69 & 42 & $-22.20$ & 0.204 & 5.3 & 9427 & F435W (7200s) & F814W (2120s) \\
   NGC4696 & Cen30/Abell 3526 & 46 & $-23.4:$ & 0.306 & 11: & 9427 & F435W (5440s) & F814W (2320s) \\
   NGC4874 & Coma/Abell 1656 & 103 & $-23.37$ & 0.025 & 20.1 & 10861,11711 & F475W (5071s) & F814W (11825s) \\
   NGC4889 & Coma/Abell 1656 & 103 & $-23.69$ & 0.025 & 15.0 & 11711,14361 & F475W (7138s) & F814W (11098s) \\
   NGC5322 & CfA122 & 27 & $-22.00$ & 0.038 & 4.4 & 9427 & F435W (3390s) & F814W (820s) \\
   NGC5557 & CfA141 & 38 & $-22.33$ & 0.016 & 5.5 & 9427 & F435W (5260s) & F814W (2400s) \\
   NGC6166 & Abell 2199 & 130 & $-23.65$ & 0.031 & 30.0 & 12238 & F475W (5370s) & F814W (4885s) \\
   NGC7049 & N7049 Group & 30 & $-21.84$ & 0.153 & 3.6 & 9427 & F435W (3480s) & F814W (1200s) \\
   NGC7626 & Pegasus/LGC473 & 44 & $-22.06$ & 0.197 & 8.2 & 9427 & F435W (7720s) & F814W (2600s) \\
   NGC7720 & Abell 2634 & 130 & $-23.39$ & 0.194 & 19.4 & 12238 & F475W (5282s) & F814W (5278s) \\
   IC4051 & Coma/Abell 1656 & 103 & $-21.68$ & 0.025 & 4.7 & 12918 & F475W (2569s) & F814W (1310s) \\
   UGC9799 & Abell 2052 & 154 & $-23.2:$ & 0.102 & 14: & 12238 & F475W (7977s) & F814W (5253s) \\
   UGC10143 & Abell 2147 & 151 & $-22.9:$ & 0.086 & 23.0 & 12238 & F475W (10726s) & F814W (5262s) \\
   ESO306-G017 & Abell S0540 & 154 & $-23.63$ & 0.090 & 18: & 10558 & F475W (7059s) & F850LP (8574s) \\
ESO325-G004  & Abell S0740 & 149 & $-23.10$ & 0.166 & 11: & 10710,10429 & F475W (5901s) & F814W (18078s) \\
   ESO383-G076 & Abell 3571 & 171 & $-24.25$ & 0.149 & 16: & 10429,12238 & F475W (10830s) & F814W (21081s) \\
   ESO444-G046 & Abell 3558 & 210 & $-23.81$ & 0.137 & 17: & 10429,12238 & F475W (20282s) & F814W (35426s) \\
   ESO509-G008 & Abell 1736 & 153 & $-23.04$ & 0.144 & 9: & 10429,12238 & F475W (10758s) & F814W (17310s) \\
   \\
\hline
\end{tabular}
\begin{tablenotes}
\item{} \textit{Key to columns:} (1) Galaxy identification; (2) Group or cluster membership;
(3) Distance $d$, calculated as $d = cz/H_0$ where $cz$ is the group redshift relative to the
CBR frame and $H_0 = 70$ km s$^{-1}$ Mpc$^{-1}$; (4) Integrated $V-$band luminosity; (5) Foreground
extinction in $V$; (6) Effective radius of the galaxy light profile in the $V-$band; (7) ID number(s) of the original HST GO program(s) from which the imaging data are
taken; (8,9) the filters and exposure times of the raw observations.  Unless otherwise noted all images are taken with the ACS/WFC camera.  Values for $M_V^T$, $A_V$, and $R_e$ are drawn from NED.
\end{tablenotes}
\end{table*}

\section{Measurement Procedures}\label{sec:data}

Photometry was carried out with DOLPHOT \citep{dolphin2000}, a widely used
package particularly designed for use with the HST ACS and WFC3 cameras.  DOLPHOT starts by using a deep `reference image' 
for the target field for the purpose of object detection and location:   the candidate stars found from
the reference image are then measured on every individual \emph{*.flc} exposure and the results
averaged to give final magnitudes.   For the galaxy fields
studied here, this deep image was constructed with \emph{astrodrizzle} in most cases from the summed
exposures in the redder filter (usually F814W or F850LP).

DOLPHOT contains a large number
of adjustable parameters for object detection, aperture photometry, PSF fitting, and other
steps in the process \citep[see, e.g.][for much more extensive discussion]{dalcanton+2012,williams+2014,cohen+2020}.
For all the galaxies studied here, many of these parameter choices are not critically sensitive
since the fields are all quite uncrowded in any absolute sense (even though the galaxy may be
surrounded by many thousands of its GCs, this still leaves typically $10^{2-3}$ resolution
elements per object).  A list of some of the key
parameters adopted here is given in Table \ref{tab:dolphot}.  These same parametric values were used for every field to maximize homogeneity.

Throughout this study, the magnitudes are in the natural ACS/WFC filter
system and on the Vegamag magnitude scale.

\begin{figure*}
    \centering
    \includegraphics[width=0.48\textwidth]{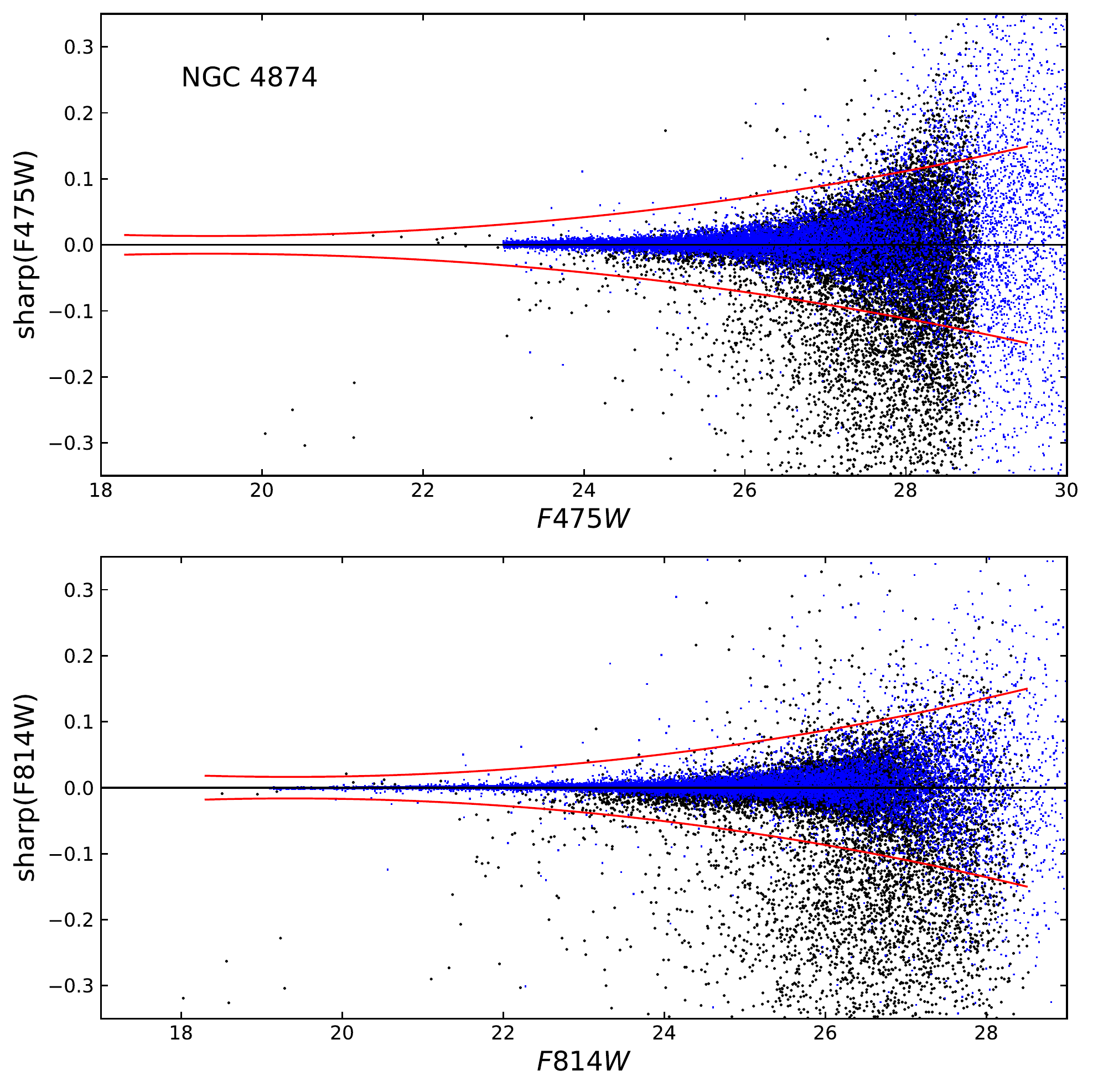}
    \includegraphics[width=0.48\textwidth]{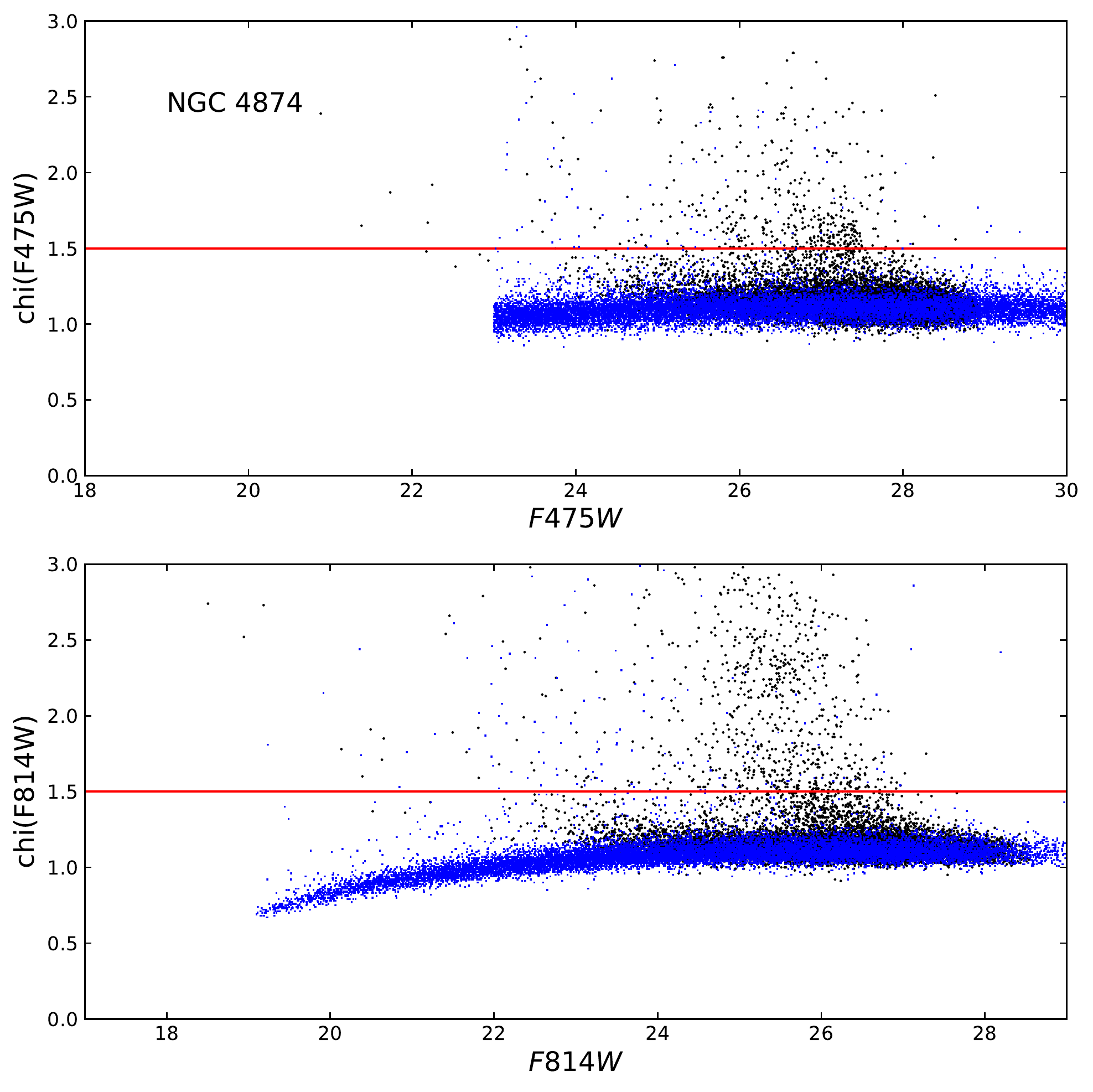}
    \caption{Use of DOLPHOT selection parameters \emph{sharp} and \emph{chi}
    for culling of nonstellar objects.  Black symbols are real objects on
    the field and superimposed blue symbols are artificial stars injected
    into the image.  Red lines mark the exclusion boundaries outside of
    which measured objects were rejected as distinguishably nonstellar.}
    \label{fig:sharp_chi}
\end{figure*}

\begin{figure}
    \centering
    \includegraphics[width=0.48\textwidth]{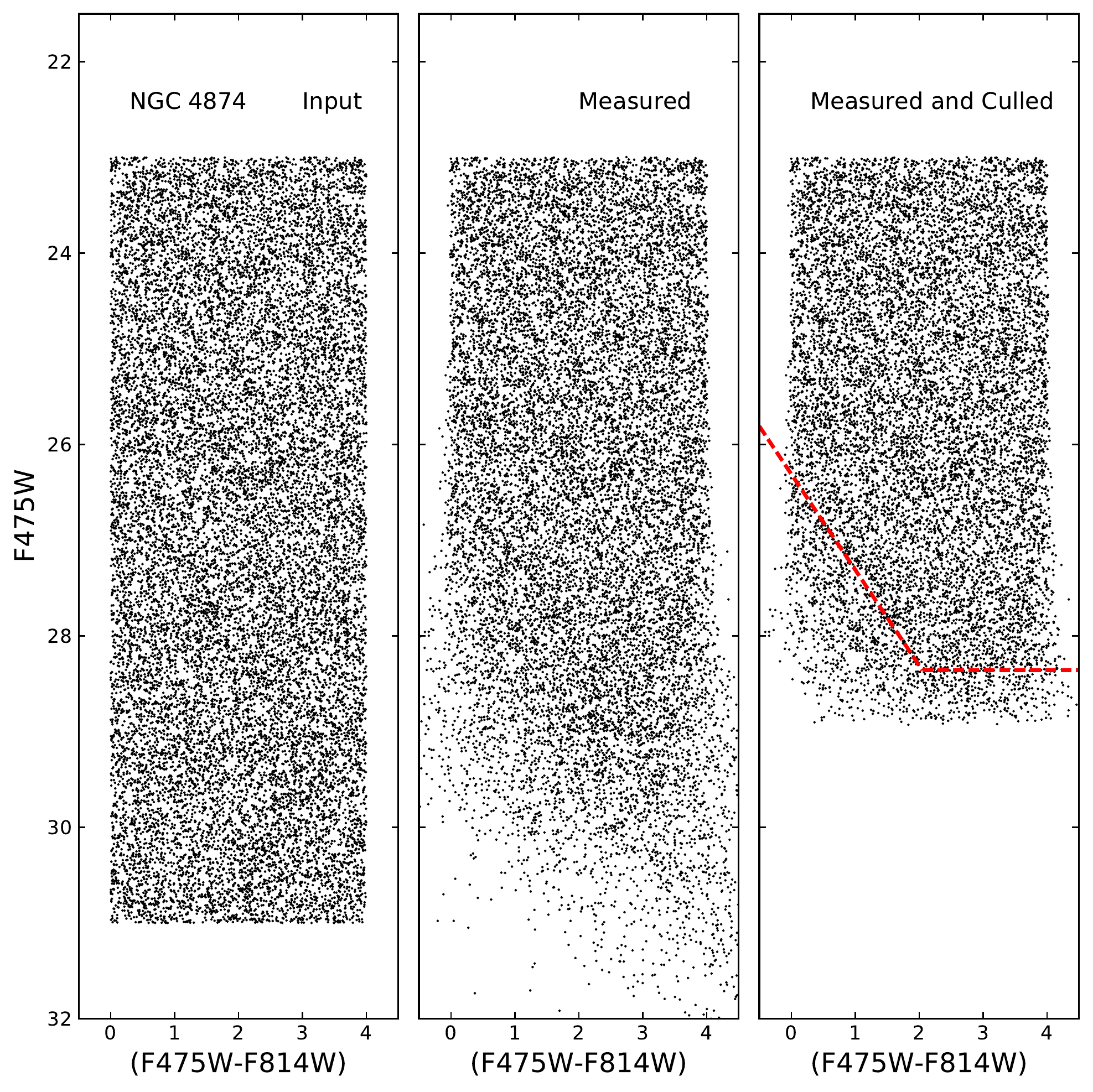}
    \caption{Artificial-star test with DOLPHOT for the NGC 4874 field.  The CMD
    for the input population of fake stars is shown in the left panel (note here the
    y-axis is the blue filter), while the CMD for the recovered stars before 
    any culling by SNR, \emph{chi, sharp} is in the middle panel.  
    The remaining objects after all culling
    steps are shown in the right panel.
    The 50\% detection completeness level is shown as the dashed red line.}
    \label{fig:cmdfake}
\end{figure}

\begin{table}[ht]
\centering
\caption{Selected DOLPHOT Parameters} 
\begin{tabular}{cc}
  \hline \hline
Parameter & Adopted Value  \\ 
   \hline
 RAper &  6.0 px \\
 RSky & 15, 25 px \\
 RPSF & 15 px \\
 aprad & 15 \\
 apsky & 25, 30 px \\
 SigFind & 3.0 \\
 FPSF & Lorentz \\
 FitSky & 2 \\
 PSFPhot & 1 \\
 SigPSF & 5 \\
 PSFStep & 0.25 px \\
                 \hline
\end{tabular}
\label{tab:dolphot}
\end{table}

Once the DOLPHOT run was finished, the list of measured objects was rigorously culled to remove
nonstellar or other unwanted objects.  In this section, the NGC 4874 field is used as a template
example to describe the details of this procedure.

In all target fields, the overwhelming source of
field contamination consists of small, faint background galaxies, whereas the GCs that are
the objects of this study are unresolved or near-starlike.  In the first round
of culling, objects classified as $n_{type} > 1$ (i.e., not `bright' stars) were
rejected, as were those with magnitudes $m = 99.999$ in either filter, or those with $SNR < 4.0$.
After these initial removals, the parameters \emph{sharp, chi} describing the goodness of fit
to the PSF were used as shown in Figure \ref{fig:sharp_chi}.  For \emph{sharp}, any objects 
lying outside the curved boundary lines shown in the figure were rejected.  A simple quadratic curve was used for
the boundary lines to reflect the gradually increasing random dispersion in the \emph{sharp} index 
towards fainter magnitude.  The artificial-star population (see below) was also used as a
check that the
limits were not excluding true starlike sources.
For \emph{chi}, a boundary line was set at a constant value and any objects falling above
the line were rejected.  In practice, \emph{sharp, chi}, and the photometric error are
well correlated (that is, objects with high {\it chi-}values usually have high {\it sharp} values
and high photometric uncertainties as well).  As can be seen in Fig.~\ref{fig:sharp_chi}, the
measured objects remaining after the first basic culling by $n_{type}$, SNR, and magnitude
are already dominated by the starlike sources that include the GC candidates, allowing the
exclusion boundaries for \emph{chi, sharp} to be placed conservatively.

\begin{figure}
    \centering
    \includegraphics[width=0.35\textwidth]{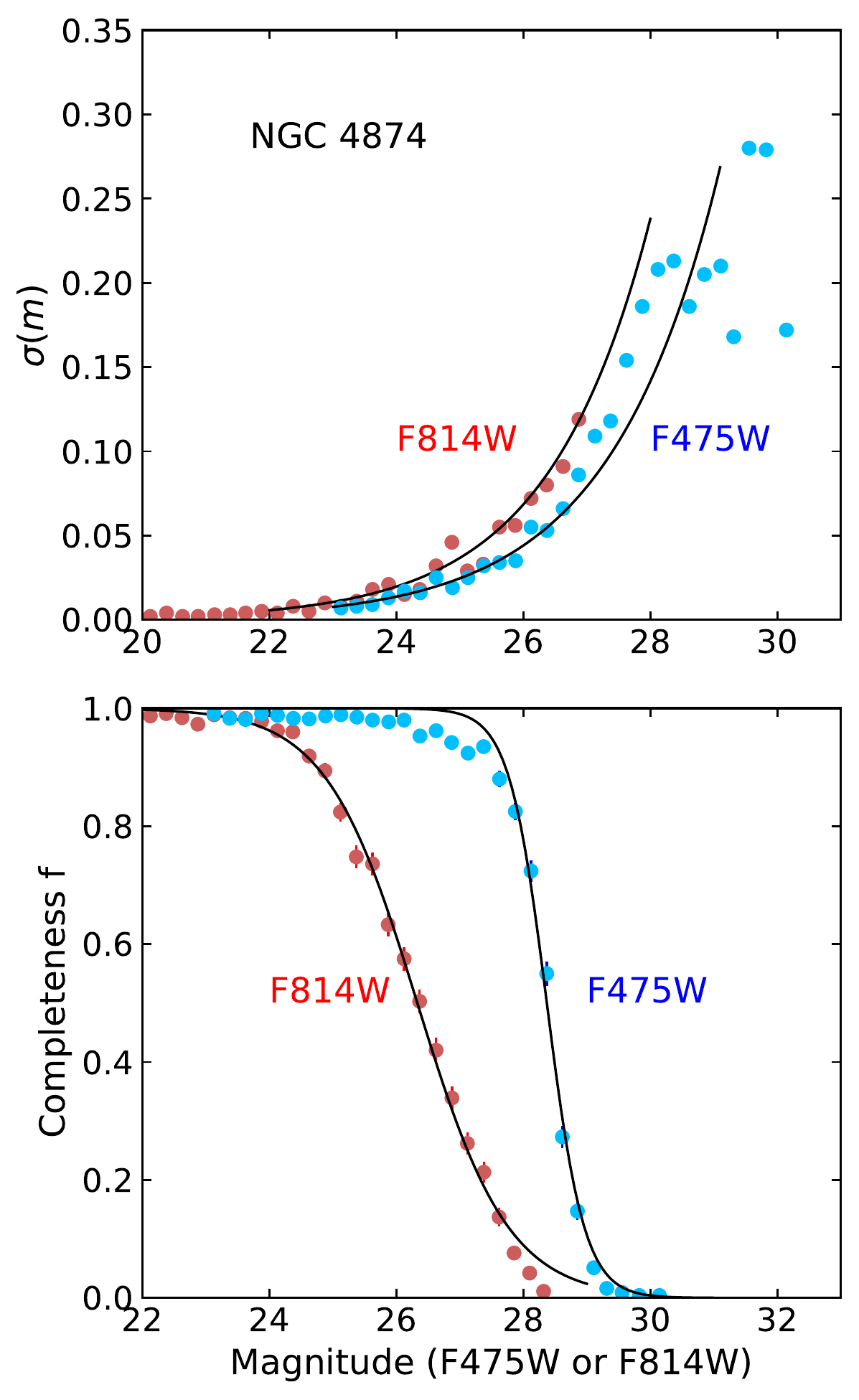}
    \caption{\emph{Upper panel:} Photometric measurement uncertainty
    versus magnitude, determined from the artificial-star tests.  The interpolation lines follow the equation given in the text.
    \emph{Lower panel:} Photometric detection completeness curves versus magnitude
    as determined from artificial-star tests.  This example is for NGC 4874.
    The sigmoid curves fitted to each of the two filters follow the equation given in the text.}
    \label{fig:completeness}
\end{figure}

\begin{figure*}
    \centering
    \includegraphics[width=0.45\textwidth]{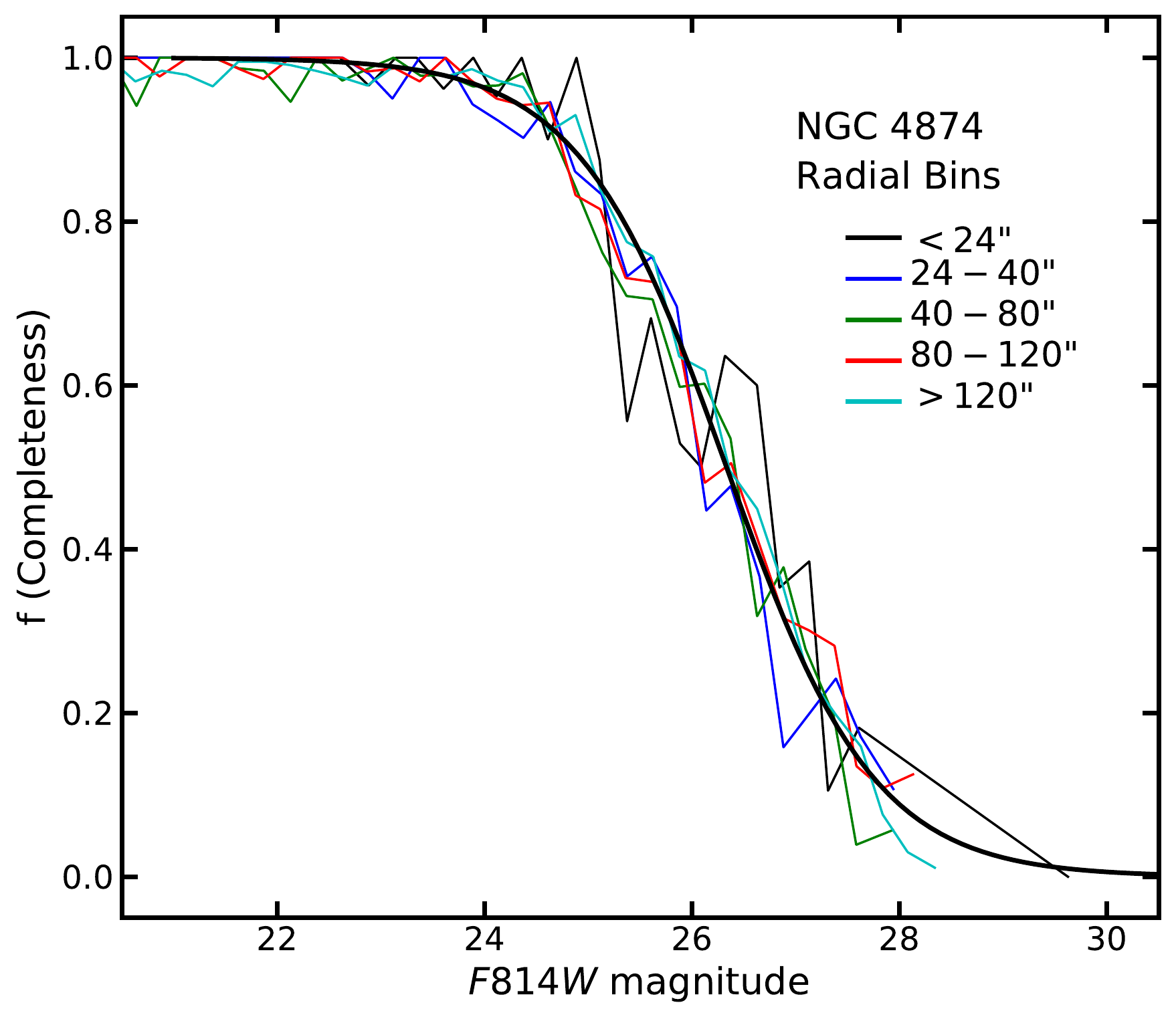}
    \includegraphics[width=0.45\textwidth]{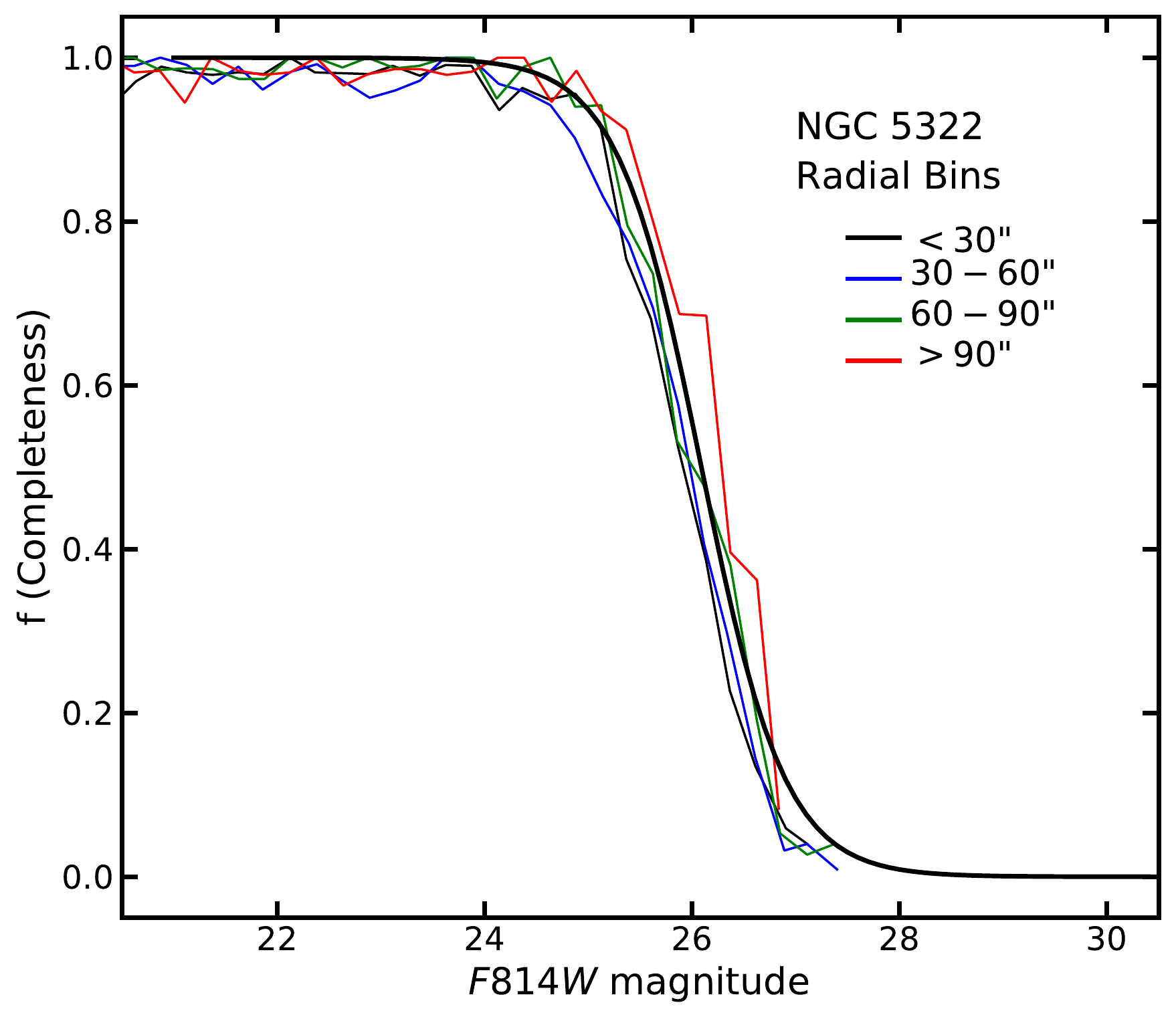}
    \caption{\emph{Left panel:} Detection completeness as a function of F814W magnitude, now plotted in five radial zones around NGC 4874.  The \emph{heavy solid curve} shows the solution for all zones combined.
     For comparison, the effective radius of the galaxy light profile is $r_e = 40''$ for NGC 4874.  \emph{Right panel:} The same material for NGC 5322, a less luminous and more centrally concentrated galaxy.  In this case, $r_e = 33''$.}
    \label{fig:f_radial}
\end{figure*}

\begin{figure*}
    \centering
    \includegraphics[width=0.8\textwidth]{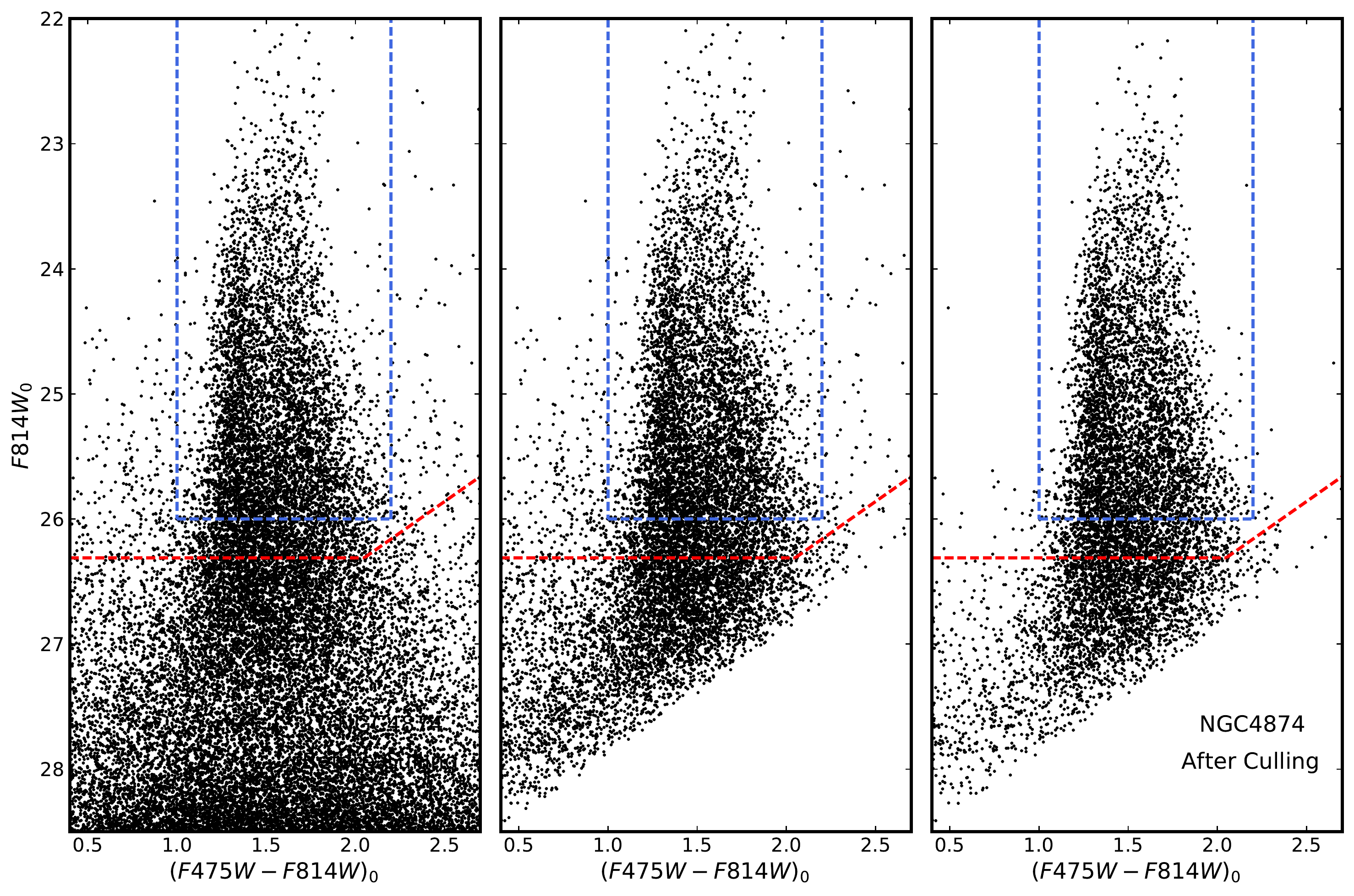}
    \caption{Color-magnitude diagram for the NGC 4874 field.  \emph{Left panel:} The CMD
    from all the DOLPHOT measurements before any culling steps.  \emph{Middle panel:} 
    The CMD after removal of objects with $ntype > 1$ (nonstellar), magnitude = 99.999 in either filter, and $SNR < 4$.
    \emph{Right panel:} The CMD after final additional culling by \emph{ chi} and 
    \emph{sharp}.  In all panels, the red dashed line
    indicates the 50\% detection completeness level, while the blue lines
    mark out the objects used to define the color and metallicity histograms
    for the globular cluster population.}
    \label{fig:ngc4874_cmd}
\end{figure*}

Photometric measurement uncertainty (rms scatter) $\sigma(m)$ and completeness of detection
$f(m)$ were determined from artificial-star tests run with the DOLPHOT tools.  Typically
10000-20000 fake stars per galaxy were used, covering the magnitude and color ranges of
the real data.  DOLPHOT adds and measures these one at a time so the degree of crowding is never affected. Figure \ref{fig:cmdfake}
shows the color-magnitude diagram (CMD) for the population of fake stars inserted into the NGC 4874 image (left panel),
the CMD for the ones that were recovered and measured before any culling (middle panel), and finally
the ones remaining after all the same culling steps that were applied to the
real data (right panel).

The measurement uncertainty $\sigma(m)$ and the detection completeness $f(m)$ were also
determined from these artificial-star tests.
A convenient and simple function that provides a good approximation to the increase of $\sigma$ with magnitude is
\begin{equation}
    \sigma(m) = \beta_1 e^{\beta_2(m - \beta_3)} .
\end{equation}
while the falloff of completeness $f$ with magnitude can be described by a sigmoid curve 
\citep{harris+2016}
\begin{equation}
    f(m) = 1 / (1 + e^{\alpha(m-m_0)})
    \label{eq:fcurve}
\end{equation}
where $m_0$ is the magnitude at which completeness is 50\% and $\alpha$ measures the steepness of decline of the $f-$curve.  Both of these trends are shown in Figure \ref{fig:completeness} for our
template case of NGC 4874:  the particular values for the various parameters 
($\beta_1, \beta_2, \beta_3, \alpha, m_0$) are listed for all the fields in 
Table \ref{tab:completeness}. These differ from field to field depending on exposure times and filters, 
but the same shapes for the fitting formulae hold.  The best-fit estimates for $\beta_2$ are consistently near $0.6 \pm 0.1$.

Completeness of detection will at some level depend on galactocentric location, since objects become progressively harder to detect against the higher background light and sky noise in the inner regions of the galaxy.  However, a feature of the giant galaxies studied here that mitigates this dependence is that they are much more diffuse in structure than smaller galaxies, along with much lower central surface brightnesses \citep[e.g.][]{kormendy+2009}.  In addition, the great majority of the GCs measured in this study are drawn from the mid- to outer-halo regions that are well outside the  $\lesssim 1 R_e$ central region where the bulge light is brightest.  Figure \ref{fig:f_radial} shows the completeness curves obtained from the artificial-star runs in NGC 4874 within five different radial zones, {\bf and also for NGC 5322, a less luminous galaxy with smaller effective radius, in four radial zones}.  Though there is a  trend for the 50\% completeness level $m_0$ to become fainter with increasing radius, the trend is small, and as will be seen below it results in no important effects on the final GC color distributions.

The measurements after various stages of culling are shown in the CMDs of Figure \ref{fig:ngc4874_cmd}, which shows dereddened colors and magnitudes.  In the left panel, all measured objects found by DOLPHOT are shown:  the
GC red and blue vertical sequences are already quite prominent, but much faint contamination is present. 
The objects remaining after rejection by $n_{type}$ and SNR are shown in the middle panel, and the ones remaining after final culling
by \emph{chi, sharp} are shown in the right panel.  The final step effectively removes only some very red or blue objects well away from the GC sequences, plus a larger number of faint blue objects
that are below the completeness limit.  With a single exception (namely NGC 1275; see below),
no young population of massive star clusters that would appear on the luminous blue side of the CMD is evident in any of the target galaxies.

\begin{figure}
    \centering
    \includegraphics[width=0.48\textwidth]{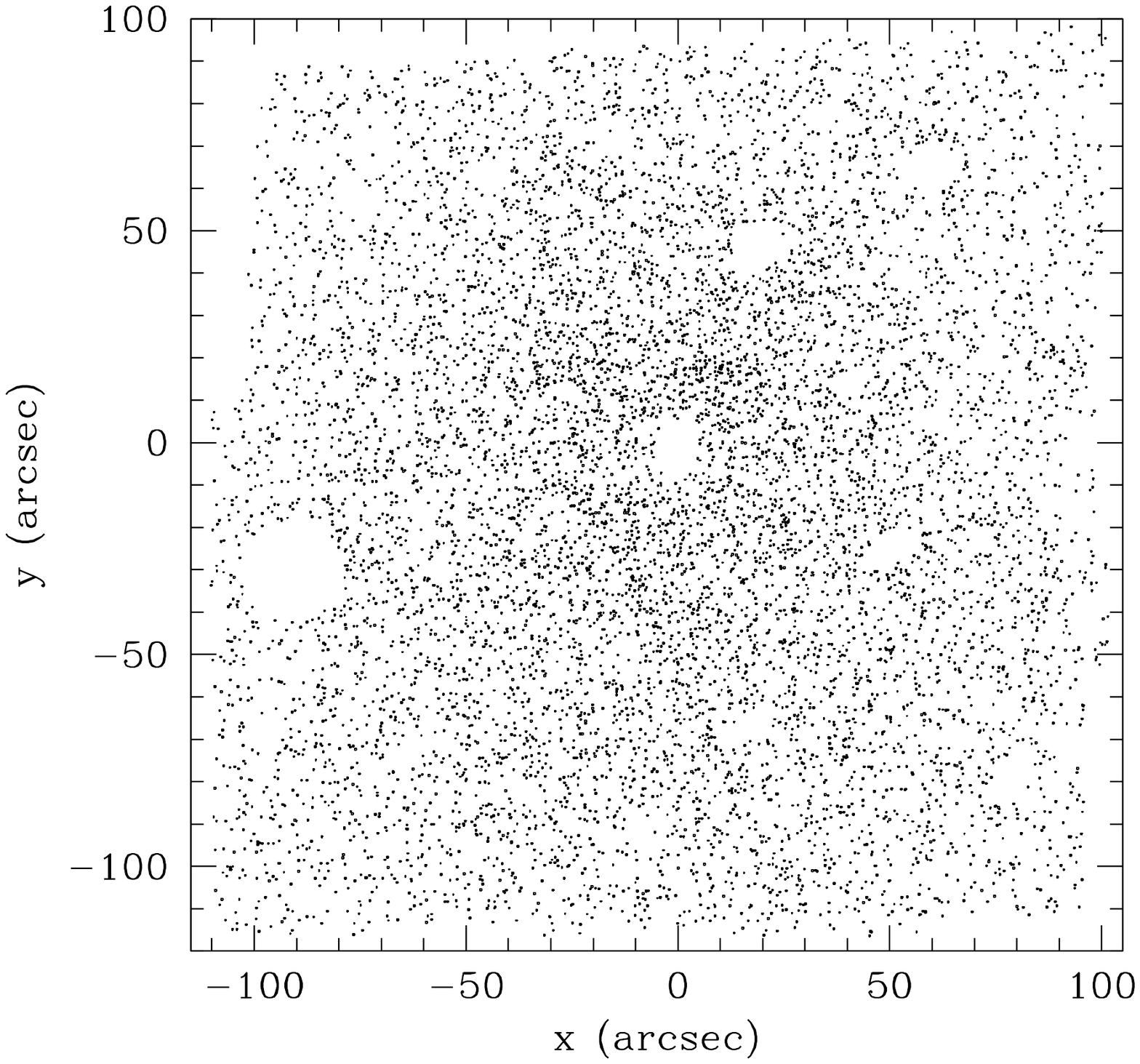}
    \caption{Distribution of GC candidates brighter than F814W=26.0
    around NGC 4874.  Holes in the distribution indicate satellite galaxies
    removed from the data list.}
    \label{fig:ngc4874_xy}
\end{figure}


The spatial distribution of GC 
candidates with $F814W < 26$ across the NGC 4874 field is shown in Figure \ref{fig:ngc4874_xy}. The GC population now totally
dominates the numbers of objects brighter than this level.

Final lists of the photometry for each galaxy include (a) object positions (x, y on the reference image, and J2000 RA and Dec); (b) magnitude and uncertainty in each filter; and (c) DOLPHOT output parameters including $\chi$, SNR, sharpness, roundness, and crowdedness in each filter.  An example is
shown in Table \ref{tab:photom} and all data, along with the reference images, can be obtained at the
DOIs listed in the Acknowledgements below, or from the author's website at
\emph{https://physics.mcmaster.ca/Fac\_Harris/Supergiants.html}.

\section{Color-Magnitude Diagrams and the Color Distribution Functions}

\begin{figure*}
    \centering
    \includegraphics[width=0.90\textwidth]{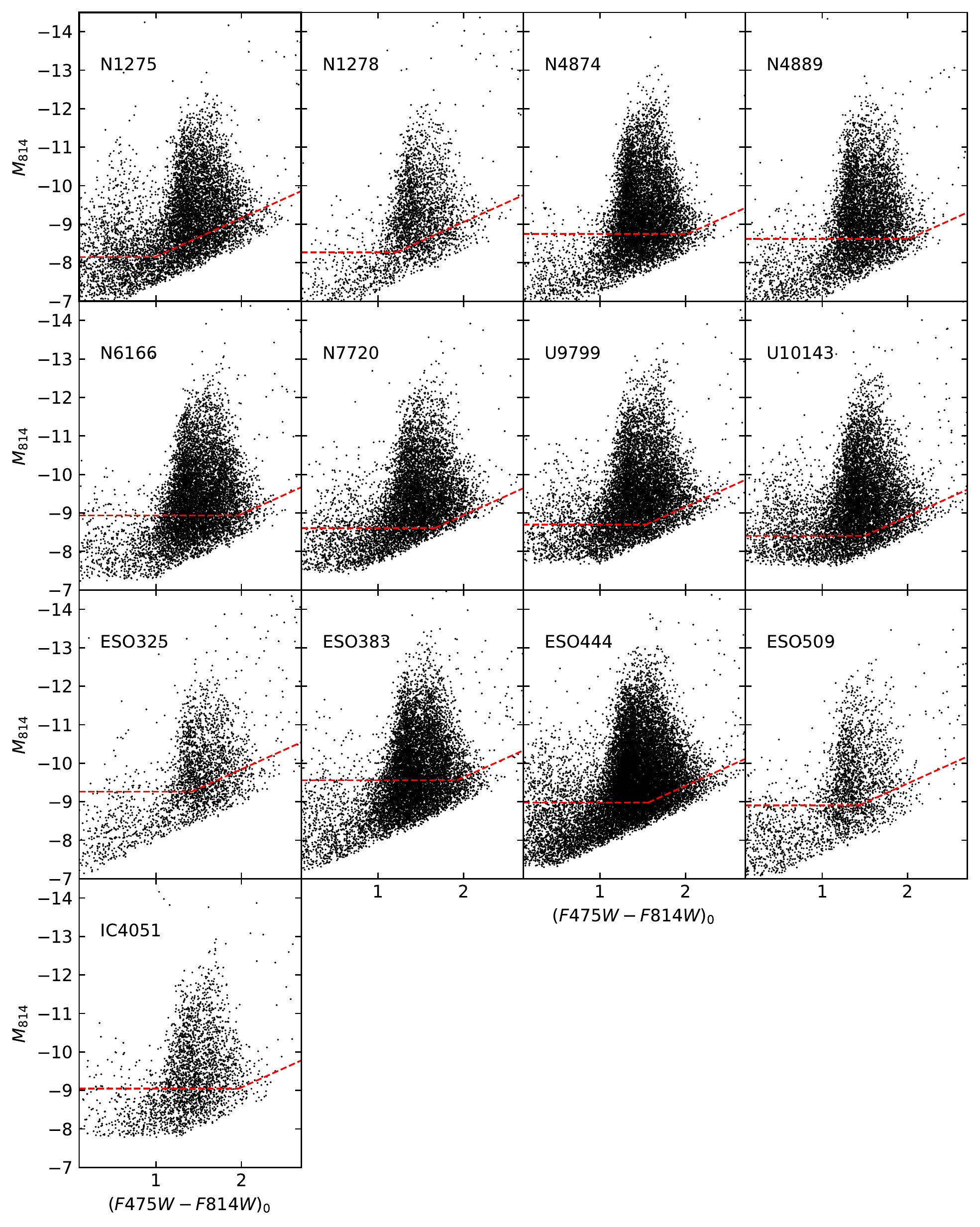}
    \caption{Color-magnitude diagrams for the GC populations around the 13 galaxies
    measured in the (F475W, F814W) filters.  Data are plotted as absolute
    magnitude $M_{814}$ versus intrinsic color for easier comparison with the next two figures.  The dashed red lines show the 50\% completeness limits as determined from artificial-star tests.} 
    \label{fig:cmd1}
\end{figure*}

\begin{figure*}
    \centering
    \hspace*{-3cm}\includegraphics[width=0.75\textwidth]{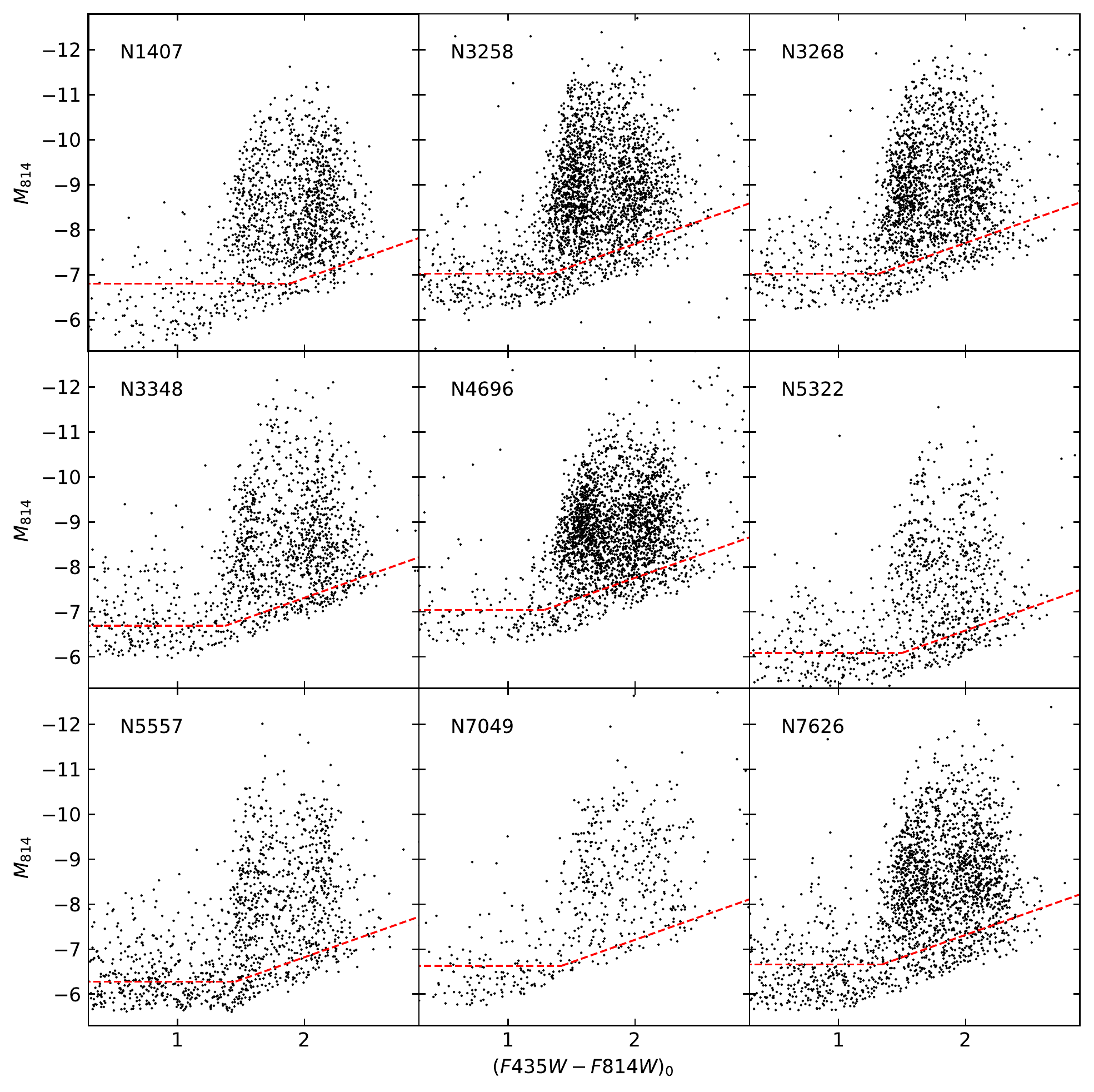}
    \caption{Color-magnitude diagrams for the GC populations around the nine galaxies
    measured in the (F435W, F814W) filters.}
    \label{fig:cmd2}
\end{figure*}

\begin{figure*}
    \centering
    \includegraphics[width=0.70\textwidth]{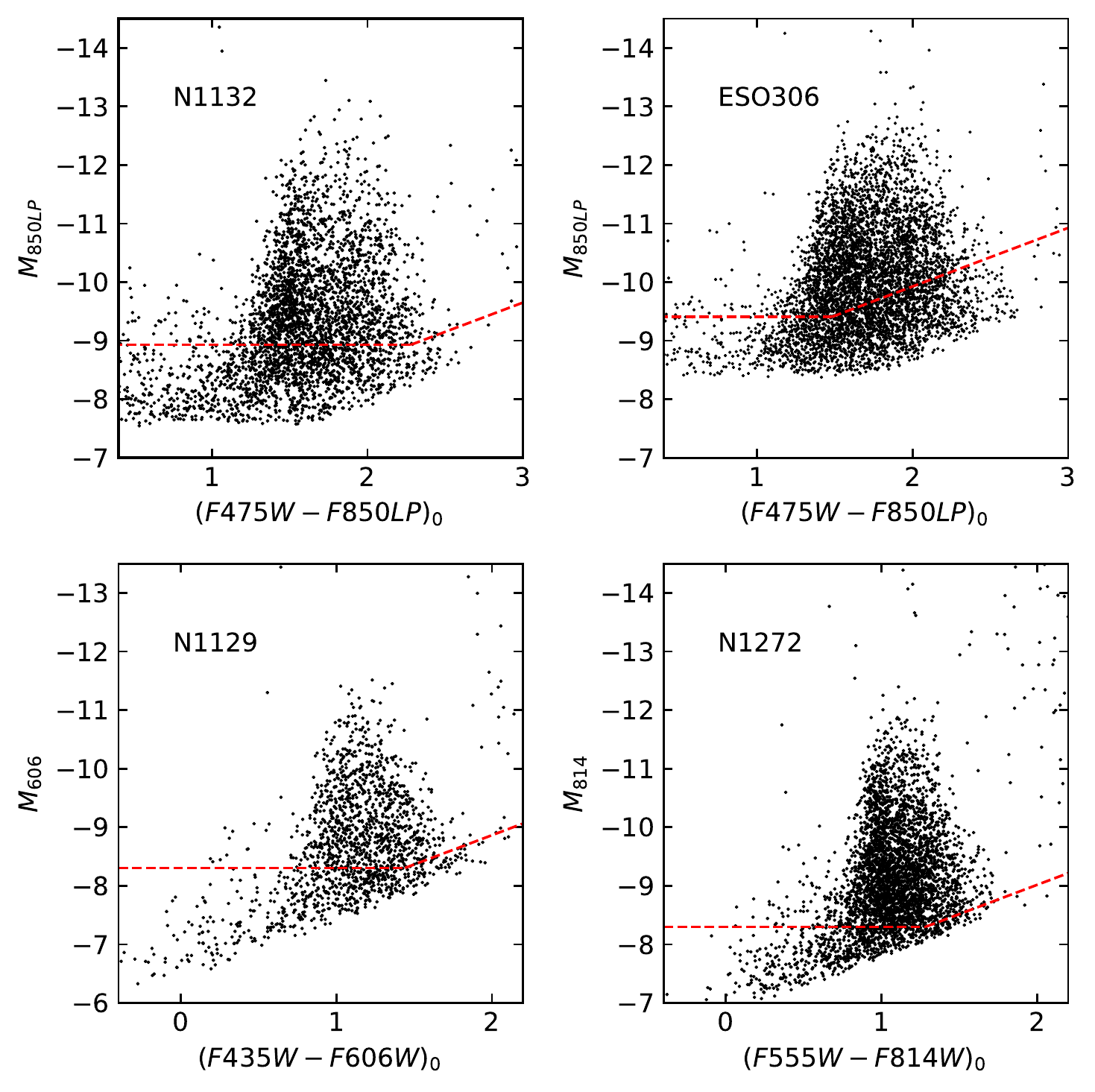}
    \caption{Color-magnitude diagrams for the GC populations around four galaxies
    measured in other filter combinations.  }
    \label{fig:cmd3}
\end{figure*}

\begin{figure*}
    \centering
    \includegraphics[width=0.80\textwidth]{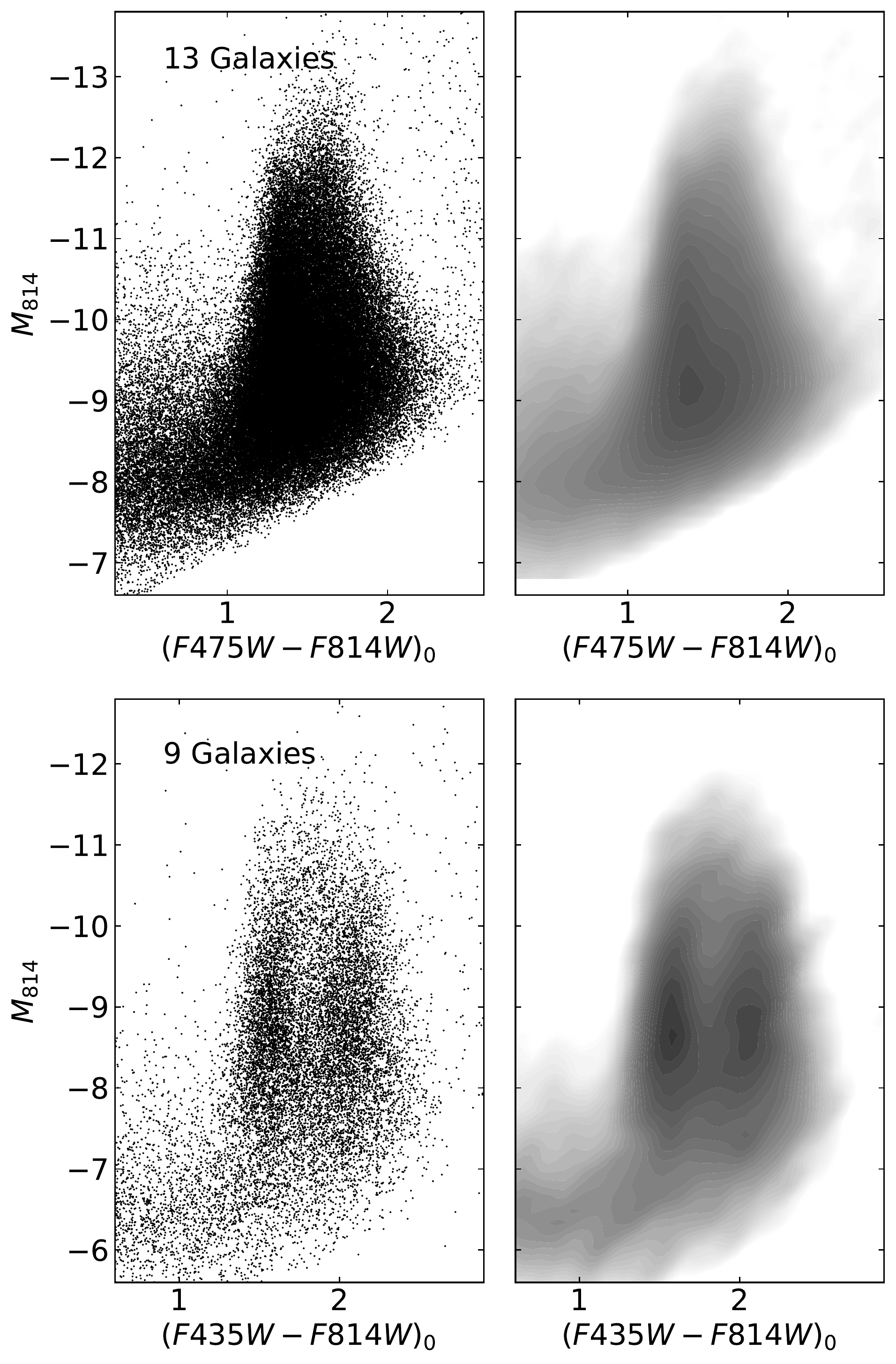}
    \caption{\emph{Upper panels:} CMD for all 13 galaxies measured
    in (F475W, F814W) combined, along with its Hess-diagram contour version.
    \emph{Lower panels:} Combined CMD for all 9 galaxies measured in
    (F435W, F814W).  For the Hess diagrams, contour levels are logarithmic.}
    \label{fig:cmd_hess}
\end{figure*}

The resulting CMDs are shown in Figures \ref{fig:cmd1}, \ref{fig:cmd2}, and \ref{fig:cmd3}.  The 26 program galaxies are grouped by color index for easier intercomparison.  Galaxies measured in the two indices $(F475W-F814W)$ and $(F435W-F814W)$ account for 22 of the 26 galaxies in the program.  In most cases, the familiar
blue and red sequences are fairly clearly visible, and once the colors are dereddened, these
sequences fall at very similar colors from one galaxy to another.  In Figure \ref{fig:cmd_hess},
the top two panels show the combined CMD for all 13 galaxies that were measured in 
$(F475W, F814W)$ in the form of absolute magnitude versus intrinsic color.  The same data
are shown in the right-hand panel in Hess-diagram form as a smoothed contour plot.
The lower pair of panels shows the same for all 9 galaxies measured in $(F435W, F814W)$.
The galaxies in the upper panels are on average the most luminous ones in the whole list,
and not surprisingly with the largest GC populations.  

One notable feature of Fig.~\ref{fig:cmd_hess}
that is also visible in several of the individual CMDs is the upward extension of the red,
more metal-rich sequence beyond $M_I \lesssim -12$.  For a mass-to-light ratio of 2, 
this range corresponds to $M \gtrsim 5 \times 10^6 M_{\odot}$ and certainly reaches
well into the luminosity range of the Ultra-Compact Dwarfs (UCDs).
This regime includes the ``superluminous'' GCs discussed in \citet{harris+2014}, who demonstrate
that there are systematically more of these objects than would be expected from a normal
statistical extension of the standard lognormal GC luminosity function.  Notably, in the
lower panels of Fig.~\ref{fig:cmd_hess}, which include less luminous galaxies, no such 
red-sequence extension is seen.  Further discussion and interpretation of these points is given
in \citet{harris+2014}.

\begin{table*}[ht]
\centering
\caption{Photometric Parameters} 
\begin{tabular}{lcccccccc}
  \hline \hline
Galaxy & $m_0$(blue) & $\alpha$(blue) & $m_0$(red) & $\alpha$(red) & $\beta_1, \beta_2, \beta_3$(blue) & $\beta_1, \beta_2, \beta_3$(red) & Color Range & $m(lim)$  \\ 
(1) & (2) & (3) & (4) & (5) & (6) & (7) & (8) & (9) \\
   \hline
   \\
   &  \multicolumn{2}{c}{F475W} & \multicolumn{2}{c}{F814W} & F475W & F814W & (F475W-F814W)$_0$ & F814W$_0$ \\  
NGC1275 & 27.72 & 1.90 & 26.45 & 1.57 & 0.15, 0.68, 27.0 & 0.05, 0.62, 24.5 & 1.0 - 2.2 & 25.0 \\
NGC1278 & 27.83 & 2.40 & 26.33 & 1.60 & 0.09, 0.61, 27.0 & 0.04, 0.59, 24.5 & 1.0 - 2.2 & 25.0 \\
NGC4874 & 28.37 & 3.38 & 26.33 & 1.39 & 0.11, 0.58, 27.5 & 0.05, 0.62, 25.5 & 1.0 - 2.2 & 26.0 \\
NGC4889 & 28.49 & 3.53 & 26.46 & 1.28 & 0.08, 0.61, 27.0 & 0.04, 0.64, 25.5 & 1.0 - 2.2 & 26.0 \\
NGC6166 & 28.64 & 2.61 & 26.65 & 1.42 & 0.10, 0.68, 27.5 & 0.04, 0.65, 25.0 & 1.0 - 2.2 & 26.0 \\
NGC7720 & 28.86 & 9.62 & 27.07 & 2.78 & 0.09, 0.66, 27.5 & 0.04, 0.68, 25.0 & 1.0 - 2.2 & 26.5 \\
UGC9799 & 28.90 & 3.14 & 27.30 & 1.36 & 0.13, 0.66, 28.0& 0.05, 0.653, 25.5 & 1.0 - 2.2 & 26.5 \\
UGC10143 & 29.09 & 3.16 & 27.54 & 1.66 & 0.11, 0.69, 28.0 & 0.05, 0.63, 25.5 & 1.0 - 2.2 & 26.5 \\
ESO325-G004 & 28.22 & 1.67 & 26.70 & 1.03 & 0.14, 0.67, 28.0 & 0.04, 0.62, 25.5 & 1.0 - 2.2. & 26.0 \\
ESO383-G076 & 28.71 & 1.96 & 26.69 & 0.94 & 0.13, 0.65, 28.0 & 0.05, 0.56, 26.0 & 1.0 - 2.2 & 26.0 \\
ESO444-G046 & 29.36 & 3.40 & 27.71 & 1.39 & 0.10, 0.65, 28.0 & 0.04, 0.61, 25.5 & 1.0 - 2.2 & 27.0 \\
ESO509-G008 & 28.62 & 2.04 & 27.10 & 1.51 & 0.10, 0.64, 28.0 & 0.03, 0.67, 25.5 & 1.0 - 2.2 & 26.5 \\
IC4051 & 28.02 & 3.26 & 26.03 & 1.66 & 0.15, 0.59, 28.0 & 0.08, 0.59, 25.5 & 1.0 - 2.2. & 25.5 \\
\\
   &  \multicolumn{2}{c}{F435W} & \multicolumn{2}{c}{F814W} & F435W & F814W & (F435W-F814W)$_0$ & F814W$_0$ \\
NGC1407 & 27.14 & 3.04 & 25.11 & 1.57 & 0.08, 0.59, 26.0 & 0.10, 0.58, 25.0 & 1.15 - 2.70 & 24.4 \\
NGC3258 & 27.87 & 2.25 & 26.37 & 2.20 & 0.05, 0.69, 26.0 & 0.06, 0.66, 25.0 & 1.15 - 2.70 & 25.5 \\
NGC3268 & 27.93 & 2.58 & 26.40 & 2.69 & 0.04, 0.62, 26.0 & 0.06, 0.60, 25.0 & 1.15 - 2.70 & 25.5 \\
NGC3348 & 28.07 & 3.27 & 26.54 & 2.49 & 0.04, 0.62, 26.0 & 0.04, 0.61, 25.0 & 1.15 - 2.70 & 26.0 \\
NGC4696 & 27.96 & 2.64 & 26.44 & 2.30 & 0.04, 0.72, 26.0 & 0.06, 0.69, 25.0 & 1.15 - 2.70 & 25.5 \\
NGC5322 & 27.62 & 2.29 & 26.09 & 2.47 & 0.05, 0.62, 26.0 & 0.06, 0.53, 25.0 & 1.15 - 2.70 & 25.5 \\
NGC5557 & 28.10 & 2.55 & 26.63 & 2.34 & 0.04, 0.60, 26.0 & 0.05, 0.57, 25.0 & 1.15 - 2.70 & 26.0 \\
NGC7049 & 27.38 & 3.61 & 25.84 & 2.00 & 0.09, 0.73, 26.0 & 0.09, 0.64, 25.0 & 1.15 - 2.70 & 25.2 \\
NGC7626 & 28.16 & 2.37 & 26.67 & 2.15 & 0.07, 0.64, 27.0 & 0.06, 0.61, 25.0 & 1.15 - 2.70 & 26.0 \\
\\
&  \multicolumn{2}{c}{F475W} & \multicolumn{2}{c}{F850LP} & F475W & F850LP & (F475W-
F850LP)$_0$ & F850LP$_0$ \\
NGC1132 & 28.58 & 2.17 & 26.17 & 1.11 & 0.06, 0.61, 27.0 & 0.07, 0.57, 25.5 & 1.0 - 2.4 & 26.0 \\
ESO306-G017 & 28.37 & 1.59 & 26.63 & 2.02 & 0.12, 0.65, 28.0 & 0.07, 0.59, 25.5 & 1.0 - 2.4 & 26.0 \\
\\
&  \multicolumn{2}{c}{F435W} & \multicolumn{2}{c}{F606W} & F435W & F606W & (F435W-F606W)$_0$ & F606W$_0$ \\
NGC1129 & 27.80 & 2.84 & 26.23 & 1.43 & 0.07, 0.64, 26.0 & 0.03, 0.65, 25.0 & 0.7 - 1.8 & 26.0  \\
\\
&  \multicolumn{2}{c}{F555W} & \multicolumn{2}{c}{F814W} & F555W & F814W & (F555W-F814W)$_0$ & F814W$_0$ \\
NGC1272 & 27.78 & 5.73 & 26.29 & 1.96 & 0.22, 0.63, 28.0 & 0.07, 0.58, 25.5 & 0.8 - 1.8 & 25.5 \\
\\
 
 \hline
\end{tabular}
\textit{Key to columns:} (1) Galaxy ID; (2,3) Completeness curve parameters for the blue filter
(Eq.2); (4,5) Completeness curve parameters for the red filter; (6) Measurement uncertainty parameters for
the blue filter (Eq.1): (7) Measurement uncertainty parameters for the red filter; (8) Selection range of dereddened
colors including GC candidates; (9) Faint limit of intrinsic (red) magnitude for selection of GCs.  The GC candidates selected for the CDF and MDF are the objects in the color-magnitude diagram brighter than $m(lim)$ and within the
given color range.
\label{tab:completeness}
\end{table*}

\begin{table*}[ht]
\centering
\caption{DOLPHOT Photometry for NGC 4874 Field} 
\begin{threeparttable}
\begin{tabular}{cccccccc}
  \hline \hline
RA & Dec & x(px) & y(px) & F475W & $\pm$ & F814W & $\pm$  \\ 
(1) & (2) & (3) & (4) & (5) & (6) & (7) & (8)  \\
   \hline
  194.8888158 & 27.9473826 & 3004.360 & 4224.360 &  24.857 &    0.010 &  20.914 &   0.002 \\
  194.8728227 & 27.9648258 &  988.080 & 4099.310 &  22.827 &    0.014 &  21.222 &   0.002  \\
  194.9314789 & 27.962022 & 4326.600 &  834.530 &  25.655 &   0.021 &  21.814 &  0.002  \\
  194.8990624 & 27.9774308 & 1562.750 & 1795.640 &  23.663 &   0.006 &  21.968 &  0.002  \\
  194.8854043 & 27.9577357 & 2134.520 & 3794.070 &  23.668 &   0.007  &  22.008 &  0.003  \\
  194.8747698 & 27.9706598 &  705.980 & 3630.200 &  23.928 &   0.007 &  22.191 &  0.003  \\
  194.9120388 & 27.9656600 & 3040.670 & 1751.930 &  23.814 &   0.007  &  22.218 &  0.003  \\
                 \hline
\end{tabular}
\begin{tablenotes}
\item{} 
 \textit{Note:} The partial table listed here illustrates the form and content of the complete data.  Final photometry for all the galaxies can be accessed through the author's website {\bf or through the DOIs listed in the Acknowledgements below}.  The final lists accessible on-line also include other DOLPHOT parameters (chi, SNR, sharpness, roundness, crowdedness).
\end{tablenotes}
\end{threeparttable}
\label{tab:photom}
\end{table*}

The distribution of GC colors (the CDF) is defined from the objects within the 
intrinsic (dereddened) color range as 
given in column (8) of Table \ref{tab:completeness}, and brighter than $m(lim)$ as given
in column (9).  The magnitude limit $m(lim)$ is conservatively brighter (by about 0.5 mag) than the 50\% completeness limit to ensure that the CDFs will not be biased by incompleteness.  In the template example of
NGC 4874, this final selection box is marked out with the blue dashed lines in Fig.~\ref{fig:ngc4874_cmd}.

\begin{figure}
    \centering
    \includegraphics[width=0.47\textwidth]{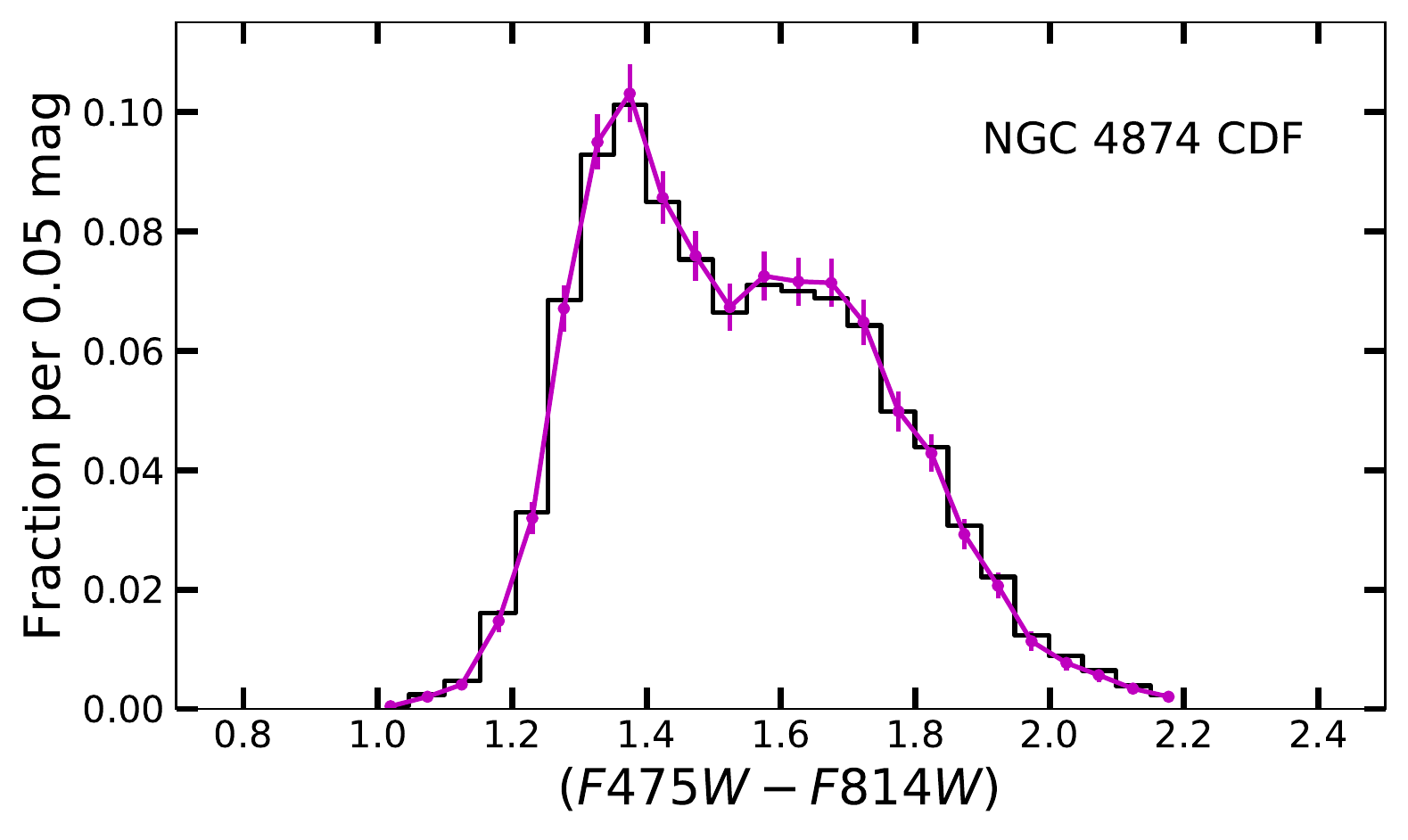}
    \caption{Color distribution function for the NGC 4874 system.  The magenta dots, errorbars, and lines show the raw CDF for the color range $1.0 < (F475W-F814W)_0 < 2.2$ and magnitude range $F814W_0 < 26.0$, plotted as the fraction of the total per 0.05-mag bin.  The black histogram shows the CDF corrected for photometric completeness, including the dependence of completeness on projected galactocentric radius; see text.}
    \label{fig:cdfcorr}
\end{figure}

The CDFs can, however, be explicitly corrected for incompleteness and the effect of doing this is illustrated in Figure \ref{fig:cdfcorr}.  Here, the CDF from the raw counts is shown as the magenta histogram, while the completeness-corrected version is shown as the black histogram.    In the completeness-corrected version, each object is counted as $(1/f)$ objects where $f$ is the completeness fraction defined by Eq.\ \ref{eq:fcurve}.  In addition, the completeness correction takes into account the projected galactocentric location of each object, since the fitted values of $(m_0, \alpha)$ in Eq.\ \ref{eq:fcurve} depend slightly on radius as shown in Fig.\ \ref{fig:f_radial}.  

In Fig.\ \ref{fig:cdfcorr} both curves have been normalized to the same total population to highlight any changes in the shape of the distribution. The \emph{total} equivalent number of objects $n_{corr} = \sum(1/f_i)$ in each color bin is correspondingly larger than the raw counts typically by about 20\%, but the two \emph{fractional} distributions across all the bins (as shown in Fig.\ \ref{fig:cdfcorr}) are very similar to within their statistical uncertainties.  This robust feature of the CDFs is partly due to a conservative magnitude cutoff as noted above, but also due to the empirical feature that the radial metallicity gradients of the GC systems in these giant galaxies are quite shallow (discussed below).  Tihat is, to first approximation the shape of the CDF is insensitive to the chosen radial range, at least within the limits of the current data.  

\begin{figure*}
    \centering
    \includegraphics[width=0.90\textwidth]{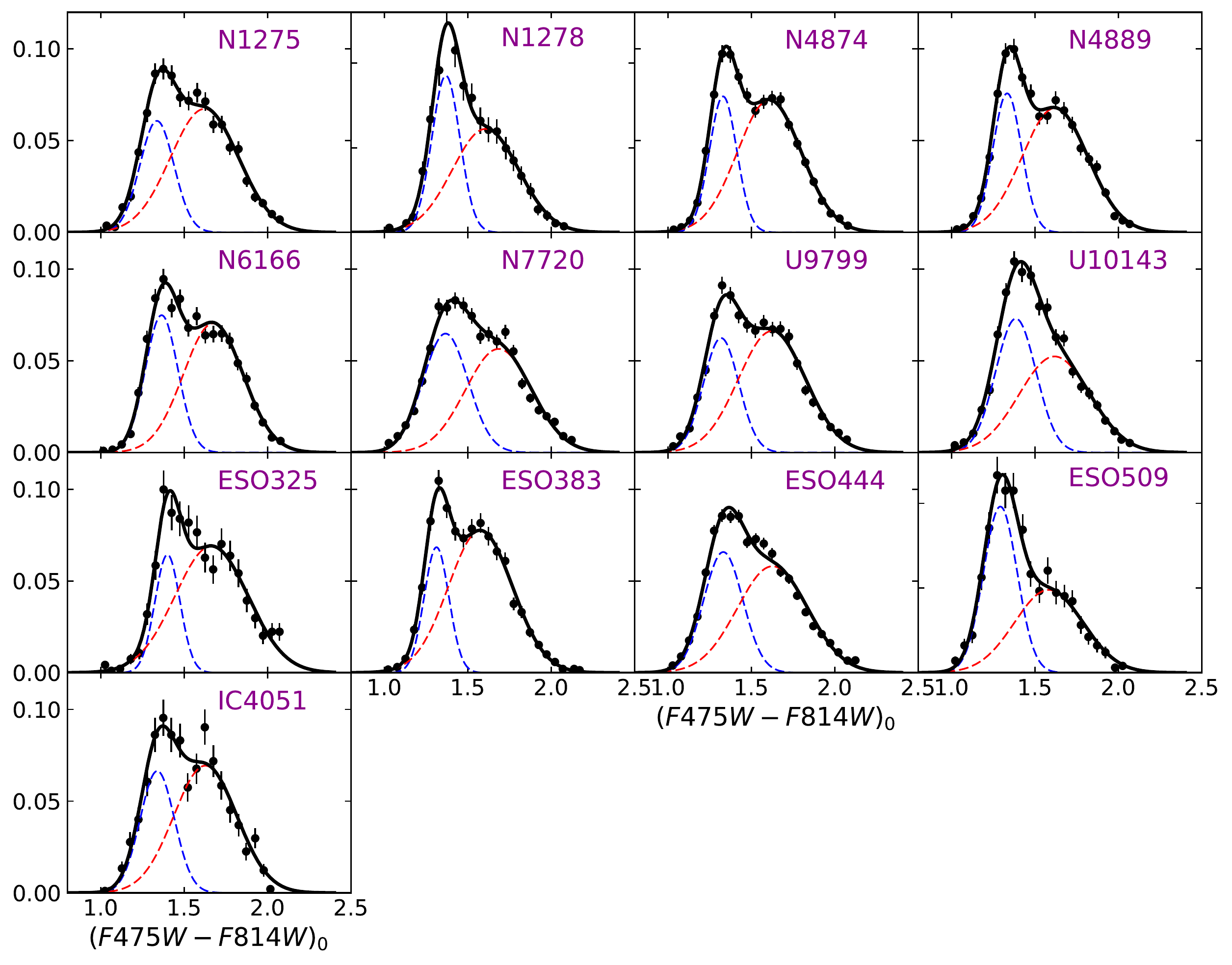}
    \caption{Color distribution functions (CDFs) for the galaxies measured
    in (F475W, F814W).  Bimodal-Gaussian fits are shown as the solid lines,
    with the blue and red subcomponents as the dashed lines.}
    \label{fig:cdf1}
\end{figure*}

\begin{figure*}
    \centering
    \includegraphics[width=0.90\textwidth]{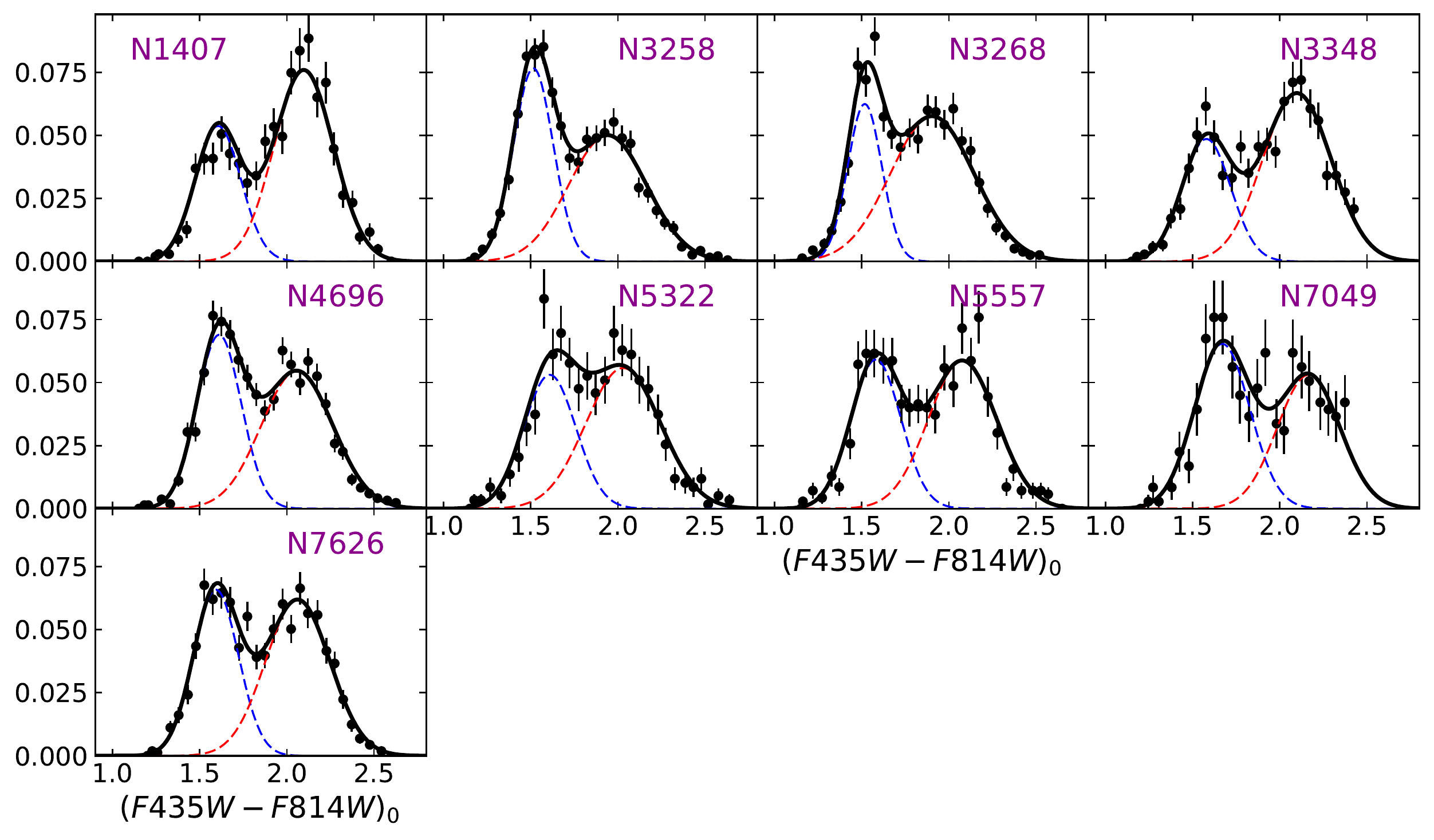}
    \caption{Color distribution functions for the galaxies measured
    in (F435W, F814W).}
    \label{fig:cdf2}
\end{figure*}

\begin{figure*}
    \centering
    \includegraphics[width=0.60\textwidth]{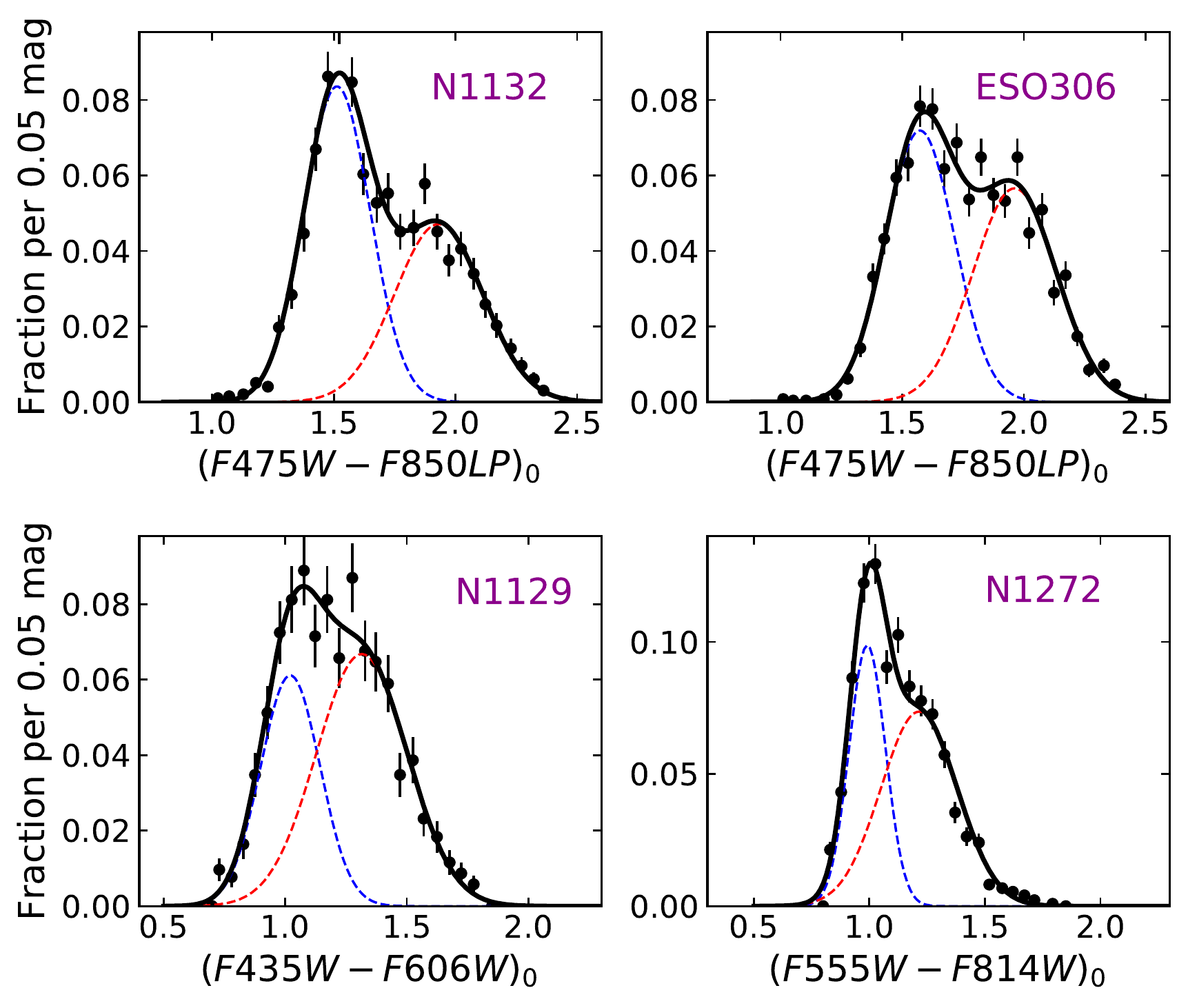}
    \caption{Color distribution functions for the galaxies measured
    in other filter combinations.  }
    \label{fig:cdf3}
\end{figure*}

The CDFs for all the program galaxies are displayed as histograms in Figures \ref{fig:cdf1}, \ref{fig:cdf2}, and \ref{fig:cdf3}.
The numbers of GCs in 0.05-magnitude bins in color, along with their Poisson errorbars ($N \pm N^{1/2}$),
are plotted versus dereddened color.  The overall shapes of these CDFs are familiar from many previous
photometric studies of GC systems in large galaxies \citep[][among numerous others]{geisler+1996,larsen+2001,rhode+2004,peng+2006,bassino+2006,harris2009b,cho+2012,kim+2013,fensch+2014,escudero+2015,harris+2017a}:  a relatively high, narrow peak on the blue side with a broader, lower distribution on the red side.

All the CDFs have been fit with a bimodal-Gaussian function with the GMM code
\citep{muratov_gnedin2010}.  The blue and red subcomponents from the solutions are shown as the
dotted lines in the figures.  In general, the bimodal-Gaussian assumption provides a good match
to all the galaxies here, and perhaps better the larger the GC population becomes.
The parameters derived from the solutions are listed in Table \ref{tab:cdf_params}, and include
the centroids of the blue and red modes ($\mu_1, \mu_2$); the standard deviations of the two
modes ($\sigma_1, \sigma_2$); the fraction of the total population taken up by the blue mode
($f_1$); and two other parameters ($DD, H_{12}$) that relate to the shape of the CDF.  $DD$ \citep{muratov_gnedin2010} measures
the separation of the two modes relative to their intrinsic widths,
\begin{equation}
    DD = {(\mu_2 - \mu_1) \over ([\sigma_1^2 + \sigma_2^2]/2)^{1/2} } .
\end{equation}
The observed $DD$ values range from a low of 1.34 (UGC 10143) to a high of 3.24 (NGC 1407). 
For clearly separated modes $DD \gtrsim 2$ is normally expected, but the sample sizes of GCs
here are so large that even for 
the galaxies in the list with $DD < 2$ a unimodal Gaussian solution is strongly rejected by GMM (at 99\% confidence and higher).
The other parameter, the \emph{height ratio} $H_{12}$, is defined as the 
maximum height (amplitude) of the blue component divided by the maximum height of the red
component and can be expressed in terms of the Gaussian fit parameters as
\begin{equation}
    H_{12} = {\sigma_2 \over \sigma_1} \cdot {f_1 \over (1-f_1)} - 1 .
\end{equation}
$H_{12} $ is zero where the modes are of equal height, positive where the blue peak is higher, and negative where the red peak is higher. 

\section{Effects of GC Intrinsic Sizes}

Most of the galaxies analyzed in this paper are at $d \gtrsim 50$ Mpc, for which the
intrinsic scale sizes of their GCs (with typical half-mass radii $r_h \sim 3$ pc)
are small enough that they are effectively unresolved.  Stellar photometry codes like DOLPHOT \citep{dolphin2000} or Daophot \citep{stetson1987} are therefore very well suited to measuring them.  However, a few of the galaxies have
$d \lesssim 40$ Mpc and for these, a significant fraction of their GCs will be
marginally nonstellar \citep{harris2009b}.  For such objects, photometry based on
PSF fitting will not be ideal.  
For the still nearer Virgo galaxies ($d \simeq 16$ Mpc) and a few other systems not studied here, the GCs are partially resolved
well enough that measurements of their half-light radii are interesting in their own right 
\citep{jordan+2005,harris+2010}.  For the galaxies in the study by \citet{harris2009b}, 
which are in the distance range $d \simeq 25-50$ Mpc, aperture photometry and 
a curve of growth technique were used to obtain the final GC magnitudes and colors, rather than
PSF fitting.  However, the initial survey of those same galaxies \citep{harris+2006} employed PSF fitting, with scarcely different results.

\subsection{ISHAPE versus DOLPHOT}

Here, a method similar to that of \citet{jordan+2005} is used to test the validity
of the photometry, and especially the color indices on which the GC
metallicity estimates are ultimately based (see below).  This test is done on
the nearest galaxy in the list, NGC 1407 at $d = 23$ Mpc.  At this distance,
a standard GC radius of 3 pc corresponds to an angular radius of $0.027''$, about
half an ACS pixel or 1/4 of the stellar fwhm.  

A fitting routine designed for exactly this task is ISHAPE \citep{larsen1999}.
For each candidate GC in the photometry list, a King-model profile \citep{king1962}
is convolved with the PSF for the image, and the assumed profile width (fwhm)
is adjusted till an optimum fit is achieved.  An ``average'' KING30
profile was assumed ($r_t/r_c = 30$).  For the PSF, following standard usage
of ISHAPE, \emph{iraf/daophot} was used to do photometry on the {\it *.drc} images in each filter,
along with its routines \emph{phot/pstselect/psf} to produce empirical
PSFs from 34 bright, uncrowded stars across the frame in both filters.  ISHAPE was then run on the entire list
of measured objects in the CMD for NGC 1407, again in both filters.  The ISHAPE parameter
FITRAD was assumed to be 5 px, or typically 10 times larger than the GC half-light
radii.  A test run with FITRAD = 7 px was also made, which produced no systematic
differences compared with the 5-px run. 

\begin{figure}
    \centering
    \includegraphics[width=0.45\textwidth]{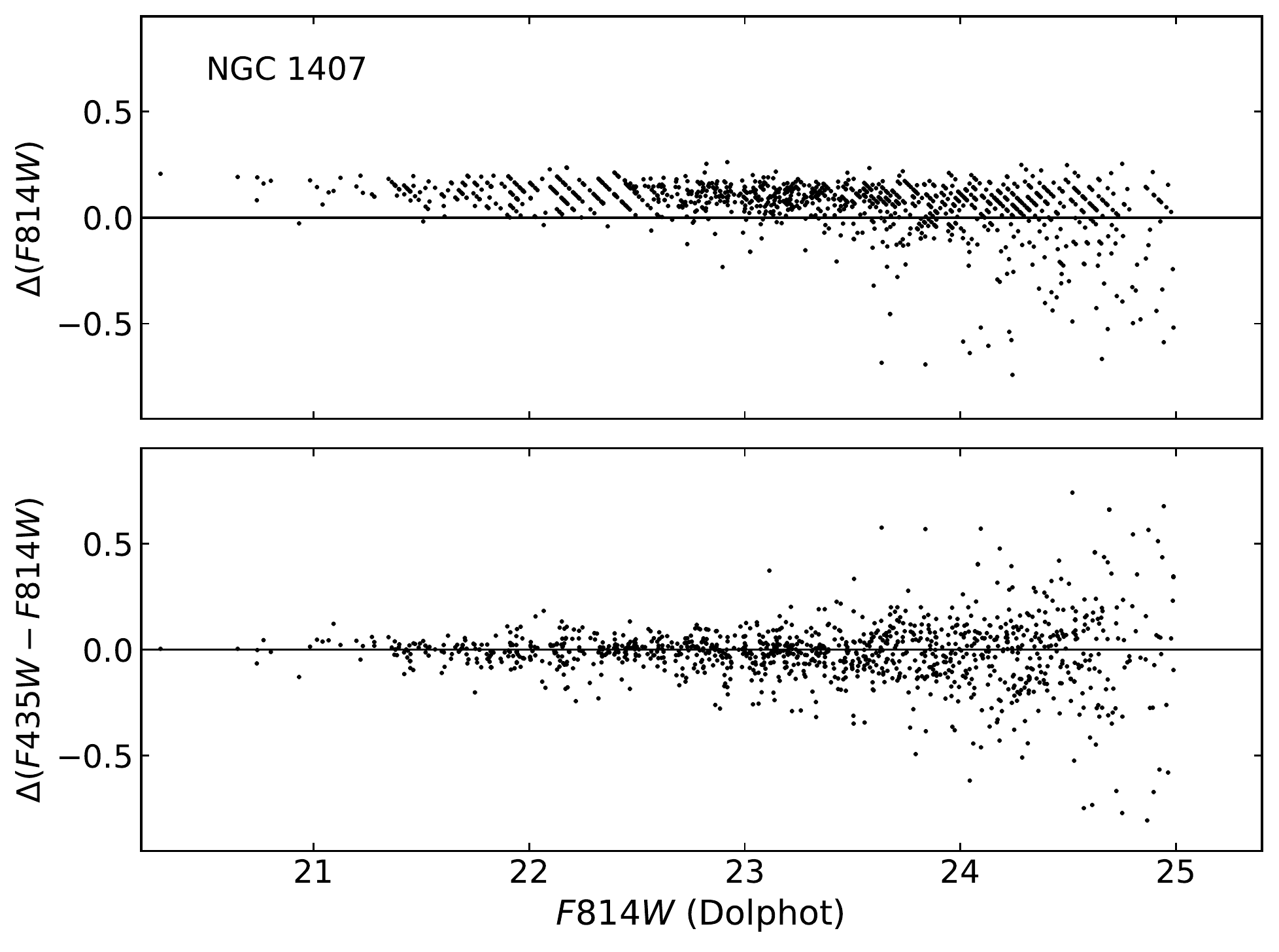}
    \caption{Comparison of photometry from DOLPHOT and ISHAPE, for the GCs in NGC 1407, the closest galaxy in the list.  \emph{Upper panel:} Difference
    between the F814W magnitudes measured by the two codes, in the sense (ISHAPE $-$ DOLPHOT).
    \emph{Lower panel:} Difference between the two color indices (F435W-F814W).}
    \label{fig:ishape}
\end{figure}

The output parameters from ISHAPE
include not just the fwhm and ellipticity of the King-model profile, but also
the total fluxes, which can then be converted to total magnitudes and colors, and finally compared directly with the DOLPHOT results.
These comparisons are shown in Figure \ref{fig:ishape}.  The upper panel shows
the difference in F814W magnitude between the two codes 
(ISHAPE $-$ DOLPHOT).  Not surprisingly, there is a nonzero offset in the sense
that DOLPHOT measures these marginally resolved GCs slightly brighter than
does ISHAPE.\footnote{The distribution of points along short diagonal lines 
in the figure is the result of some quantization of the fluxes within ISHAPE.}
The mean difference is $\langle \Delta(F814W) \rangle = +0.077 \pm 0.003$ with an
rms scatter of 0.087 mag.

The much more important test is in the comparison of color indices, shown in
the lower panel of Fig.~\ref{fig:ishape}.  Here, no systematic offset is seen;
the mean is $(-0.004 \pm 0.004)$ with rms scatter of 0.14 mag.  Clearly, net
magnitude offsets are present in both filters, but they are nearly equal
and yield color indices that are closely similar in both codes.  These tests
support the use of the measured color indices from DOLPHOT for derivation of the CDF and MDF.

\begin{figure}[h]
    \centering
    \includegraphics[width=0.49\textwidth]{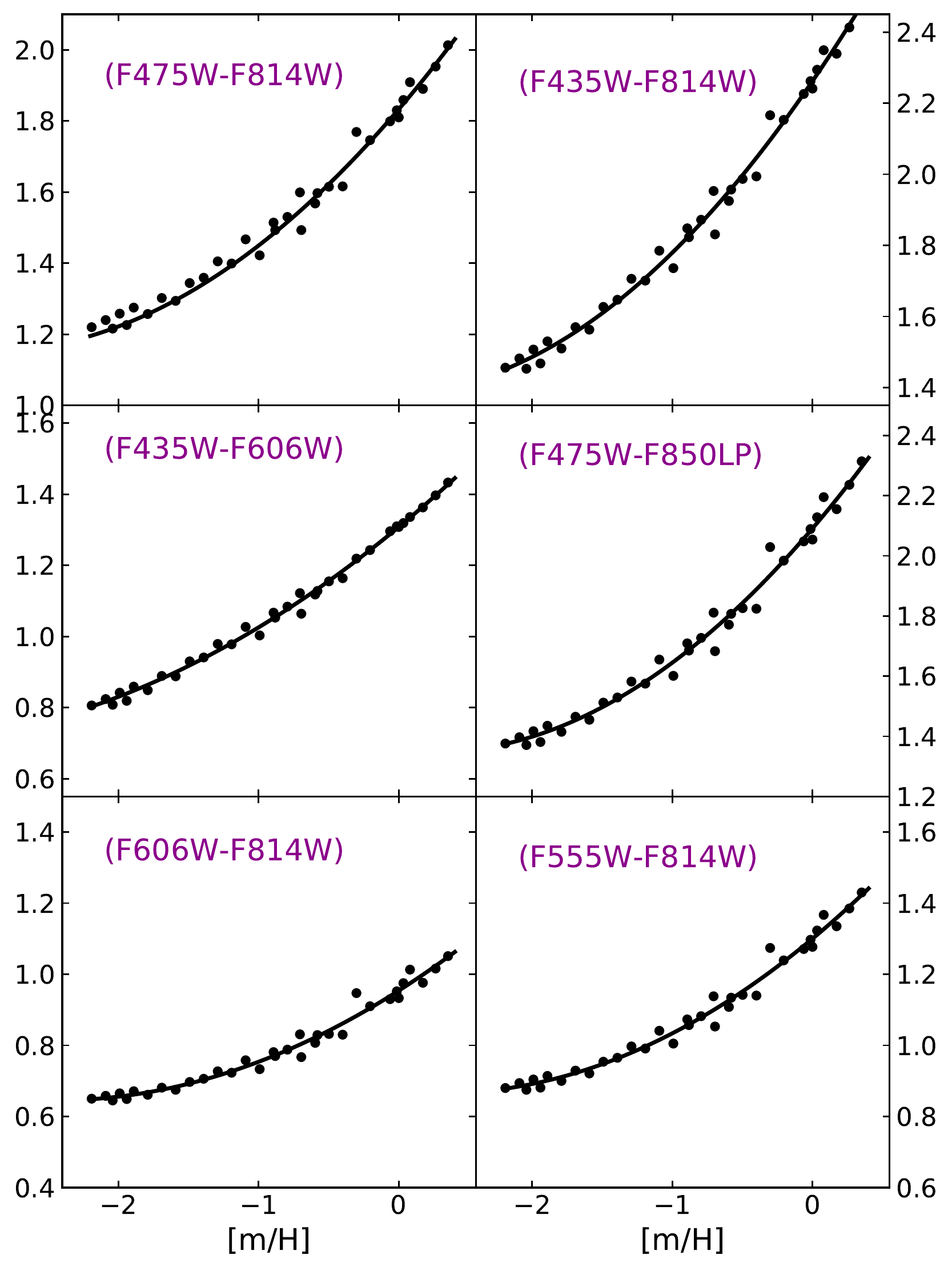}
    \caption{Predicted HST/ACS color indices calculated from the Parsec SSP models, plotted versus metallicity [m/H].
    Each dot shows one synthetic model GC for $10^5 M_{\odot}$ and a 12 Gy age.  In each panel the solid line shows the best-fit quadratic interpolation relation given in the upper half of Table \ref{tab:coefficients}. }
    \label{fig:transform}
\end{figure}

\begin{figure}3100
    \centering
    \includegraphics[width=0.49\textwidth]{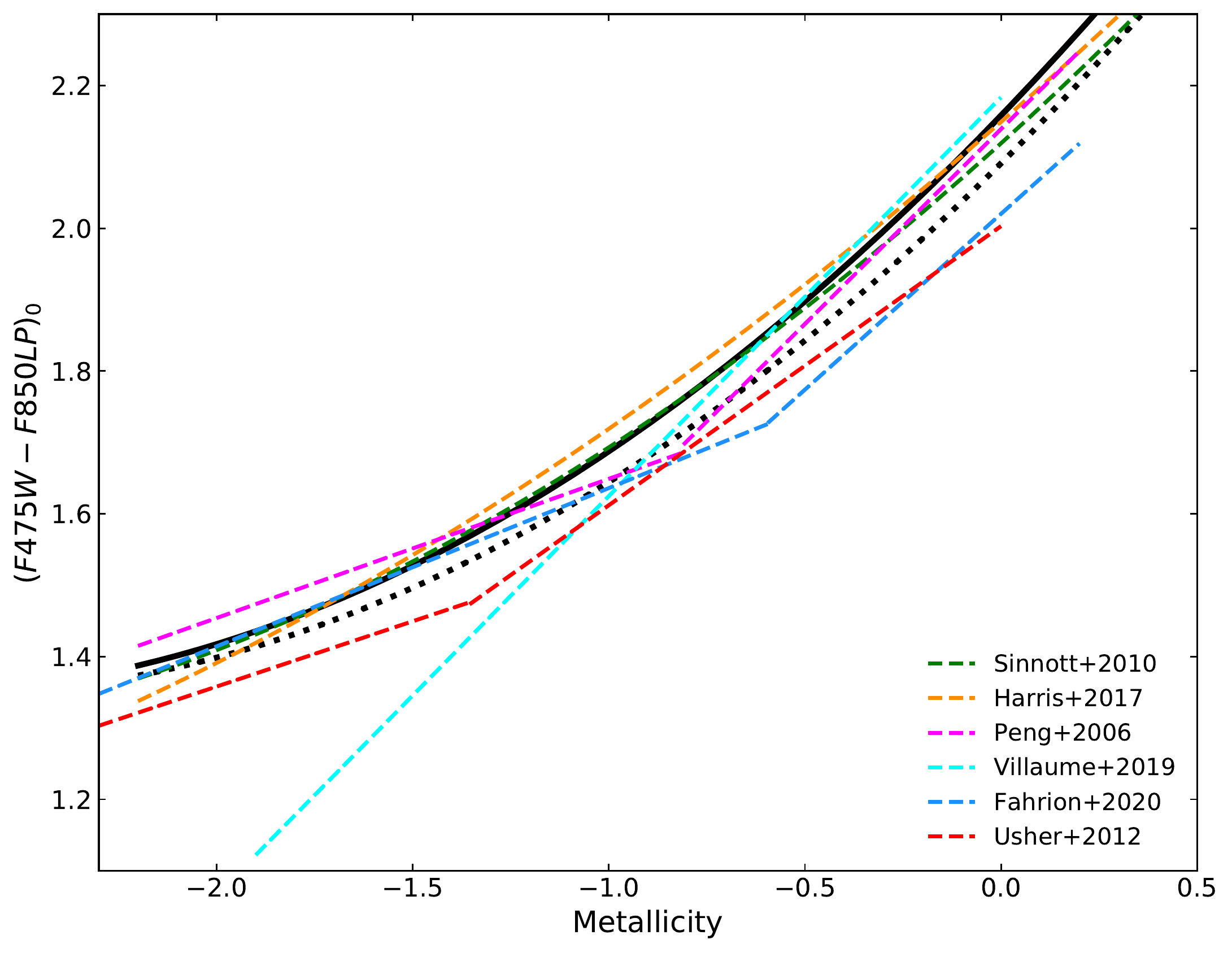}
    \caption{Color index $(F475W-F850LP)_0$ versus metallicity, from
    six recent literature sources.  The conversion relation adopted here
    is shown as the heavy solid line as given in Eq.\ \ref{eq:colorcal}.  For comparison, the black dotted line
    indicates the predicted relation from the Parsec SSP models without adjustment.}
    \label{fig:colorcal}
\end{figure}

\subsection{Notes on Individual Galaxies}

\emph{NGC 1129:} The galaxy is located near the upper right corner of the ACS field at (3600,3160) px.  Its GC system is relatively compact and a radial falloff of the GC numbers is clearly seen, but numerous objects with GC-like colors and magnitudes also populate the lower half of the field ($y < 2000$ px) that appear to be contamination from companion galaxies or possibly an intergalactic population.  For the present study, only the data with $y > 2000$ px are used.  This galaxy needs additional investigation to clarify the contamination issue as well as the puzzling result for the transformation of color to metallicity (see discussion below).

\emph{NGC 1272:} A color-magnitude array for the GC population was previously measured by \citet{penny+2012}.

\emph{NGC 1275:}  This active galaxy has a large population of young star clusters in its inner regions that has attracted much previous interest.  For HST-based photometry of the young cluster population, see, for example, \citet{holtzman+1992,carlson+1998,canning+2014,lim+2020,lim+2022}.

\emph{IC 4051:} GC photometry in this Coma giant ETG drawn from HST/WFPC2 imaging was presented in \citet{baum+1997} and \citet{woodworth_harris2000}.  This galaxy and its GC system have an unusually compact spatial central concentration.

\emph{NGC 4874, NGC 4889:} These are the two centrally dominant giants in the Coma cluster.  Previous HST photometry for their GC systems was presented in \citet{harris+2009} with WFPC2 data, and in \citet{peng+2011,harris+2017a} with the same ACS data used in the present study. 

\emph{NGC 7720:}  This galaxy has a double core, with two near-equal nuclei separated by just $12''$ (7.5 kpc); see \citet{laine+2003}.  Somewhat of necessity, the GC population is treated here as if it belongs to one combined galaxy.

\section{Transformations to Metallicity}

The major results of this study consist of the photometric database and the resulting CDFs, as laid out in the previous sections.  Following on from that, a brief discussion at the metallicity distribution functions (MDFs) is given.  For reasons to be discussed below, the MDFs and their resulting parametric descriptions, at this stage, should be viewed as only a first look.

Because the CMDs for the program galaxies are in several different filter pairs, the CDFs and their bimodal-Gaussian fit parameters cannot be directly intercompared.  A necessary first step could be to transform them all into a common filter system, for example through the use of SSP models or published empirical transformations between the various filters.  Though observationally based transformations would normally be preferred, two practical issues that make such a step uncertain are that (a) published empirical conversions \citep{sirianni+2005} were derived for single stars and not the combined light of star clusters; and (b) there are no calibrating galaxies available whose GC systems have been measured in all the five different indices in this study; hardly any that have even been measured in more than two filters.

However, a bigger issue is that the CDFs are not what we ultimately want.  What is really wanted are the MDFs that the CDFs represent, and constructing MDFs requires confronting the well known problem of transforming color to metallicity.  All contemporary SSP models show that these transformations are nonlinear to various degrees in optical/NIR bands  \citep[e.g.][]{yoon+2006,yoon+2011,peng+2006,cantiello_blakeslee2007,cantiello+2014,usher+2012}.
The physical cause of the nonlinear dependence of the color-to-metallicity relation (CMR) is simply that the integrated color of a GC is dominated by the light from its giant and subgiant stars, and the color of the red-giant branch becomes more sensitive to metallicity as [Fe/H] increases.  That is, the isochrone lines shift redward progressively faster with increasing metallicity.

Theoretical modelling exercises \citep[e.g.][]{yoon+2006,yoon+2011,cantiello_blakeslee2007,cantiello+2014} show that with appropriately constructed (very) nonlinear CMRs, a bimodal CDF can result from a flat or unimodal MDF.  (Contrarily,  an intrinsically bimodal MDF can be converted to a flat or unimodal CDF with a suitably nonlinear transformation.)  Notably, however, MDFs for the GC systems in several nearby large galaxies have been constructed directly from spectroscopic samples \citep{peng+2006,woodley+2010,alves-brito+2011,usher+2012,brodie+2012,harris+2016,caldwell_romanowsky2016,villaume+2019,fahrion+2020}.  These spectroscopic MDFs bypass the CMR issue, and importantly, they usually show bimodality.

Among the five color indices used here, one of them stands out above the others because several studies exist that connect it directly, and empirically, with metallicity.  This is (F475W-F850LP) (equivalent to  $(g-z)$ in the AB magnitude system).  For this index, CMRs built from spectroscopic metallicities in either [Fe/H] or [m/H] have been published by \citet{peng+2006,sinnott+2010,blakeslee+2010,usher+2012,vanderbeke+2014,harris+2017a,villaume+2019,fahrion+2020}.  In some of these studies a two-part piecewise linear correlation is assumed, and in other cases a continuous quadratic or other nonlinear shape is assumed, but no single definitive version can be said to exist as yet.  All these relations are limited by the observational scatter in the measured color indices and spectra, as well as by the relatively small samples of GCs being used, which may include only some dozens of clusters.  These issues of scatter and sample size have led to simple linear or piecewise-linear CMRs being used in the past for many broadband color indices  \citep[e.g.][]{barmby+2000,harris+2006,peng+2006,vanderbeke+2014,fahrion+2020}.

To go a step further, the approach used here is to define a hybrid combination of SSP models and empirical data.  First, a baseline CMR between the fiducial color index (F475W-F850LP)$_0$ and metallicity [Fe/H] is defined; and then second, SSP models are used to convert every other measured color index into (F475W-F850LP)$_0$ and thus into metallicity.

For the purposes of this preliminary investigation, the widely used ParsecV1.2S stellar models \citep{marigo+2017} are adopted here\footnote{accessible at http://stev.oapd.inaf.it/cgi-bin/cmd}.  These SSPs conveniently allow model GCs to be produced at any given metallicity, age, and cluster mass.  The input parameters adopted include the OBC bolometric corrections and a Kroupa canonical two-part powerlaw IMF.  Each model GC was computed for a total mass of $10^5 M_{\odot}$, no internal spread in metallicity, and an age of 12 Gyr (but see below for the effects of age differences).  The total luminosity of each model GC was calculated in the standard set of HST ACS broadband filters, from which all the color indices used in this study could be constructed.  The individual model GCs covered the full available range of metallicities from [m/H] $= -2.2$ to +0.4 in steps of 0.1 dex.  The resulting plots of the predicted color indices versus [m/H], without any adjustment (see below), are shown in Figure \ref{fig:transform}.  The interpolation curves shown there are simple quadratic functions, with 
 coefficients and fitting uncertainties as listed in the upper half of Table \ref{tab:coefficients}.
The residual scatter of points around the best-fit lines is the result of stochastic
differences in populating each synthetic cluster with a finite number
of stars.  Although each cluster contains $\sim10^5$ stars, the total
light is dominated by the red giants, and only some dozens of those
are present in each model cluster.  The curvature of each line shows explicitly how
the sensitivity of color to metallicity increases steadily with increasing metallicity.

These models are next tied to spectroscopically based transformations of (F475W-F850LP) to metallicity [Fe/H]. The relations from six recent studies as mentioned above are shown in Figure \ref{fig:colorcal}, including \citet{peng+2006,sinnott+2010,usher+2012,harris+2017a,villaume+2019}, and \citet{fahrion+2020}.  Where necessary, $(g-z)$ colors in the AB magnitude system have been converted to the Vegamag scale with $(F475W-F850LP) = (g-z) + 0.636$ \citep{sirianni+2005}.  Although all these studies show rather similar conversions, some use [m/H] for metallicity \citep{usher+2012,fahrion+2020} and the others use [Fe/H] \citep{peng+2006,sinnott+2010,harris+2017a, villaume+2019}.  In this study, conversions will be made to [Fe/H].

As noted above, the observational samples of GCs are small enough and show enough cluster-to-cluster scatter that a definitive transformation is difficult to define.  For the purposes of the present study, the Parsec SSPs  are used to define the \emph{shape} of the CMR, but the observational data are used to fix the \emph{zeropoint}.  The SSPs themselves give scaled-Solar [m/H] abundances.  The adopted result is shown in Fig.\ \ref{fig:colorcal} as the heavy solid line and is given by the quadratic curve
\begin{multline}
    (F475W-F850LP)_0 = (2.158 \pm 0.011) + \\
    (0.5708 \pm 0.0263) [Fe/H] + (0.1003 \pm 0.0128) [Fe/H]^2 \, .
    \label{eq:colorcal}
\end{multline}
Notably, however, the raw conversion relation from the SSPs without any adjustment already falls within the spread of the observational relations and, in the end, was shifted by only $-0.12$ dex in metallicity to yield the transformation in Eq.\ \ref{eq:colorcal}.  Given the spread of the observational calibrations, to within $\sim \pm0.1$ dex the final placement of this curve remains a matter of judgment, and the final choice was simply to fit the ensemble of calibrations that use [Fe/H] rather than [m/H].  

The second necessary step is the conversion of any of the other color indices into (F475W-F850LP).  The curves shown in Fig.\ \ref{fig:transform} are used for this purpose.  The ratio of any pair of indices is nearly constant with metallicity, and these ratios are listed in the last column of Table \ref{tab:coefficients} (upper half).  That is, the (F475W-F850LP) index is multiplied by the ratio listed in the table to give each of the other indices (for example, (F475W-F814W) = 0.884 (F475W-F850LP)). 
Combining these ratios with Eq.\ \ref{eq:colorcal} gives the final set of conversions between each color index and [Fe/H] listed in the lower half of Table \ref{tab:coefficients}.\footnote{Again, the coefficients listed in the upper half of the table give the color index as a function of [m/H] directly from the SSP models.  The coefficients in the lower half of the table give color index as a function of [Fe/H] after the empirical calibration is done.  To convert color into metallicity, these quadratic relations are numerically inverted.}

The procedure for deriving metallicities used here bears some resemblance
to the multiparameter fitting process of \citet{deMeulenaer+2017} for the star
clusters in M31.  They employ the same \citet{marigo+2017} SSP models 
to derive age, mass, metallicity, and foreground extinction, all of which 
span wide ranges for the M31 cluster population as a whole.  They
find that the integrated photometric colors from the ACS and WFC3 cameras yield metallicities
and ages in reasonable agreement with direct spectroscopy, though residual nonlinear trends
remain that resemble those found here.  The photometry for the giant ETGs studied
here represents a much simpler problem, since no young clusters or differential
extinction are present.

The hybrid procedure used here to define the CMR is clearly far from ideal.  In particular, the transformations of the different color indices into the fiducial index (F475W-F850LP) rely heavily on the SSP models, as does the adopted shape for the final CMR.  A significant and very helpful step would be to measure the GCs in the entire set of color indices for at least one of these giant ETGs and preferably a selected set of them.  Reliance on the models would be very much reduced as a result.  

A related and more fundamental issue is in the nature of the CMR itself.  The implicit assumption made here, and in most other studies, is that GCs are similar enough objects in different galaxies that a single CMR can be applied across all of them.  The possibility that GCs in different galaxies might follow different CMRs is discussed by, e.g., \citet{usher+2012,powalka+2016} and \citet{villaume+2019}.  Although the galaxies studied here are all of similar type (giant ETGs),  perhaps the most obvious single concern is the expected range of GC ages, which will differ from one galaxy to another depending on their specific history of hierarchical growth.  In the present discussion, the blanket assumption of a 12-Gyr age is made, but a test of the sensitivity of the predicted MDF to the assumed GC age is shown in the next section.   

\begin{figure}
    \centering
    \includegraphics[width=0.40\textwidth]{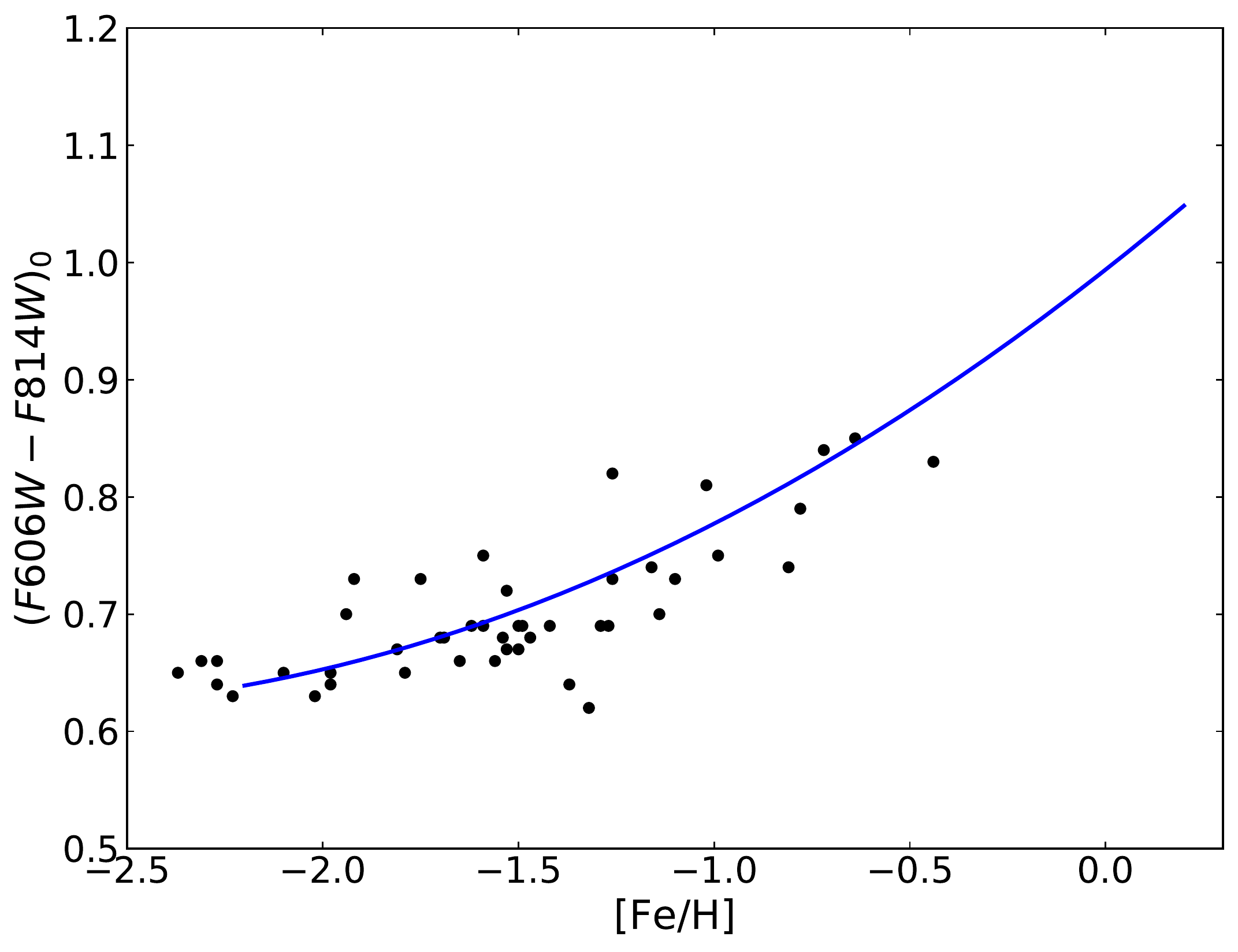}
    \caption{Color versus metallicity [Fe/H] for the ACS color index $(F606W-F814W)$. The intrinsic, dereddened colors for Milky Way globular clusters measured by \citet{bellini+2015} are shown as the solid dots.  The \emph{solid line} is the 
    adopted relation between [Fe/H] and $(F606W-F814W)_0$ as given in Table \ref{tab:coefficients}. }
    \label{fig:colorcal_VI}
\end{figure}

Nevertheless, some useful consistency checks of the procedure can be done through the available integrated photometry for the Milky Way GCs.  One example is shown in Figure \ref{fig:colorcal_VI}.  \citet{bellini+2015} provide integrated, dereddened colors for a subset of the Milky Way GCs measured directly in the ACS F606W, F814W filters.  These colors can then be plotted versus their [Fe/H] values \citep[][2010 edition]{harris1996}.  The predicted relation as given in Table \ref{tab:coefficients} matches the data well within the scatter.  

\begin{figure}
    \centering
    \includegraphics[width=0.49\textwidth]{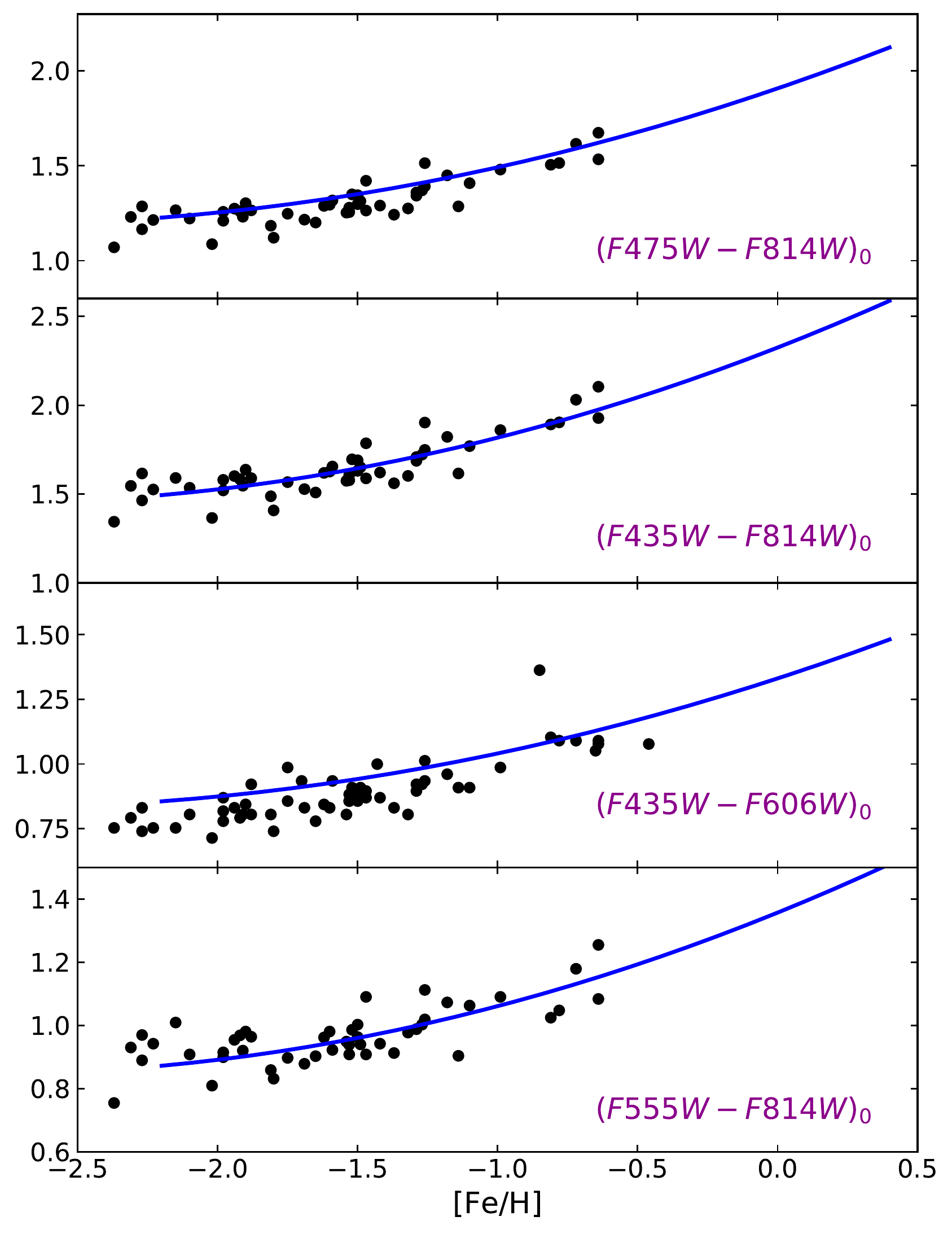}
    \caption{Color versus metallicity [Fe/H] for the additional
    color indices used in the present study.  In each case the \emph{solid line} shows the final relation between [Fe/H] and the given color index as listed in Table \ref{tab:coefficients}.  Data points (solid dots) are dereddened color indices for Milky Way GCs, transformed from $BVI$ into the Vegamag HST/ACS filter system.}
    \label{fig:colorcal2}
\end{figure}

A less direct but useful check can be made from the $BVI$ integrated colors of the Milky Way GCs listed in \cite{harris1996}.  From the conversion equations given in \citet{sirianni+2005}, their dereddened $(B-I)_0$ values were transformed into the HST/ACS color indices (F475W-F814W) and (F435W-F814W); $(B-V)_0$ was transformed into (F435W-F606W); and $(V-I)_0$ was transformed into (F555W-F814W).  These four indices are plotted versus [Fe/H] in Figure \ref{fig:colorcal2}.  The adopted transformations in Table \ref{tab:coefficients} are drawn in for comparison.  Except for (F435W-F606W), where the Milky Way GC data fall below the line by $\sim 0.1$ mag, the conversion lines and the datapoints agree reasonably well, even considering the fact that the Sirianni conversion equations are defined for single stars rather than composite stellar systems.

\begin{table*}[ht]
\centering
\caption{Bimodal Gaussian Parameters for the CDFs} 
\begin{tabular}{ccccccccc}
  \hline \hline
Galaxy & $N$ &  $\mu_1$ & $\mu_2$ & $\sigma_1$ & $\sigma_2$ & $f_1$ & DD & $H_{12}$ \\ 
(1) & (2) & (3) & (4) & (5) & (6) & (7) & (8) & (9) \\
   \hline
   \\
   \multicolumn{8}{c}{$(F475W-F814W)_0$} \\
   NGC 1275 & 2704 & 1.336 (0.009) & 1.621 (0.018) & 0.102 (0.010) & 0.204 (0.008) & 0.311 (0.049) & 1.76 (0.11) & -0.10 \\
   NGC 1278 & 1107 & 1.368 (0.009) & 1.606 (0.021) & 0.084 (0.008) & 0.199 (0.008) & 0.390 (0.058) & 1.55 (0.14) & 0.51 \\
   NGC 4874 & 4632 & 1.330 (0.005) & 1.607 (0.011) & 0.082 (0.005) & 0.192 (0.005) & 0.305 (0.029) & 1.77 (0.09) & 0.03 \\
   NGC 4889 & 3054 & 1.333 (0.005) & 1.619 (0.013) & 0.086 (0.006) & 0.198 (0.007) & 0.327 (0.034) & 1.87 (0.11) & 0.12\\
   NGC 6166 & 3162 & 1.363 (0.010) & 1.673 (0.023) & 0.098 (0.009) & 0.179 (0.011) & 0.368 (0.054) & 2.15 (0.16) & 0.06 \\
   NGC 7720 & 4242 & 1.366 (0.008) & 1.684 (0.015) & 0.135 (0.005) & 0.198 (0.006) & 0.439 (0.034) & 1.88 (0.09) & 0.15 \\
   UGC 9799 & 4094 & 1.318 (0.005) & 1.623 (0.013) & 0.106 (0.007) & 0.201 (0.006) & 0.332 (0.033) & 1.89 (0.09) & -0.06 \\
   UGC 10143 & 3623& 1.385 (0.013) & 1.617 (0.035) & 0.121 (0.013) & 0.212 (0.009) & 0.443 (0.085) & 1.34 (0.17) & 0.39 \\
   ESO325-G004 & 941 & 1.399 (0.014) & 1.663 (0.022) & 0.075 (0.014) & 0.219 (0.008) & 0.242 (0.059) & 1.61 (0.13) & -0.07 \\
   ESO383-G076 & 2906 & 1.312 (0.005) & 1.575 (0.008) & 0.074 (0.006) & 0.192 (0.004) & 0.254 (0.023) & 1.81 (0.07) & -0.12 \\
   ESO444-G046 & 7121 & 1.331 (0.007) & 1.626 (0.008) & 0.117 (0.004) & 0.211 (0.002) & 0.386 (0.024) & 1.73 (0.04) & 0.13 \\
   ESO509-G008 & 995 & 1.292 (0.008) & 1.582 (0.024) & 0.101 (0.008) & 0.205 (0.010) & 0.497 (0.054) & 1.79 (0.19) & 1.01 \\
   IC 4051 & 975 & 1.338 (0.012) & 1.626 (0.019) & 0.103 (0.007) & 0.189 (0.010) & 0.343 (0.049) & 1.90 (0.18) & -0.04 \\
   \\
   \multicolumn{8}{c}{$(F435W-F814W)_0$} \\
   NGC 1407 & 1029 & 1.605 (0.014) & 2.097 (0.011) & 0.131 (0.010) & 0.170 (0.008) & 0.353 (0.025) & 3.24 (0.12) & -0.29 \\
   NGC 3258 & 1880 & 1.515 (0.006) & 1.941 (0.017) & 0.114 (0.004) & 0.224 (0.011) & 0.438 (0.027) & 2.39 (0.15) & 0.53 \\
   NGC 3268 & 1568 & 1.517 (0.009) & 1.904 (0.031) & 0.098 (0.013) & 0.240 (0.019) & 0.307 (0.057) & 2.11 (0.26) & 0.09 \\
   NGC 3348 & 1056 & 1.578 (0.140) & 2.098 (0.015) & 0.140 (0.011) & 0.197 )0.008) & 0.341 (0.028) & 3.04 (0.11) & -0.27 \\
   NGC 4696 & 2170 & 1.611 (0.009) & 2.058 (0.026) & 0.127 (0.009) & 0.205 (0.017) & 0.439 (0.048) & 2.67 (0.21) & 0.26 \\
   NGC 5322 & 589 & 1.609 (0.018) & 2.030 (0.058) & 0.150 (0.037) & 0.215 (0.034) & 0.399 (0.118) & 2.27 (0.38) & -0.05\\
   NGC 5557 & 699 & 1.581 (0.015) & 2.078 (0.017) & 0.143 (0.011) & 0.196 (0.014) & 0.424 (0.035) & 2.90 (0.15) & 0.01 \\
   NGC 7049 & 356 & 1.669 (0.034) & 2.165 (0.059) & 0.159 (0.029) & 0.180 (0.033) & 0.521 (0.097) & 2.92 (0.36) & 0.23 \\
   NGC 7626 & 1611 & 1.593 (0.012) & 2.062 (0.015) & 0.129 (0.007) & 0.185 (0.009) & 0.426 (0.025) & 2.95 (0.11) & 0.06 \\
   \\
   \multicolumn{8}{c}{$(F475W-F850LP)_0$} \\
   NGC 1132 & 1972 & 1.513 (0.009) & 1.931 (0.021) & 0.136 (0.008) & 0.182 (0.010) & 0.570 (0.037) & 2.61 (0.15) & 0.77 \\
   ESO306-G017 & 2591 & 1.573 (0.014) & 1.962 (0.021) & 0.142 (0.008) & 0.172 (0.010) & 0.512 (0.044) & 2.47 (0.12) & 0.27 \\
   \\
   \multicolumn{8}{c}{$(F555W-F814W)_0$} \\
   NGC 1272 & 2201 & 0.991 (0.007) & 1.215 (0.008) & 0.079 (0.005) & 0.165 (0.004) & 0.391 (0.030) & 1.73 (0.07) & 0.34 \\
   \\
   \multicolumn{8}{c}{$(F435W-F606W)_0$} \\
   NGC 1129 & 1035 & 1.021 (0.024) & 1.315 (0.034) & 0.122 (0.013) & 0.187 (0.013) & 0.374 (0.092) & 1.86 (0.17) & -0.08 \\
   \\
                 \hline
\end{tabular}
\label{tab:cdf_params}
\end{table*}

\begin{table*}[ht]
\centering
\caption{Color/Metallicity Transformations for the ACS/WFC filters} 
\begin{tabular}{ccccc}
  \hline \hline
 Color Index &  $a_0$ & $a_1$ & $a_2$ & Conversion Ratio \\ 
   \hline
   \\
   \multicolumn{5}{c}{Predicted Transformations From SSP Models: color $= a_0 + a_1 [m/H] + a_2 [m/H]^2$} \\
   (F475W-F850LP) & 2.091 (0.011) & 0.5468 (0.0263) & 0.1003 (0.0128) & 1.000   \\
    (F475W-F814W) & 1.833 (0.009) & 0.4516 (0.0203) & 0.0780 (0.0099) & 0.884  \\
   (F435W-F814W) & 2.262 (0.010) & 0.5775 (0.0230) & 0.0947 (0.0112) & 1.077 \\
   (F435W-F606W) & 1.308 (0.004) & 0.3261 (0.0105) & 0.0436 (0.0051) & 0.617  \\
   (F555W-F814W) & 1.300 (0.007) & 0.3283 (0.0174) & 0.0620 (0.0085) & 0.629  \\
   (F606W-F814W) & 0.955 (0.006) & 0.2514 (0.0148) & 0.0511 (0.0072) & 0.461  \\
 \\
                 \hline
                 \\
  \multicolumn{5}{c}{Final Calibrated Transformations: color $= a_0 + a_1 [Fe/H] + a_2 [Fe/H]^2$ } \\
  (F475W-F850LP) & 2.158 (0.011) & 0.5708 (0.0263) & 0.1003 (0.0128)   \\
    (F475W-F814W) & 1.908 (0.009) & 0.5050 (0.0203) & 0.0890 (0.0099)   \\
  (F435W-F814W) & 2.324 (0.010) & 0.6148 (0.0230) & 0.1080 (0.0112)  \\
  (F435W-F606W) & 1.331 (0.004) & 0.3520 (0.0105) & 0.0618 (0.0051)   \\
  (F555W-F814W) & 1.357 (0.007) & 0.3589 (0.0174) & 0.0630 (0.0085)   \\
  (F606W-F814W) & 0.994 (0.006) & 0.2629 (0.0148) & 0.0462 (0.0072)   \\
 \\
 \hline
\end{tabular}
\label{tab:coefficients}
\end{table*}

\begin{table*}[ht]
\centering
\caption{Metallicity Gradient Parameters} 
\begin{tabular}{cccc}
  \hline \hline
Galaxy & Intercept $a$ & Slope $b$ & $(R/R_e)$ Range \\ 
(1) & (2) & (3) & (4) \\
   \hline
   \\
NGC1275 & $-0.816$ (0.025) & $-0.215$ (0.041) & $0.36 - 8.73$ \\
NGC1278 & $-0.924$ (0.023) & $-0.298$ (0.054) & $0.20 - 6.30$ \\
NGC4874 & $-0.942$ (0.011) & $-0.273$ (0.039) & $0.30 - 3.85$ \\
NGC4889 & $-0.879$ (0.016) & $-0.348$ (0.041) & $0.40 - 5.89$ \\
NGC6166 & $-0.882$ (0.011) & $-0.402$ (0.044) & $0.32 - 3.44$ \\
NGC7720 & $-0.822$ (0.012) & $-0.319$ (0.034) & $0.32 - 6.15$ \\
IC4051  & $-0.520$ (0.041) & $-0.657$ (0.057) & $1.32 - 22.5$ \\
UGC9799 & $-0.766$ (0.018) & $-0.486$ (0.037) & $0.53 - 7.55$ \\
UGC10143& $-0.939$ (0.012) & $-0.251$ (0.033) & $0.15 - 5.62$ \\
ESO325-G004 & $-0.651$ (0.031) & $-0.260$ (0.060) & $0.40 - 10.2$ \\
ESO383-G076 & $-0.956$ (0.022) & $-0.143$ (0.045) & $0.60 - 8.95$ \\
ESO444-G046 & $-0.951$ (0.015) & $-0.075$ (0.027) & $0.52 - 10.7$ \\
ESO509-G008 & $-0.987$ (0.062) & $-0.352$ (0.093) & $1.51 - 14.1$ \\
NGC1407 & $-0.778$ (0.021) & $-0.303$ (0.078) & $0.21 - 2.96$ \\
NGC3258 & $-1.088$ (0.024) & $-0.280$ (0.063) & $0.23 - 4.97$ \\
NGC3268 & $-1.072$ (0.026) & $-0.188$ (0.084) & $0.41 - 4.27$ \\
NGC3348 & $-0.786$ (0.030) & $-0.033$ (0.082) & $0.55 - 8.09$ \\
NGC4696 & $-0.913$ (0.016) & $-0.266$ (0.050) & $0.13 - 4.41$ \\
NGC5322 & $-0.910$ (0.033) & $-0.117$ (0.099) & $0.46 - 6.21$ \\
NGC5557 & $-0.838$ (0.039) & $-0.357$ (0.111) & $0.67 - 6.73$ \\
NGC7049 & $-0.697$ (0.070) & $-0.402$ (0.072) & $0.38 - 3.99$ \\
NGC7626 & $-0.891$ (0.019) & $-0.421$ (0.123) & $0.32 - 2.74$ \\
NGC1132 & $-0.971$ (0.019) & $-0.344$ (0.062) & $0.31 - 4.59$ \\
ESO306-G017 & $-0.736$ (0.021) & $-0.404$ (0.054) & $0.62 - 6.80$ \\
NGC1129 & $-0.474$ (0.030) & $+0.005$ (0.095) & $0.2 - 2.5$ \\
NGC1272 & $-0.786$ (0.014) & $-0.638$ (0.057) & $0.26 - 3.21$ \\

                \hline
\end{tabular}
\label{tab:radial}
\end{table*}

\begin{sidewaystable*}
\vspace{7cm}
\centering
\caption{Bimodal Gaussian Parameters for the MDFs} 
\begin{tabular}{cccccccccc}
  \hline \hline
Galaxy & $N$ &  $\mu_1$ & $\mu_2$ & $\sigma_1$ & $\sigma_2$ & $f_1$ & DD & $H_{12}$ & [Fe/H]$_{mid}$ \\ 
(1) & (2) & (3) & (4) & (5) & (6) & (7) & (8) & (9) & (10) \\
   \hline
   \\
   NGC 1275 & 2597 & -1.305 (0.052) & -0.407 (0.035) & 0.506 (0.022) & 0.355 (0.015) & 0.599 (0.047) & 2.05 (0.11) & 0.05 & -0.77  \\
   NGC 1278 & 1083 & -1.262 (0.044) & -0.365 (0.051) & 0.451 (0.026) & 0.314 (0.026) & 0.722 (0.052) & 2.31 (0.14) & 0.81 & -0.63 \\
   NGC 4874 & 4512 & -1.408 (0.028) & -0.467 (0.020) & 0.446 (0.014) & 0.345 (0.010) & 0.546 (0.025) & 2.36 (0.08) & -0.07 & -0.89 \\
   NGC 4889 & 2957 & -1.408 (0.034) & -0.442 (0.026) & 0.434 (0.019) & 0.350 (0.013) & 0.546 (0.030) & 2.45 (0.09) & -0.03 & -0.88 \\
   NGC 6166 & 3110 & -1.209 (0.034) & -0.311 (0.021) & 0.505 (0.017) & 0.297 (0.012) & 0.651 (0.030) & 2.17 (0.09) & 0.10 & -0.63 \\
   NGC 7720 & 4031 & -1.178 (0.027) & -0.296 (0.026) & 0.549 (0.012) & 0.345 (0.012) & 0.677 (0.027) & 1.92 (0.07) & 0.32 & -0.57 \\
   UGC 9799 & 3881 & -1.426 (0.035) & -0.459 (0.025) & 0.473 (0.017) & 0.375 (0.013) & 0.518 (0.029) & 2.27 (0.08) & -0.15 & -0.91 \\
   UGC 10143 &3124 & -1.156 (0.048) & -0.323 (0.066) & 0.518 (0.019) & 0.346 (0.026) & 0.792 (0.059) & 1.89 (0.11) & 1.54 & -0.43 \\
   ESO325-G004 & 926 & -1.095 (0.073) & -0.209 (0.068) & 0.436 (0.037) & 0.318 (0.027) & 0.625 (0.076) & 2.33 (0.16) & 0.22 & -0.55 \\
   ESO383-G076 & 2806 & -1.510 (0.040) & -0.575 (0.027) & 0.419 (0.020) & 0.368 (0.015) & 0.472 (0.033) & 2.37 (0.10) & -0.21 & -1.05 \\
   ESO444-G046 & 6703 & -1.447 (0.051) & -0.496 (0.044) & 0.479 (0.019) & 0.422 (0.017) & 0.514 (0.048) & 2.11 (0.06) & -0.07 & -0.96 \\
   ESO509-G008 & 899 & -1.590 (0.053) & -0.543 (0.063) & 0.448 (0.027) & 0.374 (0.033) & 0.634 (0.052) & 2.54 (0.14) & 0.45 & -0.96 \\
   IC 4051 & 928 & -1.323 (0.058) & -0.438 (0.042) & 0.460 (0.030) & 0.332 (0.019) & 0.573 (0.053) & 2.21 (0.17) & -0.03 & -0.81 \\
   NGC 1407 & 979 & -1.431 (0.066) & -0.357 (0.019) & 0.505 (0.031) & 0.278 (0.013) & 0.373 (0.029) & 2.63 (0.22) & -0.67 & -0.87 \\
   NGC 3258 & 1633 & -1.791 (0.031) & -0.622 (0.027) & 0.401 (0.015) & 0.376 (0.017) & 0.461 (0.023) & 3.01 (0.10) & -0.20 & -1.22 \\
   NGC 3268 & 1418 & -1.790 (0.042) & -0.645 (0.029) & 0.388 (0.020) & 0.372 (0.018) & 0.406 (0.028) & 3.01 (0.10) & -0.34 & -1.26 \\
   NGC 3348 & 985 & -1.536 (0.084) & -0.362 (0.035) & 0.465 (0.042) & 0.326 (0.021) & 0.370 (0.044) & 2.92 (0.26) & -0.59 & -0.96 \\
   NGC 4696 & 2046 & -1.465 (0.027) & -0.399 (0.018) & 0.435 (0.017) & 0.312 (0.012) & 0.514 (0.020) & 2.81 (0.10) & -0.24 & -0.88 \\
   NGC 5322 & 547 & -1.474 (0.073) & -0.462 (0.046) & 0.408 (0.041) & 0.328 (0.027) & 0.457 (0.057) & 2.74 (0.23) & -0.32 & -0.96 \\
   NGC 5557 & 649 & -1.526 (0.098) & -0.379 (0.042) & 0.496 (0.050) & 0.293 (0.037) & 0.485 (0.060) & 2.81 (0.30) & -0.44 & -0.88 \\
   NGC 7049 & 330 & -1.311 (0.135) & -0.259 (0.103) & 0.420 (0.070) & 0.270 (0.059) & 0.555 (0.102) & 2.94 (0.30) & -0.20 & -0.69 \\
   NGC 7626 & 1510 & -1.472 (0.042) & -0.390 (0.023) & 0.492 (0.021) & 0.286 (0.014) & 0.514 (0.027) & 2.69 (0.13) & -0.38 & -0.85 \\
   NGC 1132 & 1852 & -1.418 (0.036) & -0.357 (0.031) & 0.475 (0.022) & 0.314 (0.015) & 0.637 (0.031) & 2.64 (0.10) & 0.16 & -0.76 \\
   ESO306-G017 & 2521 & -1.162 (0.027) & -0.274 (0.024) & 0.532 (0.012) & 0.278 (0.013) & 0.674 (0.028) & 2.09 (0.07) & 0.08 & -0.56 \\
   NGC 1272 & 2151 & -1.200 (0.069) & -0.275 (0.068) & 0.499 (0.028) & 0.432 (0.027) & 0.588 (0.069) & 1.98 (0.11) & 0.24 & -0.65 \\
   NGC 1129 & 997 & -0.742 (0.068) & 0.167 (0.059) & 0.614 (0.025) & 0.373 (0.028) & 0.651 (0.062) & 1.89 (0.14) & 0.13 & -0.14 \\
   \\
   Mean & &-1.377 (0.044) & -0.404 (0.025) & 0.465 (0.009) & 0.336 (0.008) & 0.556 (0.021)  & 2.45 (0.07) & 0.00 (0.09) & -0.823 (0.040) \\
   
                \hline
\end{tabular}
\label{tab:mdf_params}
\end{sidewaystable*}




\begin{table*}[ht]
\centering
\caption{Additional Global Parameters for the MDFs} 
\begin{tabular}{cccccccc}
  \hline \hline
Galaxy & $\langle${[Fe/H]$\rangle$} & $\sigma$[Fe/H] & Median & Skewness & Kurtosis & IQR & IDR \\ 
(1) & (2) & (3) & (4) & (5) & (6) & (7) & (8) \\
   \hline
   \\
NGC 1275 &   -0.945 (0.012)&  0.631 (0.009)& -0.908 (0.020)& -0.200 (0.048)& -0.610 (0.096)&  0.956 (0.019)&  1.649 (0.024) \\
NGC 1278 &   -1.013 (0.018)&  0.580 (0.012)& -1.046 (0.030)&  0.038 (0.074)& -0.577 (0.149)&  0.873 (0.028)&  1.513 (0.042) \\  
NGC 4874 &   -0.981 (0.009)&  0.618 (0.007)& -0.949 (0.015)& -0.155 (0.036)& -0.703 (0.073)&  0.959 (0.013)&  1.616 (0.020) \\  
NGC 4889 &   -0.970 (0.011)&  0.625 (0.008)& -0.952 (0.022)& -0.107 (0.045)& -0.758 (0.090)&  0.981 (0.017)&  1.652 (0.024) \\  
NGC 6166 &   -0.896 (0.011)&  0.616 (0.008)& -0.856 (0.018)& -0.224 (0.044)& -0.672 (0.088)&  0.968 (0.017)&  1.611 (0.024) \\  
NGC 7720 &   -0.893 (0.010)&  0.642 (0.007)& -0.873 (0.016)& -0.206 (0.039)& -0.558 (0.077)&  0.961 (0.015)&  1.677 (0.026) \\  
UGC9799  &   -0.961 (0.010)&  0.646 (0.007)& -0.919 (0.014)& -0.192 (0.039)& -0.648 (0.079)&  0.994 (0.015)&  1.667 (0.019) \\ 
UGC10143 &   -0.983 (0.011)&  0.593 (0.008)& -0.991 (0.013)& -0.033 (0.044)& -0.442 (0.088)&  0.849 (0.017)&  1.571 (0.026) \\ 
ESO325-G004& -0.763 (0.019)&  0.584 (0.014)& -0.757 (0.025)& -0.075 (0.080)& -0.725 (0.161)&  0.926 (0.030)&  1.524 (0.044) \\ 
ESO383-G076& -1.016 (0.012)&  0.610 (0.008)& -0.967 (0.014)& -0.174 (0.046)& -0.675 (0.092)&  0.949 (0.020)&  1.597 (0.025) \\ 
ESO444-G046& -0.985 (0.008)&  0.656 (0.006)& -0.958 (0.014)& -0.108 (0.030)& -0.647 (0.060)&  0.976 (0.013)&  1.741 (0.016) \\ 
ESO509-G008& -1.207 (0.022)&  0.659 (0.016)& -1.246 (0.033)&  0.102 (0.082)& -0.761 (0.163)&  1.023 (0.033)&  1.740 (0.043) \\ 
IC 4051 &    -0.946 (0.020)&  0.600 (0.014)& -0.926 (0.038)& -0.192 (0.080)& -0.649 (0.160)&  0.913 (0.027)&  1.577 (0.048) \\ 
NGC 1407 &   -0.758 (0.021)&  0.643 (0.015)& -0.555 (0.019)& -0.772 (0.078)& -0.308 (0.156)&  0.926 (0.040)&  1.669 (0.041) \\  
NGC 3258 &   -1.161 (0.017)&  0.700 (0.012)& -1.083 (0.037)& -0.133 (0.061)& -0.988 (0.121)&  1.151 (0.020)&  1.845 (0.038) \\  
NGC 3268 &   -1.111 (0.018)&  0.678 (0.013)& -1.012 (0.030)& -0.246 (0.065)& -0.925 (0.130)&  1.129 (0.023)&  1.796 (0.039) \\  
NGC 3348 &   -0.796 (0.022)&  0.685 (0.015)& -0.612 (0.031)& -0.583 (0.078)& -0.655 (0.156)&  1.030 (0.056)&  1.841 (0.039) \\  
NGC 4696 &   -0.947 (0.014)&  0.655 (0.010)& -0.879 (0.031)& -0.228 (0.054)& -0.893 (0.108)&  1.080 (0.021)&  1.667 (0.025) \\  
NGC 5322 &   -0.925 (0.027)&  0.624 (0.019)& -0.851 (0.052)& -0.272 (0.104)& -0.811 (0.209)&  1.003 (0.038)&  1.643 (0.043) \\  
NGC 5557 &   -0.936 (0.028)&  0.702 (0.019)& -0.787 (0.046)& -0.411 (0.096)& -0.904 (0.192)&  1.165 (0.041)&  1.817 (0.050) \\  
NGC 7049 &   -0.842 (0.035)&  0.639 (0.025)& -0.801 (0.052)& -0.189 (0.134)& -1.148 (0.268)&  1.110 (0.042)&  1.700 (0.056) \\  
NGC 7626 &   -0.946 (0.017)&  0.676 (0.012)& -0.814 (0.033)& -0.380 (0.063)& -0.892 (0.126)&  1.100 (0.025)&  1.796 (0.026) \\  
NGC 1132 &   -1.033 (0.015)&  0.664 (0.011)& -1.042 (0.029)& -0.061 (0.057)& -0.852 (0.114)&  1.061 (0.020)&  1.746 (0.029) \\  
ESO306-G017& -0.872 (0.012)&  0.624 (0.009)& -0.836 (0.021)& -0.293 (0.049)& -0.672 (0.097)&  0.966 (0.019)&  1.639 (0.033) \\ 
NGC 1272 &   -0.819 (0.014)&  0.656 (0.010)& -0.812 (0.017)& -0.029 (0.053)& -0.504 (0.106)&  0.977 (0.019)&  1.698 (0.030) \\ 
NGC 1129 &   -0.457 (0.022)&  0.710 (0.016)& -0.414 (0.032)& -0.284 (0.077)& -0.530 (0.155)&  1.047 (0.035)&  1.866 (0.052) \\  
\\
Mean & -0.971 (0.033) & 0.641 (0.010) & -0.913 (0.049) & -0.226 (0.067) & -0.697 (0.057) & 1.000 (0.029) & 1.678 (0.027) \\

                \hline
\end{tabular}
\label{tab:mdf_params2}
\end{table*}

\begin{figure*}
    \centering
    \includegraphics[width=0.92\textwidth]{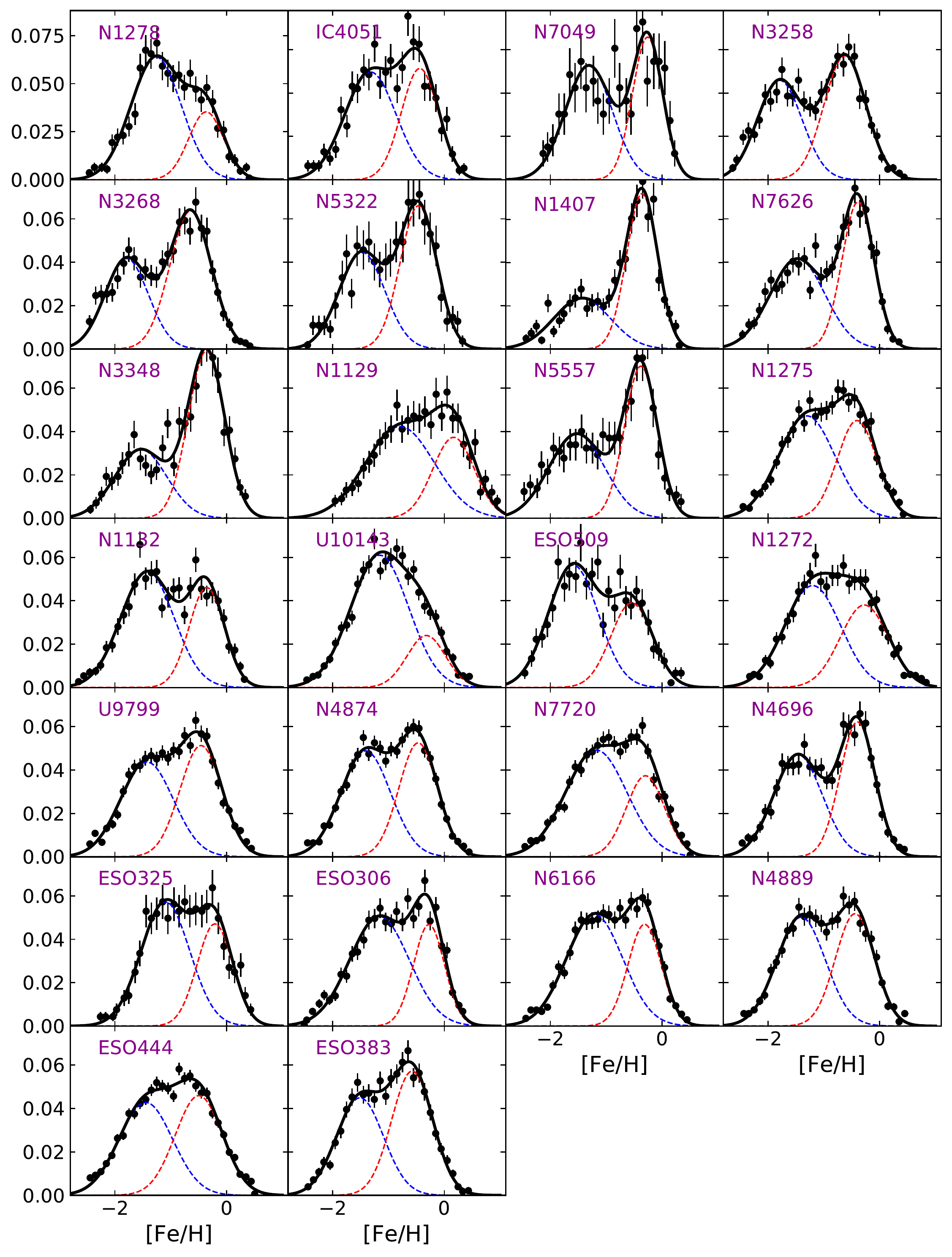}
    \caption{Metallicity distributions (MDFs) for the 26 galaxies in this study,
    displayed in order of increasing luminosity,  left to right along each row and then
    down to the next row.  
    Each panel shows the MDF transformed from the corresponding CDF shown in the
    previous figures.  In each case the best-fit bimodal Gaussian curve is shown
    as the solid line along with its blue and red components (dashed lines).}
    \label{fig:mdf_multi}
\end{figure*}

\begin{figure}
    \centering
    \includegraphics[width=0.48\textwidth]{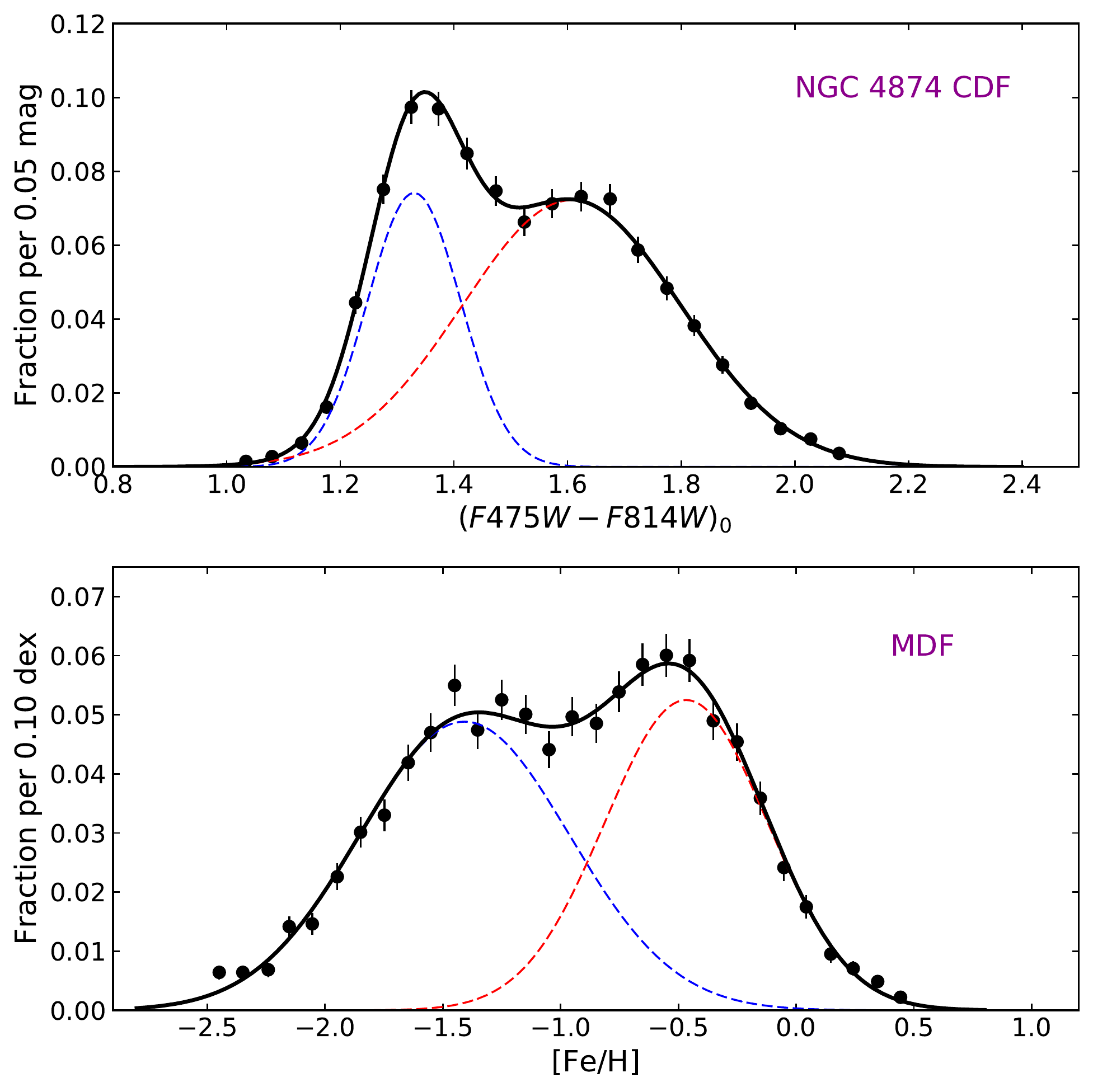}
    \caption{\emph{Upper panel:}  Color distribution function (CDF) for
    the globular clusters in NGC 4874 within the selected box shown in the CMD of Fig.\ \ref{fig:ngc4874_cmd}.  The data binned in 0.05-mag steps are plotted as the histogram, while the best-fit bimodal Gaussian curve is superimposed as the solid line.  The blue and red Gaussian subcomponents are plotted as the dashed lines.  
    \emph{Lower panel:}  Metallicity distribution function (MDF) after transformation of the color indices to [Fe/H] as described in the text. The bimodal-Gaussian fit and its two components are shown as well.}
    \label{fig:ngc4874_cdf_mdf}
\end{figure}

\begin{figure}
    \centering
    \includegraphics[width=0.37\textwidth]{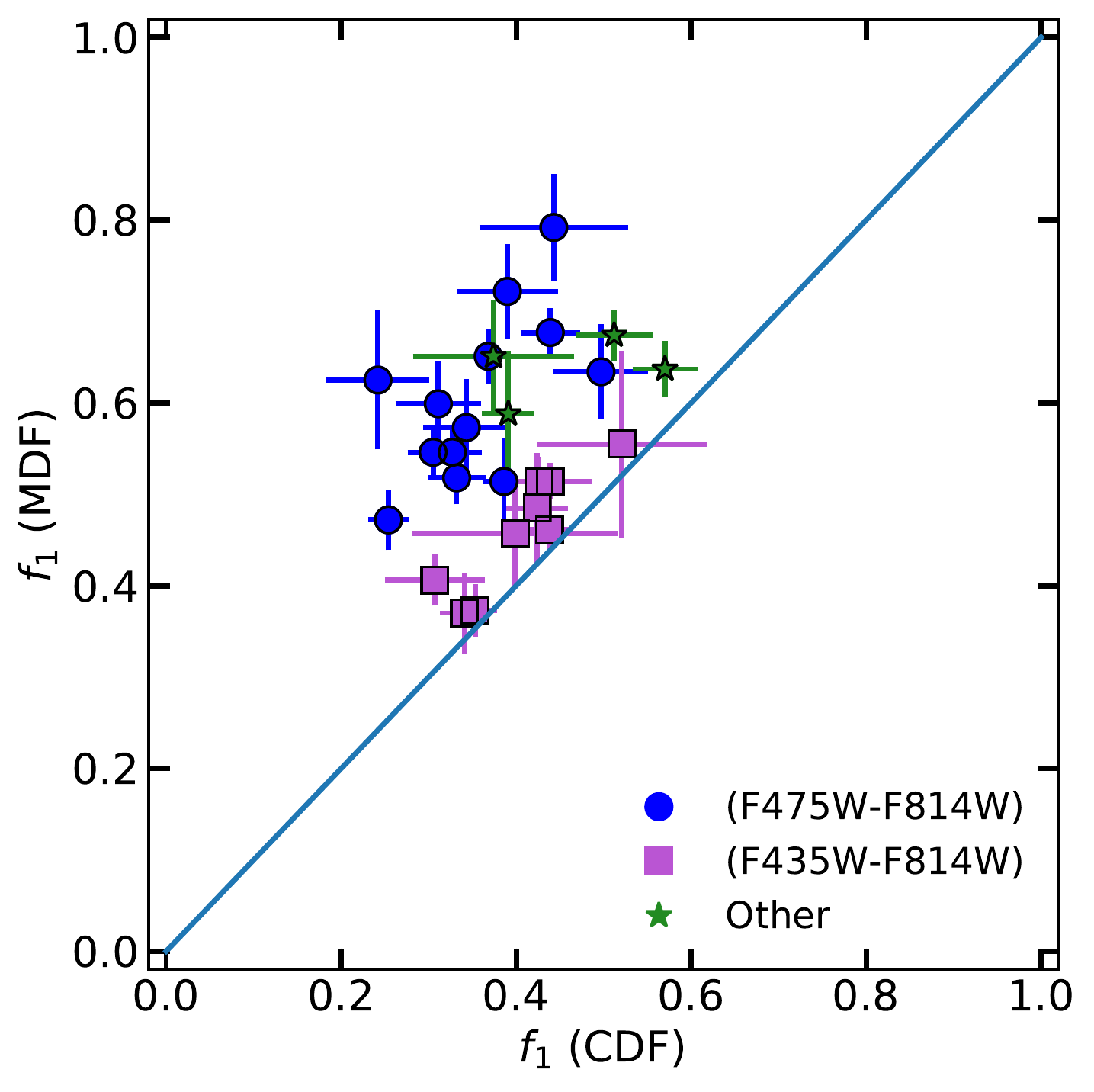}
    \caption{Blue (metal-poor) fraction of GCs from the MDFs plotted versus the same quantity from the CDFs.   Removing the nonlinearity in the color indices shifts the proportions of blue and red clusters estimated by the Gaussian Mixture Modelling fits, such that the blue component in the MDF is broader and more populated than in the CDF.
    }
    \label{fig:f1_compare}
\end{figure}

\begin{figure}
    \centering
    \includegraphics[width=0.48\textwidth]{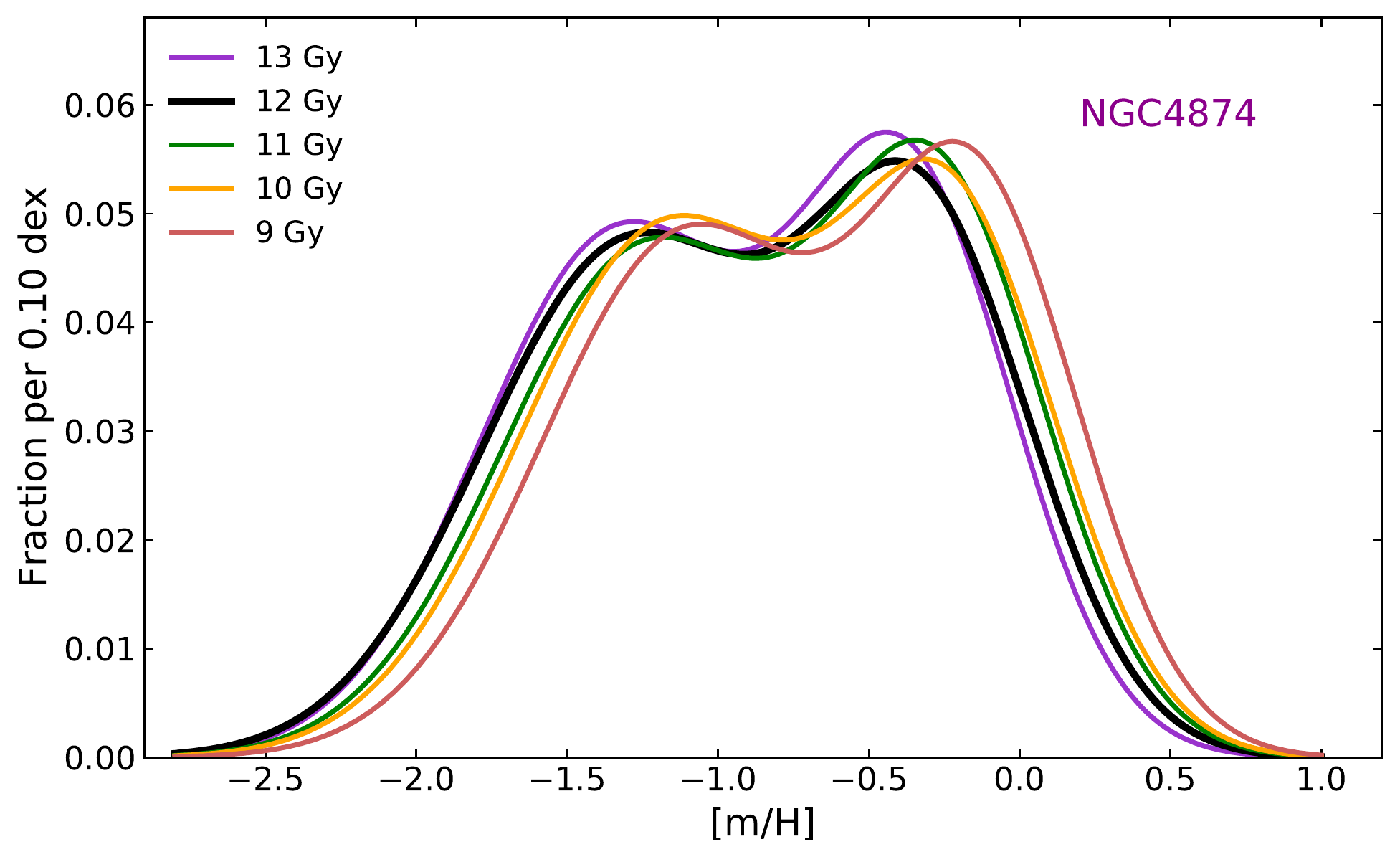}
    \caption{The calculated MDF for NGC 4874, for five different assumed
    ages of the GC population running from 9 to 13 Gy.  All the MDFs are 
    generated from the same observed CDF in $(F475W-F814W)_0$.  
    A smaller assumed age requires a higher metallicity to generate
    the same observed CDF. 
    }
    \label{fig:mdf_age}
\end{figure}

\begin{figure*}
    \centering
    \includegraphics[width=0.84\textwidth]{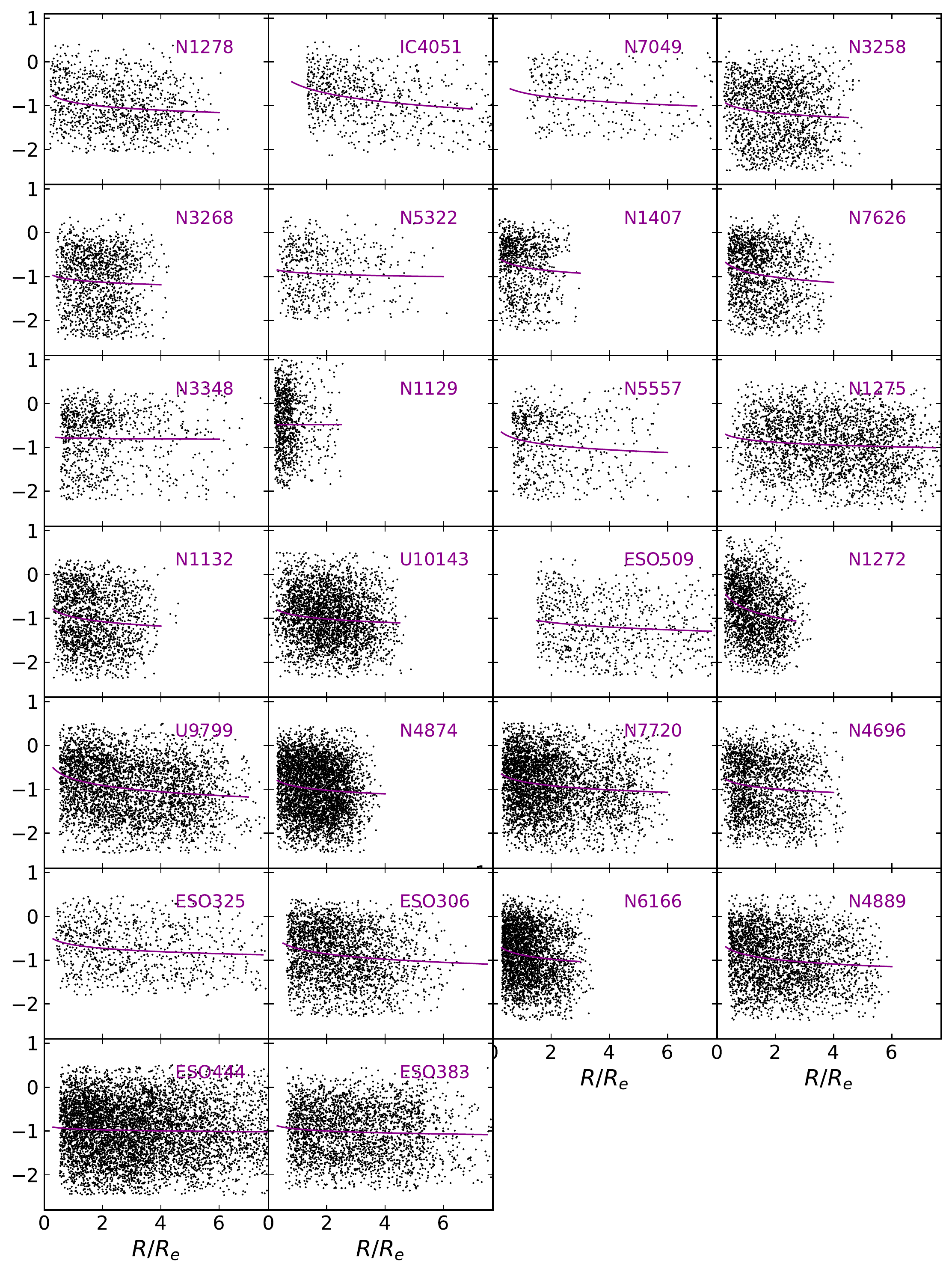}
    \caption{Metallicity versus galactocentric radius for all 26 systems.  The radius of each object is in units of the effective radius $R_e$ of the galaxy light profile.  Solid curves show the least-squares fit to the distributions as described in the text.}
    \label{fig:fehrad}
\end{figure*}

\begin{figure}
    \centering
    \includegraphics[width=0.47\textwidth]{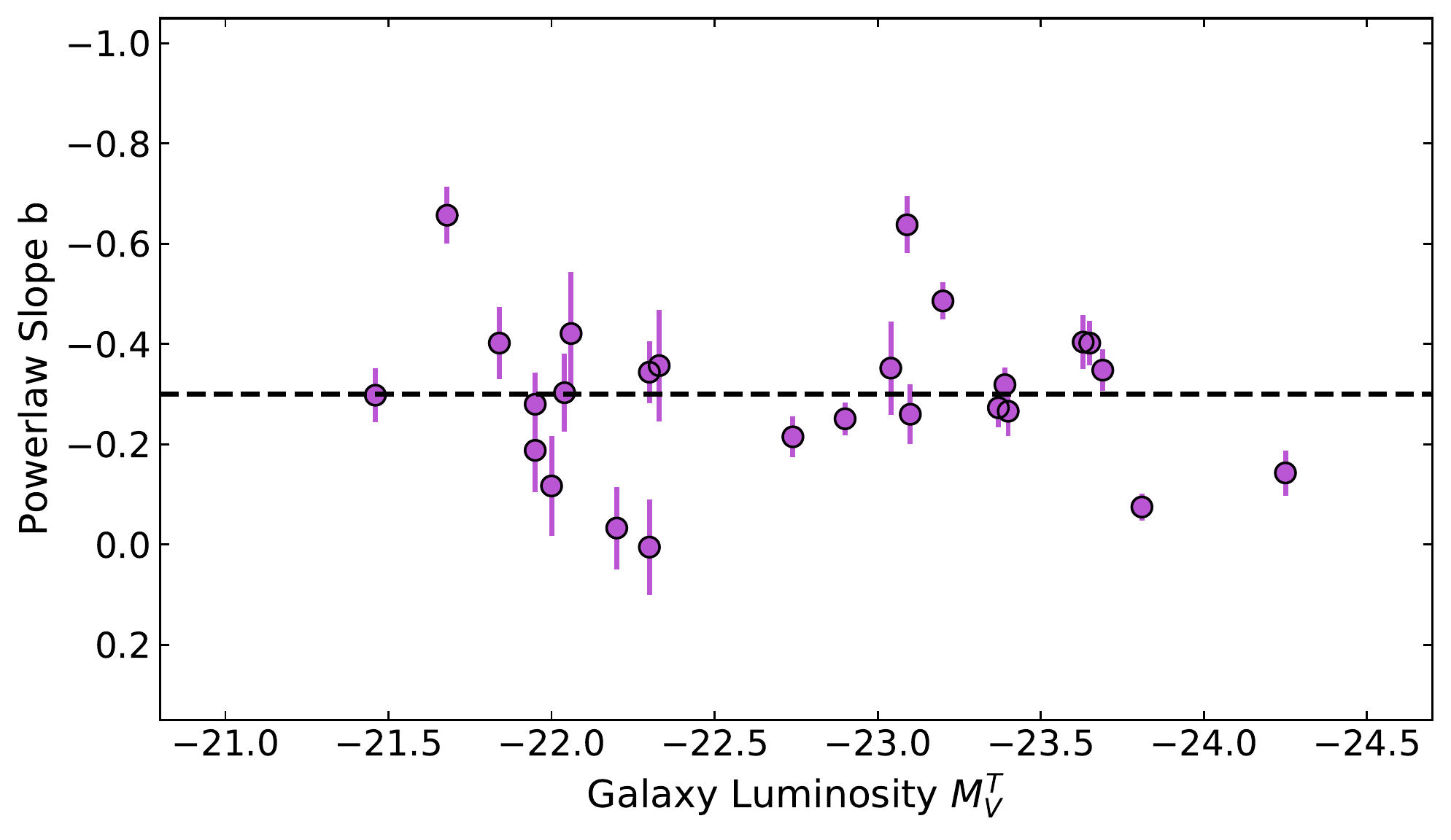}
    \caption{The powerlaw exponent $b$ defining the GC metallicity gradient for each galaxy is plotted versus galaxy luminosity.  The sample mean at $\langle b \rangle = -0.3$ is drawn in as the dashed line.  }
    \label{fig:feh_rad}
\end{figure}

\begin{figure*}
    \centering
    \includegraphics[width=0.85\textwidth]{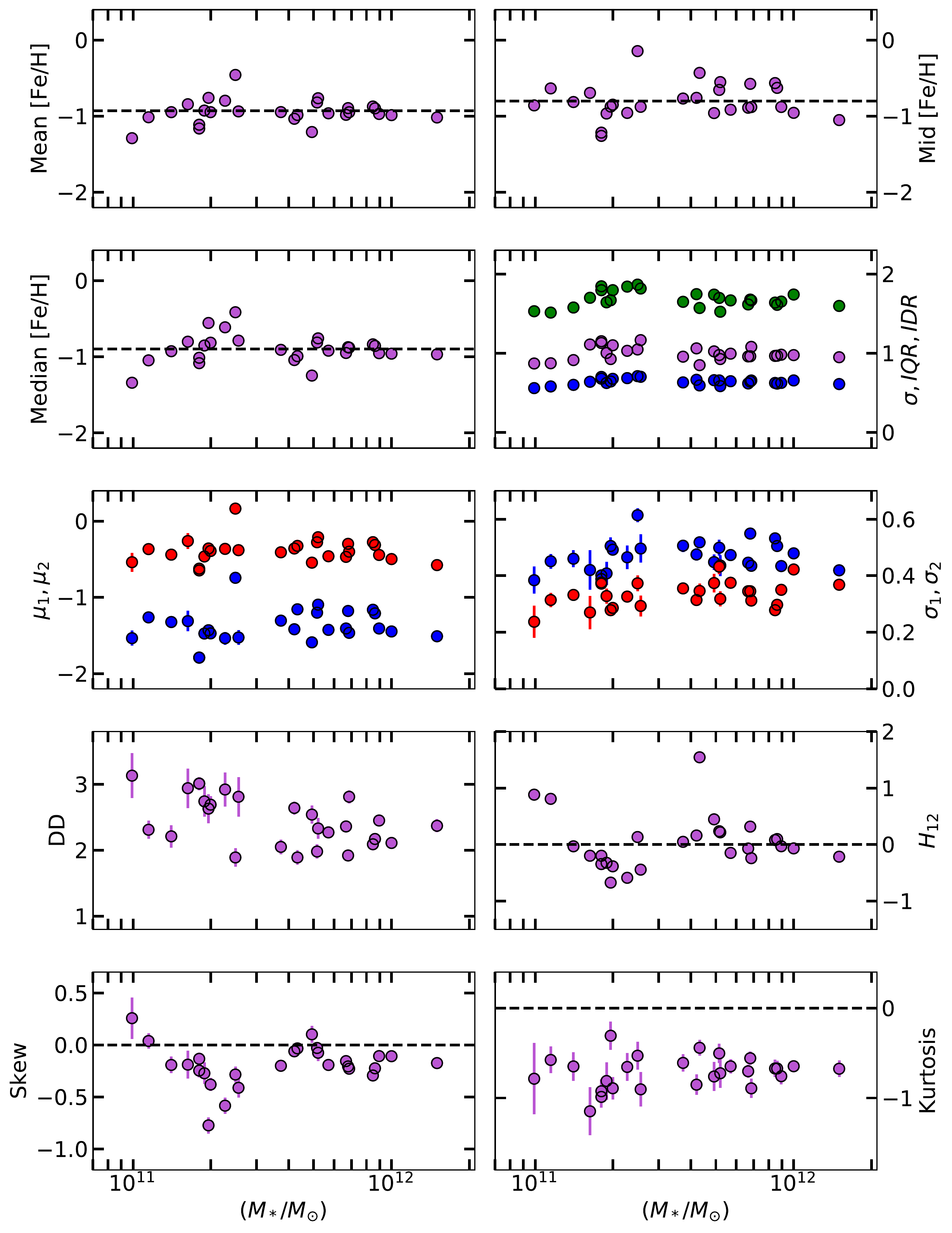}
    \caption{Parameters characterizing the MDFs, plotted
    versus galaxy stellar mass.  Successive panels show: (1) Mean metallicity, (2) [Fe/H]$_{mid}$, the metallicity at which the blue and red modes cross; (3) Median metallicity; (4) Three measures of the global spread of the MDF including the standard deviation $\sigma$ (blue points), InterQuartile Range (magenta), and InterDecile Range (green); (5) the mean metallicities $\mu_1, \mu_2$ of the blue and red modes from the bimodal-Gaussian fit; (6) the standard deviations $\sigma_1, \sigma_2$ of the two modes; (7) Peak separation parameter DD; (8) mode height ratio $H_{12}$; (9) skewness of the MDF; and (10) kurtosis of the MDF.   }
    \label{fig:mdf_params}
\end{figure*}

\begin{figure}
    \centering
    \includegraphics[width=0.48\textwidth]{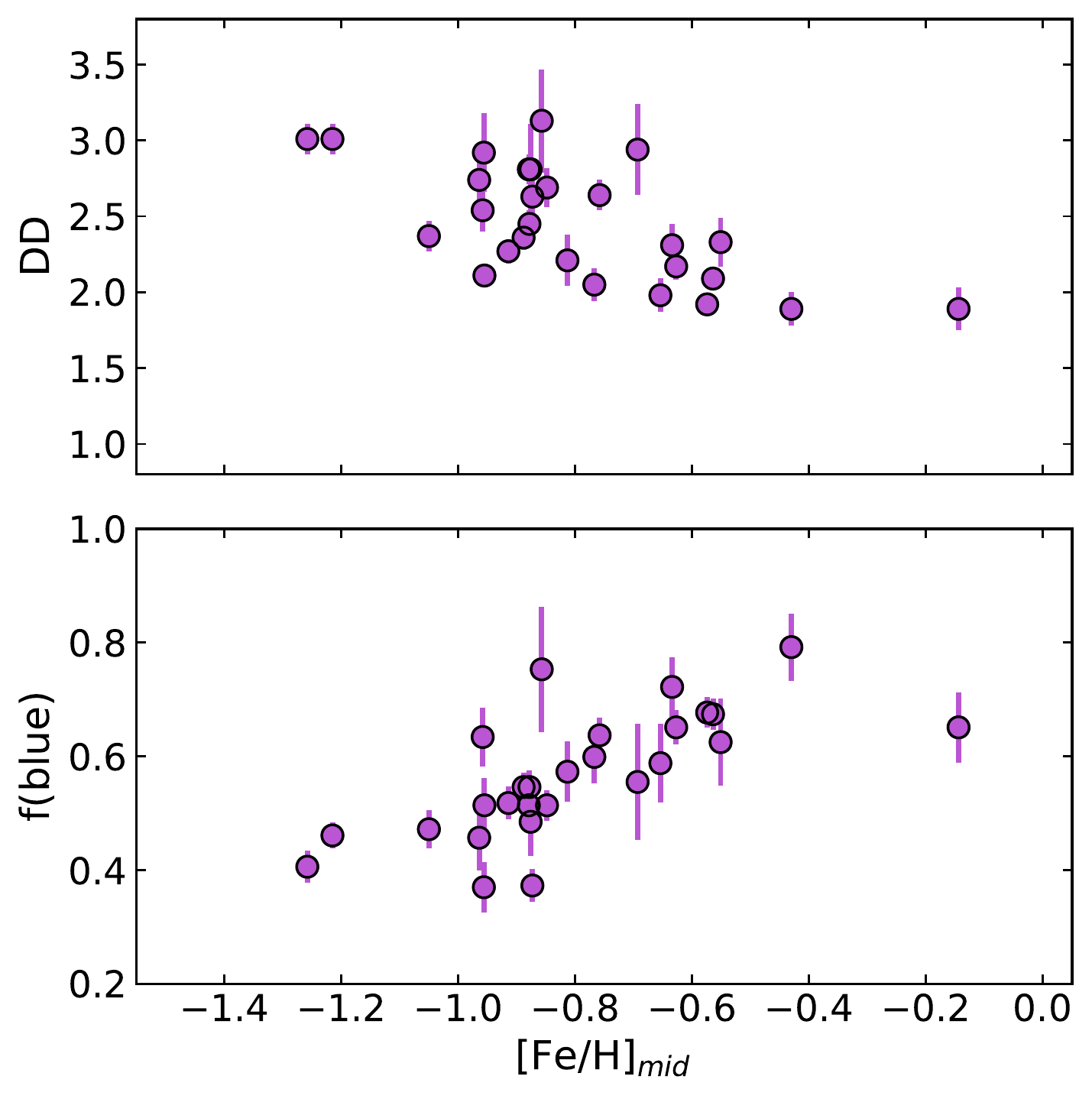}
    \caption{The peak-separation parameter DD (upper panel) and blue-GC fraction (lower panel) are plotted versus [Fe/H]$_{mid}$, the metallicity at which the blue and red mode Gaussian curves are equal. }
    \label{fig:fehmid}
\end{figure}

\begin{figure}
    \centering
    \includegraphics[width=0.44\textwidth]{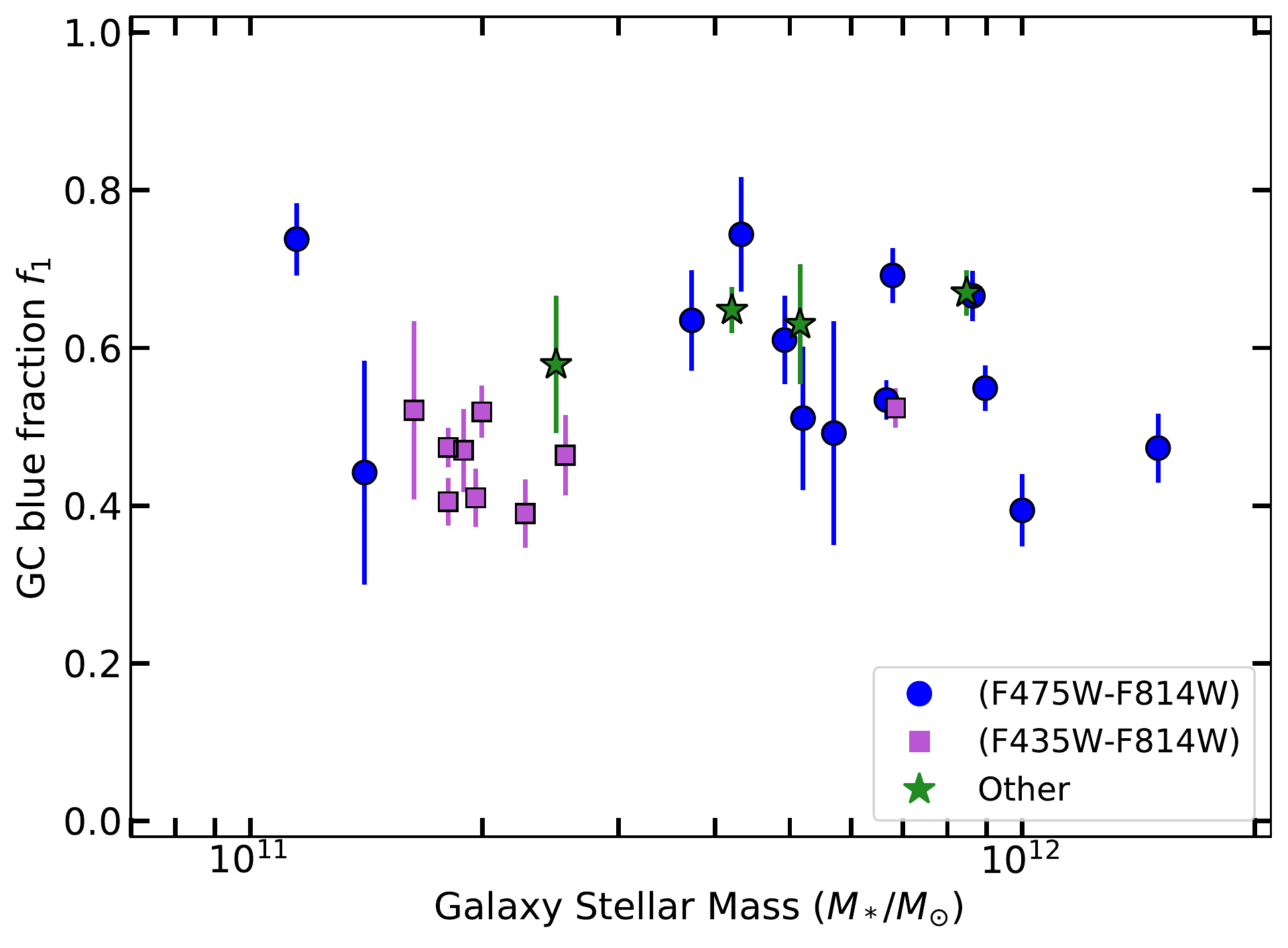}
    \caption{Blue-GC fraction $f_1 = N(blue)/N(tot)$ estimated from the bimodal-Gaussian MDF fitting, plotted versus galaxy stellar mass $M_*$. Here the samples are restricted to the radial range $ 0.5 - 5R_e$ for each galaxy to facilitate comparison.}
    \label{fig:mdf_f1}
\end{figure}

\begin{figure*}
    \centering
    \includegraphics[width=0.80\textwidth]{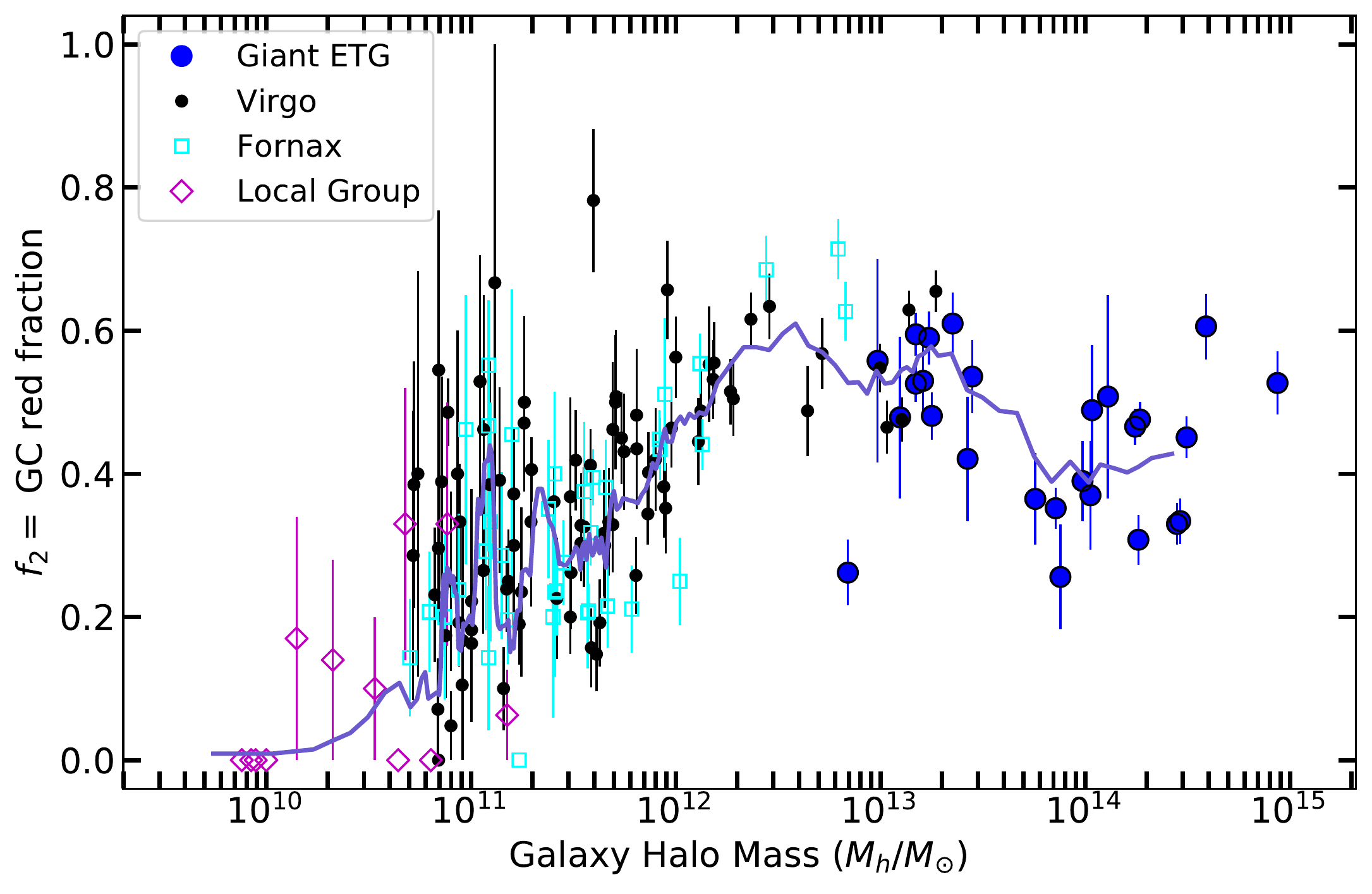}
    \caption{Red-GC fraction $f_2 = N(red)/N(tot)$ plotted versus galaxy halo mass as described in the text.  The giant ETGs studied here are plotted as large blue circles, while the other symbols are galaxies from Virgo, Fornax, and the Local Group that populate the lower-mass range of the diagram.  The solid line shows the trend of the running weighted mean, calculated with nine objects per bin. }
    \label{fig:mdf_fred}
\end{figure*}


\begin{figure*}
    \centering
    \includegraphics[width=0.80\textwidth]{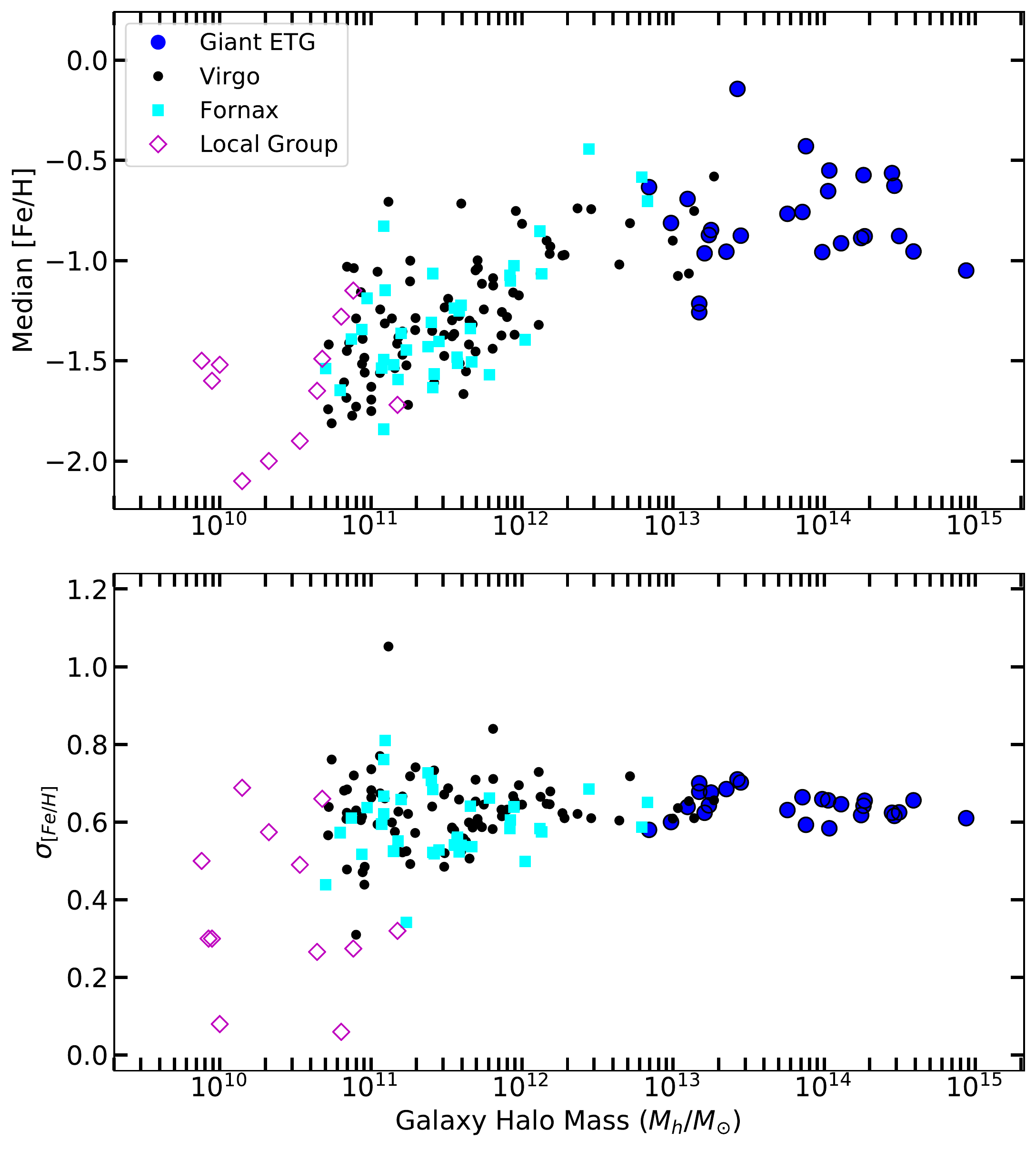}
    \caption{\emph{Upper panel:} Median [Fe/H] for the GC populations. plotted versus halo mass $M_h$.  Symbols are as in the previous figure.  \emph{Lower panel:} The standard deviation $\sigma_{[Fe/H]}$ of the metallicity distribution, plotted versus halo mass $M_h$. }
    \label{fig:mdf_2panel}
\end{figure*}

\section{The Metallicity Distributions}

For each of the 26 galaxies, the dereddened color of every individual GC within the magnitude and color limits specified above was converted to [Fe/H] by inversion of the quadratic relations in Table \ref{tab:coefficients}. From these, the MDFs were generated.  
The final MDFs for all 26 galaxies studied here are shown in Figure \ref{fig:mdf_multi}, where now every galaxy has been put onto a common system for direct comparison.  Here the galaxies are now shown in order of increasing luminosity.  

The CDF and resulting MDF for our template object, NGC 4874, are shown for direct comparison in Figure \ref{fig:ngc4874_cdf_mdf}.  The difference is striking:  both are clearly bimodal in form and a bimodal-Gaussian model fits both distributions equally well, but the numbers of clusters allocated to the `red' and `blue' subgroups are almost the mirror image of each other.  At lower metallicities, the shallower dependence of color on metallicity turns the broad blue component of the MDF into a narrow blue component in the CDF. 

Another way to display the difference between the CDF and MDF is in Figure \ref{fig:f1_compare}.  Here, the fraction $f_1$ of the total GC population assigned by the GMM fits to the metal-poor component in the MDF is plotted versus the same component in the CDF.  In all cases, removal of the nonlinearity in the color indices shifts the best-fit result to a higher blue fraction in the MDF.  It it worth noting that the (F435W-F814W) index (analogous to B-I) is closest to giving 1:1 agreement:  among the color indices used here, it has the widest range in color and is close to being linearly proportional to metallicity especially for [Fe/H] $> -1.5$.  In the consistency test shown in Fig.\ \ref{fig:colorcal2}, it also shows the closest agreement with the colors of the Milky Way GCs.

As seen in Fig.\ \ref{fig:mdf_multi}, many of the individual MDFs are unimodal in the strict sense that there is no deep inflection point at intermediate [Fe/H] dividing the components.  The same is true for the CDFs described above.  Nevertheless, a single symmetric function does not match any of the observed systems, whereas the bimodal-Gaussian model proves to fit the MDFs just as well as it did for the CDFs.  But the two modes for these massive galaxies are much more heavily overlapped than they are in less massive galaxies \citep{peng+2006}.  As pointed out by \citet{choksi_gnedin2019}, these broad MDFs are more of a continuum and the two-component division is to some extent an arbitrary procedure given that GCs are forming and being accreted almost continuously in the growth history of a giant galaxy. 

\subsection{Age Effects}

To test the sensitivity of the MDFs to the assumed GC age, SSP models were generated as described above for ages of 9 to 13 Gyr in 1-Gyr steps, and the CMR recalculated.  The resulting MDFs for the test case of NGC 4874 are shown in Figure \ref{fig:mdf_age}.  If the assumed age is lower, then the GC metallicity needs to be higher in order to yield the same observed CDF.  In this old age range, however, the differences are second-order:  going from 9 to 13 Gyr, the MDF shifts by $\sim 0.2$ dex, but the overall bimodal MDF shape stays much the same.  

In a further stage of modelling, an age/metallicity relation could be built in, such as is observed in the Milky Way \citep{dotter+2011,leaman+2013,usher+2019}, along with some level of age spread at any metallicity.  More extensive modelling of this type will be the subject of future work.

\subsection{Metallicity Gradients?}

At this stage, no selection of GCs has been made by galactocentric radius. In practice this choice has little effect on the MDFs because the radial metallicity gradients in these giant galaxies are shallow.  Figure \ref{fig:fehrad} shows the distribution of [Fe/H] versus radius $R$ normalized to the effective radius $R_e$ of the optical light profile.  In each case a simple least-squares fit was done with the form
\begin{equation}
    {\rm [Fe/H]} = a + b~ {\rm log} (R/R_e)
\end{equation}
with the solutions shown as the solid lines in the panels, and as listed in Table \ref{tab:radial}. 

Metallicity gradients are an outcome of the star formation and merger histories of galaxies.
Because of their high luminosity and easy detectability at all radii, GCs provide a way to trace gradients to much larger radii than is possible with integrated light. Systematic radial changes in mean GC metallicity can arise from intrinsic gradients in the red and blue subpopulations, or as a `population gradient' due to the radial change in the fraction of red and blue clusters, or both  \citep{geisler+1996,harris2009a,liu+2011,forbes_remus2018}.  Either \emph{in situ} star formation or later accretion dominated by low-metallicity dwarfs can leave behind gradients, along with considerable galaxy-to-galaxy differences in the outcomes.

As is seen in Fig.\ \ref{fig:fehrad}, shallow gradients in mean metallicity are present for most of the galaxies in the present sample, though the stronger impression is that of very large scatter of the individual GC metallicities at all radii.   The mean value for the powerlaw exponent across the entire sample is $\langle b \rangle = -0.301 \pm 0.031$ with an rms scatter of $\pm0.16$ and no clear trend with galaxy luminosity. Figure \ref{fig:feh_rad} shows the individual results plotted versus galaxy luminosity. In other words, the mean GC metallicity in these large galaxies scales as $Z \sim R^{-0.3}$ and with significant individual variation.  

\citet{liu+2011} give an extensive discussion of the color and metallicity gradients of the GCSs in 76 Virgo and Fornax Survey member galaxies.  They find that the mean color gradient is $\Delta (g-z)/\Delta {\rm log} R = -0.112$ with rms scatter $\pm 0.077$, which from the \citet{blakeslee+2010} transformation of $(g-z)$ to [Fe/H] (a quartic polynomial form) converts to a mean exponent $\langle b \rangle = -0.387 \pm 0.034$ with rms scatter of $\pm0.28$.  The conversion relation used here for (F475W-F850LP), however, would give a mean $\langle b \rangle \simeq -0.3 \pm 0.02$ with rms scatter $\pm 0.21$, highly consistent with the results here.  Almost all of the Virgo and Fornax members are less luminous than the giants studied here, extending down into the dwarf regime as low as $M_V \sim -16$.  There is some indication that for the smallest dwarfs in their sample the mean gradients become shallower or even absent.  Gradients in the range $b \sim -0.3$ are also found in the SLUGGS sample of galaxies \citep{pastorello+2015}.

\citet{forbes_remus2018} give results for a collated sample of large galaxies and find somewhat smaller slopes in the range $b \simeq -0.1$ to $-0.2$, though these are obtained with a different (linear) transformation from colors.  GC samples built on purely spectroscopic data are more rare \citep[e.g.][]{woodley+2010,pota+2015,pastorello+2015,caldwell_romanowsky2016,villaume+2019,ko+2022} but have generally shown negative gradients in line with the photometrically based ones shown here.  
Future work will discuss more comprehensively the gradients associated with the blue and red subpopulations, and links with contemporary formation models (cf.\ the references cited above).

\subsection{MDF Parameters}

Just as for the CDFs, bimodal Gaussian fits were made to all of the MDFs in Fig.\ \ref{fig:mdf_multi}, with the quantitative results listed in Table \ref{tab:mdf_params}.  In addition to the peak [Fe/H] values $(\mu_1, \mu_2)$ and dispersions $(\sigma_1, \sigma_2)$  of the two subcomponents, the table also includes the fraction $f_1$ (number of blue GCs relative to the total), the mode separation parameter $DD$, the mode height ratio $H_{12}$, and finally the metallicity [Fe/H]$_{mid}$ at which the blue and red modes are equal (i.e.~ the metallicity at which the two Gaussian curves cross).  
The sample means of each quantity are listed at the end of the table.

Table \ref{tab:mdf_params2} includes a further list of other parameters characterizing
the MDFs that do not depend on any particular assumption about the intrinsic
shape of the distribution.  These include the global mean metallicity $\langle$[Fe/H]$\rangle$; the
global standard deviation $\sigma$[Fe/H]; the median metallicity; the skewness
and kurtosis of the MDF\footnote{The definition of kurtosis used here is the ``excess kurtosis'', which
equals zero for a Gaussian distribution.}; and the InterQuartile and InterDecile Ranges (IQR, IDR).

The correlations of most of these parameters versus galaxy stellar mass are shown in Figure \ref{fig:mdf_params}.  The stellar mass is calculated as
\begin{equation}
    M_* = (M/L)_V \cdot 10^{0.4(4.83-M_V^T)}
\end{equation}
where the mass-to-light ratio is adopted as $(M/L)_V = 3.5$ from the calibrations of \citet{bell+2003}, assuming a Chabrier IMF and a mean intrinsic color $(B-V)_0 = 0.9$ for these giant ETGs.

In Fig.\ \ref{fig:mdf_params}, many of the parameters are remarkably uniform across a factor of about 20 in mass range.  These near-constant parameters include the mean, median, and mid-[Fe/H] metallicities, the central locations $\mu_1, \mu_2$ of the Gaussian components and their dispersions $\sigma_1, \sigma_2$, and the measures of the total spread of the MDF ($\sigma$, IQR, and IDR).  The mode dispersions show systematic increases with mass at a barely significant level:  $\Delta \sigma_1 / \Delta {\rm log} M_* = 0.035 \pm 0.028$, and $\Delta \sigma_2 / \Delta {\rm log} M_* = 0.043 \pm 0.027$.

For comparison, for lower-mass galaxies in Virgo \cite{peng+2006} found that the mean metallicities of the  blue and red components increase as $Z_1 \sim M_*^{0.16}, Z_2 \sim M_*^{0.26}$ (albeit derived from a slightly different CMR as discussed above).  The results here for the more massive ETGs suggest that this scaling does not continue upward, instead settling to a more uniform level as expected from recent simulations such as \citet[][particularly their Figure 4]{choksi_gnedin2019} and \citet{el-badry+2019}.

The three measures of MDF total width ($\sigma$, IQR, IDR) are extremely well correlated with each other and essentially equivalent.  Quantitatively, there is also little difference in their ratios of mean value to rms scatter.  The one practical disadvantage of the IDR (or even the IQR) is simply it they will become undefined for dwarf galaxies with very small GC numbers.

There is one obvious outlier in the graphs of mean metallicity and $\mu_1, \mu_2$, which is NGC 1129.  Its transformed [Fe/H] values are more metal-rich than expected by a surprising 0.6 dex, which is equivalent to the colors being too red by more than 0.2 mag.  NGC 1129 is the only CDF measured in $(F435W-F606W)$, for which the predicted CMR was seen in Fig. \ref{fig:colorcal2} to be significantly offset relative to the Milky Way GCs by about the same amount.  The reason for this discrepancy is not clear at present and will require further investigation, and ultimately probably remeasurement in different filters.

Theoretical work predicting the MDFs for GC systems has been quite limited up until recent years but now growing.  The common bimodal shape of the \emph{metallicity} distribution (as opposed to the color distribution) in large galaxies has been shown by several authors to be a natural and frequent outcome of hierarchical growth \citep{cote+1998,muratov_gnedin2010,tonini2013,choksi+2018,choksi_gnedin2019,el-badry+2019,kruijssen+2019,halbesma+2020}.  To first order, high-metallicity GCs are formed \emph{in situ} in the massive potential well of the main progenitor branch while low-metallicity GCs originate \emph{ex situ} within the many metal-poor dwarfs that are accreted throughout its growth history \citep{beasley2020,choksi+2018,forbes+2018,forbes_remus2018,el-badry+2019}.  This is not an exclusive division, though, since blue GCs also form early along the main progenitor branch within the small potential wells that populate the base of the merger tree, and red GCs can be accreted later any time a merger with another large galaxy takes place \citep{choksi+2018}.

One of the simplest tests of recent theoretical predictions is the mean or median metallicity of the entire GC system.  Observationally (Table \ref{tab:mdf_params2}) the data give a mean $\langle$Fe/H$\rangle = -0.97$, and median $\overline{{\rm{[Fe/H]}}} = -0.91$.  A common theme of the models at present seems to be overestimation of the mean, though for differing reasons.
\citet{pfeffer+2018} predict $\overline{{\rm{[Fe/H]}}}$ from the E-MOSAICS simulations almost 0.5 dex higher than the observations give for large galaxies, which they ascribe to insufficient disruption of small disk clusters that have high metallicities.  
\citet{halbesma+2020} show MDFs for their Milky Way and M31 analogs derived from the Auriga simulations that show much individual variation but are often too metal-rich, suggested to be a result of the way the star particles are assigned to GCs or field stars.
The models from \citet{choksi+2018} and \citet{choksi_gnedin2019} built on the Illustris simulations \citep{pillepich+2018} come close at present to matching the data:  they show that when the full effect of merging and accretion is accounted for, the mean metallicity of the GC system increases steadily with galaxy mass up to $M_* \sim 10^{11} M_{\odot}$ and then roughly levels off at $\langle$Fe/H$\rangle \simeq -0.8$, about $0.1-0.2$ dex higher than the observations in Fig.\ \ref{fig:mdf_multi} suggest.  Their definition of GC metallicity, however, uses a theoretical prescription for the mass-metallicity relation between halo mass and its star-forming gas, and so it is not clear if this small offset is important.

The Choksi et al.\ models also predict that the overall dispersion $\sigma_{\rm{[Fe/H]}}$ shows a shallow increase with galaxy mass but again levels off at $\simeq 0.5-0.6$ dex at high mass, whereas the observations in Table \ref{tab:mdf_params2} give $\sigma_{\rm{[Fe/H]}} = 0.64 \pm 0.01$.  As noted above, their predicted metallicities for the red and blue modes considered separately as functions of present-day galaxy mass are also within $\sim 0.1$ dex of the observations.

Finally, two other correlations between MDF parameters are shown in Figure \ref{fig:fehmid}.  These plot the mode separation parameter DD and the blue fraction $f_1$ versus [Fe/H]$_{mid}$, the metallicity at which the blue and red Gaussian components give equal probabilities.  Both show well defined trends that arise from the MDF shape.  If $f_1$ is low, the fact that the component dispersions $\sigma_1, \sigma_2$ stay almost constant means that as [Fe/H]$_{mid}$ shifts to lower metallicity, the sharper red mode stands out more clearly, and the mode separation DD increases. Consequently, the height ratio $H_{12}$ also increases, as can be seen from inspection of Table \ref{tab:mdf_params}. 

\subsection{Trends Over a Wider Mass Range}

One comparison with theory that has interesting new potential for constraining models \citep{choksi_gnedin2019,el-badry+2019} is the correlation shown in Figure \ref{fig:mdf_f1}, the metal-poor GC fraction $f_1$ versus galaxy mass $M_*$.  An advantage of focussing on $f_1$ or $f_2 = (1-f_1)$ is that it should be relatively insensitive to the zeropoint of the metallicity scale from different color indices, as long as the CMR transformation has successfully removed the nonlinearity in the colors.\footnote{The case of NGC 1129 discussed above provides some evidence for this.  It is an obvious outlier in the graphs of $\mu_1, \mu_2$ and mean metallicity, but for $f_1$ and the metallicity dispersions it fits well with the distributions of the other galaxies.}  In Fig.\ \ref{fig:mdf_f1}, to minimize any biases from sampling different parts of the halo in different galaxies, $f_1$ has been recalculated within the restricted radial range of $(0.5 R_e - 5 R_e)$ rather than the entire range of the data (though in practice, constraining the radial range this way has very little effect given the shallowness of the metallicity gradients in all the systems).  

This correlation hints at some intriguing behavior with increasing galaxy mass that is anticipated from current simulations of hierarchical growth.  However, the mass range shown is not much more than a factor of 10, and the numbers of datapoints are still limited.   To gain a broader look at the trend of red/blue fractions over a wider mass range, the BCG sample here has been supplemented by GC metallicity data for smaller galaxies from the Virgo and Fornax ACS imaging surveys \citep{peng+2006,villegas+2010}.  From the photometric catalogs for both these programs \citep{jordan+2009,jordan+2015}, the $(g-z)$ colors for the individual GCs in each galaxy were extracted, the de-reddened indices converted to (F475W-F850LP) = $(g-z)_0 + 0.636$, and then into metallicity through Eq.~\ref{eq:colorcal}.  As in \citet{liu+2019}, only GC candidates with probability $p \geq 0.5$ were adopted to reduce possible sample contamination.  Where the total numbers of GCs were large enough, GMM bimodal-Gaussian fits were then done for each Virgo or Fornax galaxy to extract the same MDF parameters as described above for the BCGs.  For the smaller dwarfs where the numbers of clusters were too low to permit a bimodal-Gaussian fit, the list of GCs was simply divided at [Fe/H] = -1 with lower-metallicity ones counted as `blue' and higher-metallicity ones as `red'.

Finally, to extend the galaxy mass range down to the lowest possible levels, the Local Group dwarfs containing GCs were added, with spectroscopic and photometric data from a range of individual sources \citep{forbes+2018a,forbes2020,colucci+2011,piatti+2018,law_majewski2010,massari+2019,dalessandro+2016,eadie+2022,kruijssen+2019a,dacosta_mould1988,deboer_fraser2016,pace+2021,cole+2017,georgiev+2010,larsen+2014,veljanoski+2013,veljanoski+2015,beasley+2019,wang+2019,caldwell+2017,crnojevic+2016,cusano+2016}.  For these dwarfs the numbers of clusters are too small for GMM solutions, so again the samples were simply divided at [Fe/H] = -1 as above.

The combined sample is plotted in Figure \ref{fig:mdf_fred}.  Here, in order to compare more easily with relevant theory \citep{choksi_gnedin2019,el-badry+2019} the numbers have been recast as $f_2$ (the red-GC fraction) versus galaxy halo mass $M_h$.  Conversion of $M_*$ to $M_h$ has been done through the SHMR (stellar-to-halo mass ratio) prescription in \citet{hudson+2015}, which uses weak lensing as the basis of calibration.  

Fig.\ \ref{fig:mdf_fred}, despite the internal scatter, reveals a complex and non-monotonic overall trend.   For the range between $M_h \simeq 10^{10} M_{\odot}$ and  $3 \times 10^{12} M_{\odot}$, the red fraction scales roughly as $f_2 \simeq 0.25 \cdot {\rm log} (M_h/{10^{10} M_{\odot})}$. 
At higher mass, $f_2$ reverses and declines again, reaching a local minimum of $f_2(min) \simeq 0.4$ at $M_h \sim 10^{14} M_{\odot}$.  Past that point, however, $f_2$ begins to rise again, reaching $\sim 0.6$ for the very most massive BCGs in the observed sample.  The solid line shows the mean trend of $f_2$ versus galaxy mass, calculated as a weighted mean with nine consecutive points per bin. 

In the dwarf regime, very large scatter is present near $10^{11} M_{\odot}$. Some of this spread is due to small-number statistical scatter, but this is also the low-luminosity and low-N range where the relative effects of field contamination would be worst \citep[see the discussions in ][]{jordan+2009,jordan+2015,liu+2019}, since for the Virgo and Fornax dwarfs the numbers of GCs are determined only after subtraction of local number densities of objects that match GCs in size and morphology.  For the still smaller Local Group members the same issue does not come up, since their GCs can be individually identified.

The trend of $f_2$ seen here is expected from current understanding of hierarchical growth.  Figures 6 and 7 of \citet{choksi_gnedin2019} and Figure 9 of \citet{el-badry+2019} indicate the predicted trends.  At first, starting from the smallest dwarfs, $f_2$ increases with increasing halo mass because more massive halos contain more enriched gas, and the mass fraction accreted from satellites  stays relatively low. For these smaller galaxies, their GC population is dominated by the \emph{in situ} component. But at higher mass the accreted mass fraction (mostly from metal-poor dwarfs) rises strongly, 
and eventually accreted GCs dominate the \emph{blue} GC population, forcing the red fraction $f_2$ to decrease again.  Then, at still higher mass, 
accreted GCs also start to dominate the \emph{red} GC population because of mergers with other large galaxies that bring in metal-rich GCs. This effect slows the decrease of $f_2$ or even drives it back upward again depending on the details of the merger history and the total numbers of clusters involved.

More quantitatively, from the simulations the first reversal point at $\sim 3 \times 10^{12} M_{\odot}$ is where 50\% of the metal-poor GC population is from accreted satellites, whereas the second reversal point at $\sim 5 \times 10^{13} M_{\odot}$ is where 50\% of the metal-rich GCs are from accreted systems \citep{choksi_gnedin2019}.  
These two transition masses are quite close to what is seen in Fig.\ \ref{fig:mdf_fred}.  Different individual merger histories will generate scatter in the overall trend, as is evident in the data.  But an additional quantitative point of agreement with the predictions of both \citet{choksi_gnedin2019} and \citet{el-badry+2019} is that $f_2$ reaches a maximum near $\simeq 0.6$ across the entire mass range.  Interestingly, this maximum is reached twice:  once at a few $\times 10^{12} M_{\odot}$ and again beyond $10^{15} M_{\odot}.$


In brief, the trend of red/blue GC fraction (Fig.\ \ref{fig:mdf_fred}) allows us to identify three important mass transition points ($M_0, M_1, M_2$) that connect the observed data with current theory:
\begin{enumerate}
    \item $M_0 \simeq 10^{10} M_{\odot}$, the galaxy halo mass below which no galaxies have metal-rich GCs ($f_2 \rightarrow 0$).  Dwarf galaxies below this limit may have produced some star clusters, but most of their clusters would have been too low-mass to survive to the present day, and the small initial gas mass in the halo was disrupted after feedback from SNe and radiative heating \citep[cf.\ the discussions of][]{kruijssen2019,eadie+2022}.
    \item $M_1 \simeq 3 \times 10^{12} M_{\odot}$, the halo mass at which accreted GCs start to dominate the metal-poor GC component.
    \item $M_2 \simeq 10^{14} M_{\odot}$, the halo mass at which accreted GCs also start to dominate the metal-rich GC component.  $M_1$ and $M_2$ both mark reversal points in the proportions of red and blue GCs.
\end{enumerate}

Figure \ref{fig:mdf_2panel} (upper panel) shows the trend of the observed median metallicity for the same combined sample of galaxies.  For smaller galaxies, the median rises steadily with the increasing presence of red GCs, but levels off past $M_h \sim 10^{13} M_{\odot}$ as the total GC population becomes dominated by accreted systems with a wide mixture of both metal-rich and metal-poor clusters.  Again, this behavior of the mean or median [Fe/H] matches the predictions of the current suites of simulations cited above.

Lastly, Figure \ref{fig:mdf_2panel} (lower panel) shows the rms spread (standard deviation) of the MDF considered as a whole, plotted versus galaxy mass.  This is seen to stay nearly constant at $\sigma \simeq 0.6$ dex for $M_h > 10^{11} M_{\odot}$, scattering to lower values only for some of the lowest-luminosity systems that contain only metal-poor GCs.  The models discussed in \citet[][see their Figure 5]{choksi_gnedin2019} predict a $\sigma$ versus mass trend starting near 0.4 dex at $M_h = 10^{11} M_{\odot}$ and then gradually rising to $\simeq 0.6$ at higher mass.  Thus in this case, the simulations underestimate the observational spread.
\section{Overview and Conclusions}\label{Summary and Following Work}

A summary of the main results of this study is as follows:
\begin{enumerate}
    \item From the HST MAST archive, data for 26 giant ETGs with large globular cluster populations and imaged with the ACS/WFC camera have been reduced homogeneously with the DOLPHOT photometry code.  All of these systems are distant enough that their GCs appear near-starlike, and are imaged with long enough exposure times to cover at least the bright half of the GC luminosity distribution.
    \item From the photometry, the color distribution function (CDF) of the GC population is generated for each galaxy.  A total of 13 galaxies are measured in (F475W-F814W), nine in (F435W-F814W), two in (F475W-F850LP), and one each in (F435W-F606W) and (F555W-F814W).  All the photometry is publicly available.
    \item Bimodal-Gaussian fits prove to be accurate descriptions of every CDF. The only system in this set of galaxies showing evidence of a very young cluster population and recent star formation is the Perseus giant NGC 1275, well known to be an active giant with large amounts of gas.  The final CDFs are shown to be insensitive to corrections for galactocentric radius, photometric completeness, and (for the closest targets) marginal nonstellar morphology of the GCs.
    \item For the most luminous galaxies, the red (metal-rich) GC sequence extends upward to distinctly higher luminosities than does the blue sequence, reaching predicted masses well above $5 \times 10^{6} M_{\odot}$ and into the UCD regime.
    \item The measured GC colors are converted to metallicity [Fe/H] through a through a simple quadratic color-to-metallicity relation, built on a combination of $12-$Gyr SSP models and empirical spectroscopic calibrations of the index (F475W-F850LP) that are in the previous literature.  From the CMRs for every color index,  metallicity distribution functions (MDFs) are generated on a common scale.  Among the six optical/near-IR color indices tested here, the most satisfactory one for these purposes appears to be (F435W-F814W) for its dynamic range in color.
    \item All the program galaxies exhibit shallow metallicity gradients versus projected galactocentric distance:  on average, heavy-element abundance scales as $Z \sim R^{-0.3}$ though with significant individual variation.  This result agrees well with typical gradients found in a wide range of less luminous galaxies.
    \item Bimodal-Gaussian fits also prove to accurately match all the MDFs as well, though with the important difference that the blue (metal-poor) mode is much broader and has a higher proportion of the GC totals in the MDF than in the CDF.  Age differences in the range $9-13$ Gyr will make second-order differences in the [Fe/H] scale, in the sense that a younger assumed age needs slightly higher metallicity to yield the same color.
    \item Several parameters of the MDFs are calculated including the mean and dispersion of the blue and red modes, the fraction of the total in each mode, and several indicators of the total mean, range, and shape of the MDF as a whole.  The intrinsic dispersions of the blue and red modes show shallow systematic increases with galaxy stellar mass $M_*$, but most of the other parameters remain remarkably uniform.
    \item A plot of blue/red GC fraction versus galaxy mass shows interestingly complex behavior that has the potential to put new constraints on current theoretical predictions of hierarchical assembly of these large galaxies.  For $M_h \gtrsim 3 \times 10^{12} M_{\odot}$, the blue GC component is expected to become dominated by accreted systems, driving a decrease in the red-GC fraction $f_2$; but at still higher mass above $\sim 10^{14} M_{\odot}$  the red GCs also become dominated by accreted systems, causing $f_2$ to increase again.   This double reversal in the trend of red/blue fraction versus galaxy mass is now tentatively seen across the full mass range of observed galaxies, with the reversal points in approximate agreement with recent theory.
\end{enumerate}

 One obvious limitation of the discussion here is the assumption of a single age (12 Gyr) for all GCs in the conversion of color to metallicity.  In a next stage of analysis, following what is known about the Milky Way GCs and other nearby galaxies, sample age/metallicity relations can be built in along with some intrinsic scatter.  In addition, different SSP models should be used to test the consistency and reliability of the color transformations used here.

On the observational side, new photometry of the GCs in one or more of these giant galaxies should be carried out that would measure the GCs in the entire set of color indices, not just one or two.  Reliance on the models to define the conversions of one color index to another would be drastically reduced, increasing the confidence in the eventual metallicity distributions.

Much else can be done with the current dataset in other directions.  The radial distributions of the GCs and the GC luminosity distributions can be measured, leading to total GC population estimates, specific frequencies, and a stronger definition of the correlation of total GC system mass versus galaxy halo mass at its high-mass end.   The shape of the MDF can also be traced as a function of GC luminosity (mass) and projected location in the halo.  These directions will be the subject of followup work.



\section*{Acknowledgements}

The complete set of reduced photometry, along with the reference images that define the local xy coordinate systems for each galaxy field, can be obtained at this DOI: 
\dataset[10.5281/zenodo.7342871]{https://doi.org/10.5281/zenodo.7342871}.

The imaging data used in this paper were obtained from the Mikulski Archive for Space Telescopes (MAST) at the Space Telescopes Science Institute.  The spcific observations analyzed can be accessed via these DOIs: 
\dataset[10.17909/8c69-1q86]{https://doi.org/10.17909/8c69-1q86}, 
\dataset[10.17909/36nb-z070]{https://doi.org/10.17909/36nb-z070},
\dataset[10.17909/pvn9-ya02]{https://doi.org/10.17909/pvn9-ya02},
\dataset[10.17909/64m6-1c38]{https://doi.org/10.17909/64m6-1c38},
\dataset[10.17909/bk3t-td27]{https://doi.org/10.17909/bk3t-td27},
\dataset[10.17909/3x7e-fj90]{https://doi.org/10.17909/3x7e-fj90},
\dataset[10.17909/qssg-zb94]{https://doi.org/10.17909/qssg-zb94},
\dataset[10.17909/qafy-dv12]{https://doi.org/10.17909/qafy-dv12},
\dataset[10.17909/fk8b-jv82]{https://doi.org/10.17909/fk8b-jv82}.

The author acknowledges the financial support of NSERC (Natural Sciences and Engineering Research Council of Canada).  For very helpful practical advice during early stages of this work, I am grateful to Pat Durrell, Doug Welch, Guillaume Hewitt, and Andy Dolphin.
\facility{HST (ACS)}
\software{DOLPHOT \citep{dolphin2000}, IRAF \citep{tody1986,tody1993}, Daophot \citep{stetson1987}, ISHAPE \citep{larsen1999}, Astrodrizzle \citep{hack+2012}}

\bibliographystyle{apj}
\bibliography{bcg}

\label{lastpage}

\end{document}